\documentclass[twocolumn]{aastex631}
\usepackage{svg}
\usepackage{bm}
\usepackage{gensymb}
\usepackage{hyperref}
\usepackage{array}
\usepackage{latexsym, graphicx, amssymb, longtable, epsf}
\usepackage{amsmath}
\usepackage{amssymb}
\usepackage{mathtools}
\usepackage{mathrsfs}
\usepackage{threeparttable}
\usepackage{soul}

\newcommand{\STU}{\texttt{ST-U}}
\newcommand{\STPDT}{\texttt{ST+PDT}}
\newcommand{\PDTU}{\texttt{PDT-U}}
\newcommand{\Msun}{$M_{\odot}$}

\begin{document}

\title{A NICER view of the $1.4$\,\Msun\ edge-on pulsar PSR~J0614$-$3329}

\correspondingauthor{Lucien Mauviard}
\email{lucien.mauviard-haag@irap.omp.eu}

\author[0000-0002-3408-2759]{Lucien~Mauviard}
\affil{IRAP, CNRS, 9 avenue du Colonel Roche, BP 44346, F-31028 Toulouse Cedex 4, France}
\affil{Universit\'{e} de Toulouse, CNES, UPS-OMP, F-31028 Toulouse, France}
\author[0000-0002-6449-106X]{Sebastien~Guillot}
\affil{IRAP, CNRS, 9 avenue du Colonel Roche, BP 44346, F-31028 Toulouse Cedex 4, France}
\affil{Universit\'{e} de Toulouse, CNES, UPS-OMP, F-31028 Toulouse, France}

\author[0000-0001-6356-125X ]{Tuomo~Salmi}
\affil{Department of Physics, University of Helsinki, P.O. Box 64, FI-00014 University of Helsinki, Finland}

\author[0000-0002-2651-5286]{Devarshi~Choudhury}
\affil{Anton Pannekoek Institute for Astronomy, University of Amsterdam, Science Park 904, 1098XH Amsterdam, the Netherlands}

\author[0000-0002-9407-0733]{Bas~Dorsman}
\affil{Anton Pannekoek Institute for Astronomy, University of Amsterdam, Science Park 904, 1098XH Amsterdam, the Netherlands}

\author[0000-0001-5848-0180]{Denis~Gonz\'alez-Caniulef}
\affil{IRAP, CNRS, 9 avenue du Colonel Roche, BP 44346, F-31028 Toulouse Cedex 4, France}
\affil{Universit\'{e} de Toulouse, CNES, UPS-OMP, F-31028 Toulouse, France}

\author[0009-0005-8019-0426]{Mariska~Hoogkamer}
\affiliation{Anton Pannekoek Institute for Astronomy, University of Amsterdam, Science Park 904, 1098XH Amsterdam, the Netherlands}

\author[0000-0002-1169-7486]{Daniela~Huppenkothen}
\affiliation{Anton Pannekoek Institute for Astronomy, University of Amsterdam, Science Park 904, 1098XH Amsterdam, the Netherlands}

\author[0009-0008-3894-4783]{Christine~Kazantsev}
\affil{IRAP, CNRS, 9 avenue du Colonel Roche, BP 44346, F-31028 Toulouse Cedex 4, France}
\affil{Universit\'{e} de Toulouse, CNES, UPS-OMP, F-31028 Toulouse, France}

\author[0000-0002-0428-8430]{Yves~Kini}
\affil{Anton Pannekoek Institute for Astronomy, University of Amsterdam, Science Park 904, 1098XH Amsterdam, the Netherlands}

\author{Jean-Francois~Olive}
\affil{IRAP, CNRS, 9 avenue du Colonel Roche, BP 44346, F-31028 Toulouse Cedex 4, France}
\affil{Universit\'{e} de Toulouse, CNES, UPS-OMP, F-31028 Toulouse, France}

\author[0009-0005-7766-5638]{Pierre~Stammler}
\affil{IRAP, CNRS, 9 avenue du Colonel Roche, BP 44346, F-31028 Toulouse Cedex 4, France}
\affil{Universit\'{e} de Toulouse, CNES, UPS-OMP, F-31028 Toulouse, France}

\author[0000-0002-1009-2354]{Anna~L.~Watts}
\affiliation{Anton Pannekoek Institute for Astronomy, University of Amsterdam, Science Park 904, 1098XH Amsterdam, the Netherlands}

\author[0000-0002-5250-0723]{Melissa~Mendes}
\affil{Technische Universit\"at Darmstadt, Department of Physics, 64289 Darmstadt, Germany}
\affil{ExtreMe Matter Institute EMMI, GSI Helmholtzzentrum f\"ur Schwerionenforschung GmbH, 64291 Darmstadt, Germany}
\affil{Max-Planck-Institut f\"ur Kernphysik, Saupfercheckweg 1, 69117 Heidelberg, Germany}

\author[0000-0002-9626-7257]{Nathan~Rutherford}
\affiliation{Department of Physics and Astronomy, University of New Hampshire, Durham, New Hampshire 03824, USA}

\author[0000-0001-8027-4076]{Achim~Schwenk}
\affil{Technische Universit\"at Darmstadt, Department of Physics, 64289 Darmstadt, Germany}
\affil{ExtreMe Matter Institute EMMI, GSI Helmholtzzentrum f\"ur Schwerionenforschung GmbH, 64291 Darmstadt, Germany}
\affil{Max-Planck-Institut f\"ur Kernphysik, Saupfercheckweg 1, 69117 Heidelberg, Germany}

\author[0000-0002-9211-5555]{Isak~Svensson}
\affil{Technische Universit\"at Darmstadt, Department of Physics, 64289 Darmstadt, Germany}
\affil{ExtreMe Matter Institute EMMI, GSI Helmholtzzentrum f\"ur Schwerionenforschung GmbH, 64291 Darmstadt, Germany}
\affil{Max-Planck-Institut f\"ur Kernphysik, Saupfercheckweg 1, 69117 Heidelberg, Germany}

\author[0000-0002-9870-2742]{Slavko~Bogdanov} 
\affil{Columbia Astrophysics Laboratory, Columbia University, 550 West 120th Street, New York, NY 10027, USA}

\author[0000-0002-0893-4073]{Matthew~Kerr}
\affil{Space Science Division, U.S. Naval Research Laboratory, Washington, DC 20375, USA}

\author[0000-0002-5297-5278]{Paul~S.~Ray}
\affil{Space Science Division, U.S. Naval Research Laboratory, Washington, DC 20375, USA}

\author[0000-0002-9049-8716]{Lucas~Guillemot}
\affil{LPC2E, OSUC, Universit\'{e} d’Orl\'{e}ans, CNRS, CNES, Observatoire de Paris, F-45071 Orl\'{e}ans, France}
\affil{Observatoire Radioastronomique de Nançay, Observatoire de Paris, Universit\'{e} PSL, Universit\'{e}
d’Orl\'{e}ans, CNRS, 18330 Nançay, France}

\author[0000-0002-1775-9692]{Isma\"{e}l~Cognard}
\affil{LPC2E, OSUC, Universit\'{e} d’Orl\'{e}ans, CNRS, CNES, Observatoire de Paris, F-45071 Orl\'{e}ans, France}
\affil{Observatoire Radioastronomique de Nançay, Observatoire de Paris, Universit\'{e} PSL, Universit\'{e}
d’Orl\'{e}ans, CNRS, 18330 Nançay, France}

\author[0000-0002-3649-276X]{Gilles~Theureau}
\affil{LPC2E, OSUC, Universit\'{e} d’Orl\'{e}ans, CNRS, CNES, Observatoire de Paris, F-45071 Orl\'{e}ans, France}
\affil{Observatoire Radioastronomique de Nançay, Observatoire de Paris, Universit\'{e} PSL, Universit\'{e}
d’Orl\'{e}ans, CNRS, 18330 Nançay, France}
\affil{Laboratoire Univers et Th\'{e}ories LUTh, Observatoire de Paris, Universit\'{e} PSL, CNRS, Universit\'{e} de Paris, 92190 Meudon, France}

\begin{abstract}
Four neutron star radius measurements have already been obtained by modeling the X-ray pulses of rotation-powered millisecond pulsars observed by the Neutron Star Interior Composition ExploreR (NICER). We report here the radius measurement of PSR~J0614$-$3329 employing the same method with NICER and XMM-Newton data using Bayesian inference. For all different models tested, including one with unrestricted inclination prior, we retrieve very similar non-antipodal hot region geometries and radii. For the preferred model, we infer an equatorial radius of $R_{\rm eq}=10.29^{+1.01}_{-0.86}\,$km for a mass of $M=1.44^{+0.06}_{-0.07} \, M_{\odot}$ (median values with equal-tailed $68\%$ credible interval), the latter being essentially constrained from radio timing priors obtained by MeerKAT. A more complex model, fitting the data equally well, resulted in a consistent inferred radius. We find that, for all different models, the pulse emission originates from two hot regions, one at the pole and the other at the equator. The resulting radius constraint is consistent with previous X-ray and gravitational wave measurements of neutron stars in the same mass range. Equation of state inferences, including previous NICER and gravitational wave results, slightly soften the equation of state with PSR~J0614$-$3329 included and shift the allowed mass-radius region toward lower radii by $\sim 300\,$m, which is compatible with previous analyses to within less than one standard deviation.
\end{abstract}


\section{Introduction} \label{sec:intro}

Neutron stars (NSs) are known to harbor ultra-dense matter in their cores (up to $10^{15}\,\text{g/cm}^3$), whose microscopic state can be constrained by the macroscopic properties of NSs. The Neutron Star Interior Composition ExploreR (NICER, \citealt{gendreau_neutron_2016}) is a fast X-ray timing mission, whose primary goal is the measurement of the radii of fast rotating NSs, known as millisecond pulsars (MSPs), to yield constraints on their inside matter.

MSPs are produced through a recycling process, where an NS accretes matter from a companion star, inducing its spin-up to millisecond rotation periods via transfer of angular momentum \citep{alpar_new_1982}. Such accelerated MSPs have a magnetic field strong enough ($\sim 10^{8-9}\,$G) to cause the acceleration of charged particles in the magnetosphere, which results in complex and pulsed multi-wavelength emission as the NS rotates. In particular, the magnetic field induces an electron-positron particle shower near the magnetic poles of the NS surface. This phenomenon deposits energy at the magnetic polar caps of the NS, which locally heats the surface to $\sim 10^6\,$K, consequently emitting thermal radiation in the soft X-ray \citep{ruderman_theory_1975, arons_pair_1981, harding_pulsar_2001,harding_pulsar_2002} observed as pulsed emission as the NS rotates.

Measurements of MSPs radii are carried out by estimating their compactness through the imprint of their strong gravitational field on the pulse profile from the aforementioned surface emission. Contrary to other kinds of NS emission (e.g., radio, non-thermal hard X-rays or $\gamma$-rays), such surface emission is strongly affected by spacetime curvature due to the vicinity of the emission region to the center of the gravitational potential ($\sim 10\,$km). Other relativistic effects, such as Doppler boosting and angle aberration, also carry information on the radius. In practice, pulse profile modeling (PPM) is used to model these effects and to infer the source parameters, including its radius, and their uncertainties using statistical methods \citep{pavlov_mass--radius_1997,zavlin_soft_1997,bogdanov_nearest_2013,watts_constraining_2019,bogdanov_constraining_2021}. This method has now been extensively tested and used on four different MSPs observed by NICER, with different datasets \citep{riley_nicer_2019,riley_nicer_2021,miller_psr_2019, miller_radius_2021,salmi_radius_2022,salmi_atmospheric_2023,choudhury_nicer_2024,dittmann_more_2024,salmi_radius_2024,salmi_J1231_2024,vinciguerra_updated_2024,hoogkamer_cross-comparison_2025}. Such methods enable precise modeling of the surface emission at the cost of high computation requirements.

Mass and radius measurements of NSs lead to constraints on the equation of state (EOS) of dense nuclear matter by informing about its stiffness \citep{Watts:2016uzu,Chatziioannou:2024tjq}. Around nuclear densities, the EOS is informed by chiral effective field theory ($\chi$EFT, see, e.g., \citealt{Hebeler2013,Tews:2012fj,drischler_chiral_2019,Keller2023,Tews:2024owl}). The electromagnetic wave measurements obtained with PPM, along with gravitational wave (GW) measurements (GW170817, \citealt{abbott_gw170817_2018}; GW190425,  \citealt{abbott_gw190425_2020}), currently favor EOSs on the stiffer side of a $\chi$EFT informed prior distribution (e.g., \citealt{Raaijmakers21,rutherford_constraining_2024,Koehn:2024set}). While nucleonic EOS are the simplest explanation, more exotic EOS families also exist (see \citealt{lattimer_equation_2016,oertel_equations_2017,burgio_neutron_2021} and references therein for examples). However, none of these EOS models are significantly favored by observations, which highlights the need for more numerous and tighter observational constraints.

The MSP analyzed in this study is PSR~J0614$-$3329, an intermediate mass MSP rotating at $318$ Hz discovered from its radio pulsations \citep{ransom_three_2011}. It has also been found to exhibit soft X-ray pulsations, likely of thermal origin, with no sign of non-thermal pulsations as all the pulsed signal was below $\sim 1.5\,$keV \citep{guillot_nicer_2019}. A white dwarf (WD) companion with a hydrogen atmosphere has been discovered \citep{bassa_cool_2016}, allowing for distance estimates using the photometric temperature of the companion. This distance was later confirmed with radio parallax \citep{shamohammadi_meerkat_2024}. Tight constraints on the mass and orbital inclination also exist from strong Shapiro delay caused by the edge-on view of the PSR~J0614$-$3329 system \citep{reardon_gravitational-wave_2023, miles_meerkat_2025}. Such prior knowledge greatly decreases possible degeneracies with PPM and provide important insights to yield a tight and robust radius measurement, which proves more complicated without prior mass and inclination information  \citep{vinciguerra_updated_2024,salmi_J1231_2024}. PSR~J0614$-$3329 has been observed by both NICER and the X-ray Multi-Mirror Mission (XMM-Newton, XMM hereafter), with a deep NICER accumulated exposure of $\sim 1.6\,$Ms. It is hence a good candidate for a new mass-radius measurement using PPM.

In Section~\ref{sec:data} we describe the X-ray data used for this analysis and its extraction. Section~\ref{sec:model} outlines the modeling procedure adopted throughout this analysis. In Section~\ref{sec:inference} we present the results from the inference runs. In Section~\ref{sec:Validation} we assess the validity of our exploratory runs. In Section~\ref{sec:discussion} we discuss the results and their implication on the EOS of dense matter. We conclude in Section~\ref{sec:conclusion} and summarize our findings.

\section{X-Ray Event data} \label{sec:data}

This section summarizes the reduction of the X-ray data used for inference. All the data products used for this analysis can be found in the Zenodo repository doi: \href{https://doi.org/10.5281/zenodo.15603405}{10.5281/zenodo.15603405} \citep{zenodo}.

\subsection{NICER} \label{subsec:NICER_data}

This analysis of PSR~J0614$-$3329 makes use of $\sim$1.6\,Ms of raw NICER data acquired in the 2018 March 3rd -- 2023 May 15 period (OBSIDs $0030050101$ to $6030050328$), comprised of all the observations of this source antecedent to the NICER light leak\footnote{\href{https://heasarc.gsfc.nasa.gov/docs/nicer/analysis_threads/light-leak-overview/}{https://heasarc.gsfc.nasa.gov/docs/nicer/analysis\_threads/light-leak-overview/}}. The few available observations dating after the light leak occurrence ($\sim$19\,ks) are characterized by a high noise level caused by optical loading and were not included due to the difficulty of properly screening them alongside their short additional exposure. 

The screening procedure, to remove time intervals with high background, generally follows the one adopted in \cite{Bogdanov2019a} and by the other NICER PPM analyses listed above. Calibration was handled by the \textsc{Heasoft} software \citep{2014ascl.soft08004N}, version \textsc{6.33.2}, which includes \textsc{NICERDAS}v12 along with the calibration file \texttt{xti20240206}. Initial screening was produced with the \texttt{nicerl2} task. Unlike previous analyses, no \textit{``hot"} focal plane modules (FPMs) were consistently screened out since, as part of the \texttt{niautoscreen} task, they are now individually removed when exhibiting count rates $3\sigma$ above the median of active FPMs. The final ancillary response file (ARF) is automatically scaled down accordingly, based on the average number of FPMs activated during the whole exposure. This allows to extract more events while limiting background.

Good time intervals (GTIs) were further selected with the \texttt{psrpipe} script from \textsc{NICERsoft}\footnote{\href{https://github.com/paulray/NICERsoft}{https://github.com/paulray/NICERsoft}}, by removing time intervals which included any of the following:
\begin{itemize}
    \item High optical photon flux causing considerable FPM reset: median undershoot rate $\ge 200\,$cts/s.
    \item High particle flux causing excess high-energy detections: overshoot rate $\ge 1.5\,$cts/s/FPM.
    \item Strong solar activity, when the geomagnetic storm index ($K_p$ index) is high: $K_p \ge 5$.
    \item Spacecraft in a region of low shielding to particle flux, i.e., with low cut-off rigidity:\\ COR\_SAX $\le 1.5\,$GeV/$c$.
    \item Strong remnant background after having applied previous criteria: count rate $\ge 1.0\,$cts/s for $16\,$s bins in the $2-10\,$keV band. As we do not expect any source count in this energy range, this is a good proxy for background count rate.
\end{itemize}
These values are chosen to be similar to previous NICER MSP analyses. This procedure resulted in a final filtered dataset with $1.090\,$Ms of effective exposure time.

\begin{figure}[!b] 
\includegraphics[width=\linewidth]{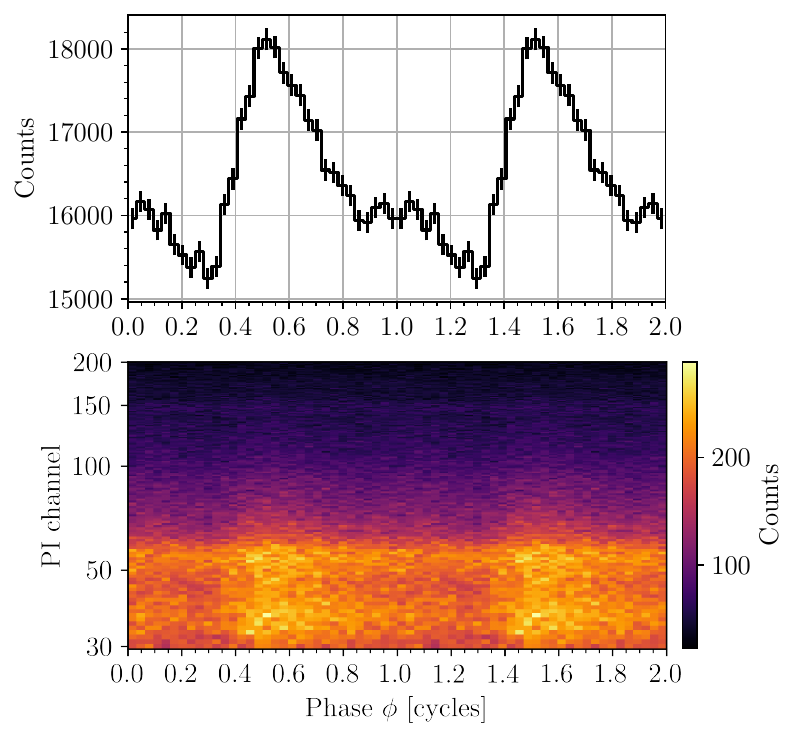}
\caption{Reduced and phase-folded NICER event data for PSR~J0614$-$3329 duplicated over two rotational cycles for visualization purpose. The top panel displays the $0.3-2.0$\,keV bolometric pulse profile with its associated Poisson uncertainty. The bottom panel shows the energy resolved pulse profile. Phases are separated into $32$ regular sized bins and energy by registered PI channel, as used for the inference. Energy resolved pulse profiles in previous publications had their number of counts per bin divided by the number of cycles shown, which is not the case here.}
\label{fig:J0614_NICER}
\end{figure}

Extracted events were phase-folded with the \textsc{PINT} task \texttt{photonphase} \citep{luo_pint_2021}, using a joint Nançay Radio Telescope (NRT) and Fermi timing solution spanning the entire NICER observation window, to produce a phase-resolved spectrum (Figure~\ref{fig:J0614_NICER}). The resulting bolometric pulse profile has a strong main peak and a weaker secondary peak, which is consistent with previous findings \citep{guillot_nicer_2019}. Finally, the associated ARF and response matrix file (RMF) were extracted using \texttt{nicerl3-spect} task.

The analyses were performed on pulse invariant (PI) energy channels $[30,200]$, corresponding to an energy band of $0.3-2.0\,$keV. While the lower energy bound is motivated by the NICER effective area sudden drop as well as residual optical photon flux background below $0.3\,$keV, the upper bound is motivated by the low source flux and absence of pulsed emission over $2.0\,$keV, making the total flux background dominated above this energy. A Kolmogorov-Smirnov test comparing the phase distribution of photons with energy over $2.0\,$keV to a uniform distribution did not find any significant deviation ($p\simeq0.71$). All inferences pointed toward a contribution of the source to the total NICER flux lower than $2\%$ at $2.0\,$keV (see Section~\ref{sec:inference}), validating the choice of this upper limit. 

The NICER phase-averaged spectrum shows two prominent features near $0.35\,$keV and $0.57\,$keV respectively (see Section~\ref{subsec:Headline}). Both are phase-independent peaks, the latter appearing mostly in observations after mid-2022. We identify it as a fluorescence line of oxygen~VII at $577\,$eV in the Earth atmosphere, which is excited by the stronger solar activity at later dates. 
This line also already appeared prominently in NICER spectra of some analyses \citep{salmi_radius_2024,salmi_J1231_2024} and subtly for other analyses that did not include any observations after the solar activity started rising (mid-2022), and for which the source flux was higher \citep{choudhury_nicer_2024,vinciguerra_updated_2024}.

MSP surface emission models cannot explain such time-independent narrow spectral features. Hence, such an excess is accounted for as a phase-independent background, which is fitted alongside the modeled pulse profile (more details in Section~\ref{subsec:Background}). We ensured that these spectral features were well reproduced by the fitted phase-independent background by inspecting the inferred background for all runs. Such peaks being also inferred as background in previous analyses further supports that PPM accounts for them in the phase-independent background, even in cases where the O~VII line was less prominent \citep{vinciguerra_updated_2024}.

Although NICER background models such as Space Weather (SW) or 3C50 \citep{remillard_empirical_2022} exist, none has been used in this study and the NICER background is inferred directly from the data. This approach is facilitated by the available XMM observations, which inform on the phase-averaged source spectrum and flux, which in turn inform the NICER expected source spectrum, and hence its background spectrum. \cite{salmi_radius_2022} has shown that including background constraints from XMM for PSR~J0740$+$6620 gives comparable results to using SW or 3C50 without relying on any of the assumptions that either of them makes. Moreover, 3C50 relies on observations from before mid-2020, when the O~VII line was not as prominent, which can lead to underestimation of the contribution of this background component to the flux. As one XMM observation was available, we decided to use it instead of one of the background models.

\subsection{XMM-Newton} \label{subsec:XMM_data}

The XMM data for this source consist of a single observation taken on 2010 October 4 (OBSID $0653190101$) with the imaging mode. All EPIC instruments were used (PN, MOS1, MOS2), each providing phase-averaged source and background spectra. These were extracted using the XMM Science Analysis Software \textsc{SAS} version \texttt{21.0.0}. The data were screened with the recommended filters (FLAG $=0$ and PATTERN $\le4$ for PN and PATTERN $\le12$ for MOS1/2). No flaring particle background occurred during the observation and the entire exposure was kept. This resulted in a final 11.36~ks effective exposure for PN and 14.42~ks for MOS1/2. 

\begin{figure*}[t]
    \centering
    \includegraphics[width=\linewidth]{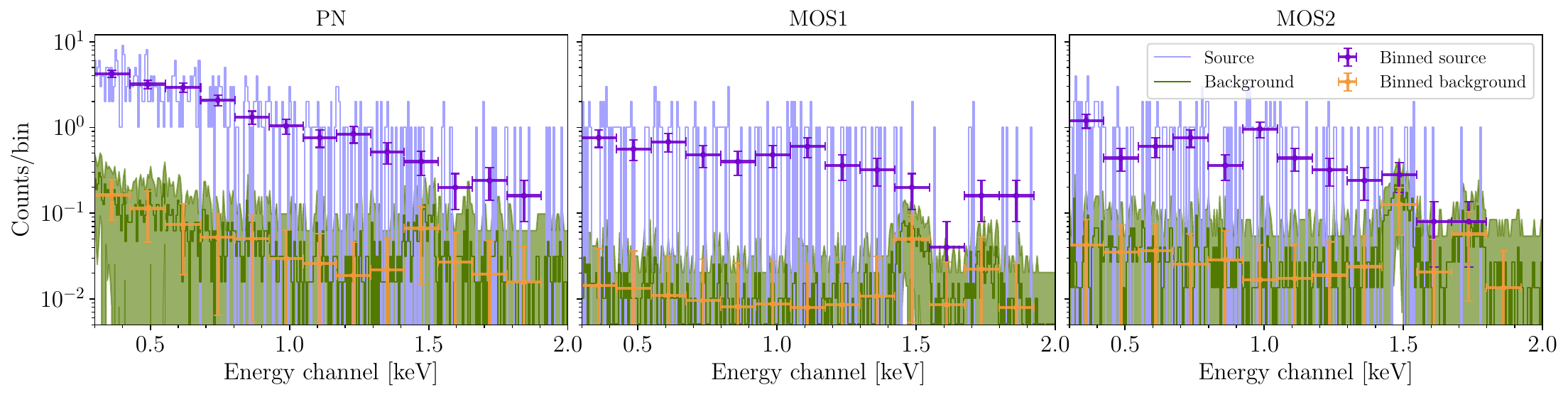}
    \caption{Extracted source (blue) and background (green) spectra from all EPIC instruments used for inference. The green shaded area represents the uniform prior range for the background ($\pm 3 \sigma$, see Section~\ref{subsec:priors}). Backgrounds are rescaled using the source and background \textsc{backscal} values. For visualization purpose only, the rescaled source and background spectra grouped in 13 uniformly sized bins are shown, respectively, in purple and orange.}
    \label{fig:EPIC_countSpectra}
\end{figure*}

The source spectra were extracted from a $\sim$35\arcsec\ circular region, while the backgrounds were extracted from a large $\sim$35-550\arcsec\ annulus region centered on the source, and limited to the same charge-coupled device. Other X-ray sources were excluded and low vignetting regions avoided. The exact regions differ between EPIC instruments. The \textsc{backscal} parameter and respective exposure times written in the processed observation files were used to scale the background spectra to the sources spectra. The ARF and RMF were extracted using the \texttt{arfgen} and \texttt{rmfgen} tasks.

For our analyses, we used the XMM counts within the same $0.3-2.0\,$keV energy band as for NICER (Figure~\ref{fig:EPIC_countSpectra}). Moreover, the observed source counts fall to background level at around 2.0\,keV, making the flux above this energy undistinguishable from background.

\section{Modeling} \label{sec:model}
The modeling procedure as a whole is similar to that of \cite{riley_nicer_2021,salmi_radius_2022,choudhury_nicer_2024,salmi_radius_2024,vinciguerra_updated_2024}. This section summarizes the PPM, the parameter inference method and adopted settings. All relevant files for modeling and inference are provided in the Zenodo repository \citep{zenodo}.

\subsection{X-PSI settings} \label{subsec:xpsi}

We model the energy dependent pulse profile using the X-ray Pulse Simulation and Inference package (X-PSI, \citealt{xpsi}) \texttt{v2.2.7}. For post-processing of the inference runs only, we used X-PSI \texttt{v3.1.0}.

The resolution settings of X-PSI (see Appendix~\ref{app:compare_resolution} for explanations on the different settings) have below $\sim$1\% effects on the computed waveform for simple geometries and can reach a few percent under extreme cases with thin and elongated spot structures \citep{choudhury_exploring_2024}. Due to the limited count statistics of our dataset, we deemed such complex geometries to be difficult to distinguish from simpler ones. Hence, these were not investigated during our exploration runs which were performed using the low resolution (LR) X-PSI settings from \cite{choudhury_exploring_2024}. This allowed a lower computational cost and the use of a better sampling resolution instead. However, our headline result uses higher X-PSI resolution to compute accurate waveforms for those extreme cases and mitigate the effects of a limited X-PSI resolution on the sampling procedure (see Appendix \ref{app:compare_resolution}). These effects are further investigated in Section~\ref{sec:Validation}.

\subsection{Sampling} \label{subsec:sampling}

The posterior distributions of model parameters were obtained with nested sampling \citep{skilling2004nested}. Among other Bayesian inference algorithms, nested sampling has the advantage of providing thorough sampling over a large prior parameter space and easy model comparison with limited number of hyper-parameters. In practice, sampling was carried out using \textsc{PyMultiNest} \citep{PyMultiNest}, a Python wrapper of the nested sampling implementation \textsc{MultiNest} \citep{MultiNest_2009}. Although \textsc{UltraNest} \citep{buchner_ultranest_2021} is in theory more robust, it is also much slower and performed equally well as \textsc{MultiNest} for PPM on PSR~J0740$+$6620 \citep{hoogkamer_cross-comparison_2025}. We hence opted to use \textsc{MultiNest} for its lower computational cost.

All exploratory runs referenced here used the same set of hyper-parameters, chosen to ensure good sampling of the parameter space, while limiting computation time and allowing for easy and fair comparison between runs. \textsc{MultiNest} has three hyper-parameters, which were chosen to be $10^4\,$ live points (\textsc{LP}), a sampling efficiency (\textsc{SE}) of $0.1\,$ and an evidence tolerance (\textsc{ET}) of $0.1$ for the exploratory runs (see \citealt{MultiNest_2009} for a description of these hyper-parameters). We ran all inferences with multi-modal exploration enabled. This allocates distinct sampling regions to the different modes identified, allows for individual post processing of the latter, and may speed-up the computations \citep{MultiNest_2009}. Such hyper-parameters grant a more complete sampling (for exploratory runs) than in previous studies. Our headline result was run with $2 \times 10^4\,$ \textsc{LP}, $0.05$ \textsc{SE} and $0.1$ \textsc{ET}, which allows for a higher sampling accuracy. The exploration of the parameter space completeness is further discussed in Section~\ref{sec:Validation}.

\subsection{Priors} \label{subsec:priors}

Unless stated otherwise, we settled on the following common prior configuration for all exploratory runs. The prior equatorial radius samples are drawn from a uniform distribution $R_{\rm eq} \sim \mathcal{U}(6,16)\,$km, with the surface gravity and compactness conditions rejecting samples with high compactness (see Appendix~\ref{app:headline}). This prior is different than those in previous analyses, which drew samples down to $\sim 4.4\,$km.

\cite{bassa_cool_2016} discovered a WD companion to PSR~J0614$-$3329. Its observed temperature from optical photometry resulted in an estimated distance $D$ in the range $[540,630]$\,pc, consistent with radio timing parallax measurements ($D=670^{+250}_{-140}$\,pc, \citealt[][estimated values and uncertainties reported here and all throughout our work are median values with $68.3\%$ equal-tailed credible intervals]{shamohammadi_meerkat_2024}). For our inference, we used a distance prior following a normal distribution $D\sim \mathcal{N}(\mu_D=585\,\rm{pc},\sigma_D=45\,\rm{pc})$, corresponding to the optical photometry constraint as a $1\sigma$ range, truncated at $\pm5\sigma$. 

Tight constraints on mass and inclination were obtained by MeerKAT from radio timing thanks to the detection of a strong Shapiro delay \citep{miles_meerkat_2025}. These resulted in a mass and orbital inclination of, respectively,\footnote{These values are derived from the radio solution provided in the $4.5$-years MeerKAT PTA data release.} $M=1.44 \pm 0.07\, M_\odot$ and $\sin(i)=0.99955927\pm0.00011665$, used in our inference as Gaussian priors, with the sine of the inclination truncated at one. These are consistent with the measurements from \cite{reardon_gravitational-wave_2023}. As in previous X-PSI analyses, we considered the spin axis of PSR~J0614$-$3329 to be aligned with its orbital momentum, which is expected to be the case at the end of the recycling process \citep{Yang_inclination_2023}. The unconstrained inclination hypothesis was also tested and resulted in similar posterior distributions (see Section~\ref{subsec:ST+PDT}).
 
The interstellar absorption was modeled with the \texttt{Tbabs}\footnote{\href{https://pulsar.sternwarte.uni-erlangen.de/wilms/research/tbabs}{https://pulsar.sternwarte.uni-erlangen.de/wilms/research/tbabs/}} model, depending only on the neutral hydrogen column density parameter $N_{\rm H}$. The $N_{\rm H}$ value from \cite{hi4pi_collaboration_hi4pi_2016} in the direction of PSR~J0614$-$3329 is $N_{\rm H} \sim 3 \times 10^{20}\text{cm}^{-2}$. This value is possibly overestimated as it is integrated along the whole line of sight. Using the relation between radio dispersion measure (DM) and $N_{\rm H}$ from \cite{he_correlation_2013} with $\rm{DM} = 37.054(2) \, \rm{cm}^{-2}$ yields $N_{\rm H} \simeq 10^{21}\text{cm}^{-2}$, but this relation is empirical and has a large scatter associated with it. Since there are uncertainties on those measurements, we adopted a wide uniform prior of $N_{\rm H} \sim \mathcal{U}(0.001,10)\times 10^{20}\text{cm}^{-2}$.

\subsection{Atmosphere model}
Knowledge of the atmosphere of the NS is also crucial for PPM, as the inferred radius and geometry can depend on the atmospheric composition \citep{salmi_atmospheric_2023}. The hydrogen atmosphere of the DA spectral type WD companion, from which the NS likely accreted material to spin-up, heavily favors the hydrogen atmosphere hypothesis. Because of gravitational settling, the lighter elements rise to the top of the atmosphere, making the accreted hydrogen the only visible component of the NS atmosphere. Using a model with a different composition implies the assumption of a total depletion of hydrogen in the NS atmosphere. This could in principle happen due to diffuse hydrogen burning, but the timescale for this process to complete is uncertain \citep{chang_diffusive_2003, chang_evolution_2004}. We chose to use a fully ionized \texttt{NSX} hydrogen atmosphere \citep{ho_atmospheres_2001} for our exploratory runs, as done in all our previous analyses. A single inference run with a partially ionized \texttt{NSX} atmosphere table (used previously in \citealt{miller_radius_2021}) was also carried out for our most favored model (see Section~\ref{subsec:ST+PDT}).

\subsection{Background}
\label{subsec:Background}
A phase-independent background is inferred by fitting a background parameter for every instrumental energy channel. Priors of these parameters can be informed by background measurements, models, or left free to vary from $0$ up to the total number of observed counts when a more precise prior value is not available. However, to save on the computational expense from sampling hundreds of supplementary parameters, these parameters are marginalized rather than being sampled directly. The procedure is as follows: for a set of source parameters, X-PSI computes the expected source phase-resolved spectrum and a supplementary contribution to the otherwise Poisson likelihood is evaluated. This supplementary contribution is based on the marginalization of background parameters over the background prior, which is computed in practice within a region of the parameter space around the best fitting background parameters (for a precise derivation of this contribution, see \citealt{riley_neutron_2019}, Appendix~B.2). These best fitting background parameters associated with the source parameters are the background parameters used for plotting in Section~\ref{subsec:Headline}. This method allows for much faster computations.

Previous works made use of 3C50 background estimates as priors for the NICER background \citep{salmi_radius_2022,choudhury_nicer_2024}, but we chose here to leave it uninformed and use phase-averaged XMM data instead. The information from the phase-averaged source spectrum derived by XMM allows X-PSI to infer constraints on the source spectrum in the NICER data, hence resulting in NICER background constraints without relying on 3C50 assumptions (see Section~\ref{subsec:NICER_data}).

The same background marginalization procedure is applied to the phase-averaged XMM data. However, contrary to NICER, the background prior for XMM is tighter as it is informed by the data. As in previous analyses \citep{riley_nicer_2021,salmi_radius_2022,choudhury_nicer_2024,salmi_radius_2024,vinciguerra_updated_2024}, the backgrounds of the different EPIC instruments were left to vary uniformly in a $\pm n \sigma$ range around the measured EPIC background spectra, where $\sigma$ is the uncertainty on the number of counts generated by a Poisson process (Figure~\ref{fig:EPIC_countSpectra}). We adopted $n=3$ throughout all our analyses, whereas previous analysis of PSR~J0740$+$6620 with XMM data used a value of $n=4$ to account for possible biases in the background, which was not extracted from the same exposures but rather from blank-sky exposures \citep{salmi_radius_2022,salmi_radius_2024}. The XMM background spectra for this analysis were extracted from the same exposures as the source spectra, hence this constraint was tightened. This did not prove problematic as, for the EPIC instruments, the contribution of the background to the flux ($\lesssim0.3$\,cts/bin) is negligible compared to Poisson uncertainty (Figure~\ref{fig:EPIC_countSpectra}).

\subsection{Instrument responses} \label{subsec:response}

We ran joint NICER and XMM fits for all our inference analyses, hence the cross-calibration uncertainties between the instruments had to be taken into account. Each instrument ARF was multiplied by an energy-independent fitted scaling factor parameter. These parameters are $\alpha_{\text{NICER}}$, $\alpha_{\text{PN}}$, $\alpha_{\text{MOS1}}$, $\alpha_{\text{MOS2}}$, and were all fit using the same uncorrelated Gaussian prior $\alpha \sim \mathcal{N}(\mu=1.0,\sigma=0.104)$. This choice relaxes assumptions adopted in previous analyses, which used a single factor for all XMM instruments. 

An exploratory run with uniform priors instead, $\alpha \sim \mathcal{U}(0.5,1.5)$, resulted in broader scaling factor posteriors still centered around $1.0$. Interestingly, it showed a strong positive correlation across XMM instruments and no visible correlation with NICER. However, maximum likelihood cross-calibration scaling factors pointed toward the edge of the prior distribution, which is highly unlikely based on current knowledge of the uncertainty on cross-calibration between instruments \citep[$<10\%$, ][]{marshall_concordance_2021}, hence resulting in biased best fit values. Hence, we opted to keep the Gaussian priors of the scaling factors for all inference runs.

\section{Inference} \label{sec:inference}

In this section, we present inference results for the different hot region geometry models that have been tested, and their respective concordance to the data. The taxonomy of different models is the same as in \cite{riley_nicer_2019}. Results and posterior distributions for all the models can be found in the Zenodo repository \citep{zenodo}.

\subsection{Exploratory ST-U} \label{subsec:ST-U}

\textbf{Model definition} -- The observed double peaked bolometric pulse profile hints at the presence of at least two hot regions. A simple \STU\, model (\textit{Single Temperature - Unshared}) was tested. It is composed of two non-overlapping circular hot regions with independent positions, sizes, and temperatures.

\textbf{Geometry and modes} -- The inference analysis results in two modes with similar configurations of the hot regions, where one spot is located near the equator while the other is on either of the rotational poles. This arises from the edge-on viewing geometry of PSR~J0614$-$3329, which results in a North/South degeneracy on the location of the polar hotspot. Such an effect was already observed during the analysis of PSR~J0740$+$6620, which is also seen edge-on \citep{riley_nicer_2021,salmi_radius_2022,salmi_radius_2024}. Both modes have highly similar ($<1\%$) marginal 1D posteriors on non geometric parameters, are equally likely, and neither is favored by the data.

\textbf{NICER Background} -- The phase-independent inferred background spectrum has features at the energies of the peaks identified in the data (see Section~\ref{subsec:NICER_data}). As expected, X-PSI picked up these contributions as part of the phase-independent background. The inferred spectrum of the NS is made of two thermal components and has no peak around this energy. The background ranges from $\sim$75\% of the signal at low energies up to $\gtrsim98\%$ of the signal at $2.0\,$keV, hence the chosen upper energy limit of the fitted data is well suited. 

\textbf{Inferred parameters} -- The main peak in the NICER pulse profile (see Section~\ref{subsec:NICER_data}) at phase $\phi \sim 0.50$ (Figure~\ref{fig:J0614_NICER}) is caused by the equatorial spot, going in and out of view during rotation. The weak peak is caused by the polar hotspot at $\phi \sim 0.94$, which never fully goes out of view. The phases and temperatures of both hotspots are well constrained for both modes (respectively, $\pm2\%$ and $\pm0.5\%$). Spot sizes are only constrained to about $\pm 20 \%$. The colatitude of the equatorial hotspot is not tightly constrained ($\theta \sim 88^{\circ} \pm 17^{\circ}$). Finally, the radius is rather small and tightly constrained, with $R_{\rm eq}^\STU = 9.1 \pm 0.7 \,$km. The full \STU\, posterior is available in the Zenodo repository \citep{zenodo}.

\textbf{Model performance metrics} -- Values of log-evidence and maximum log-likelihood are given in Table~\ref{tab:model_metrics}. The 2D residuals of all the tested models can be found in the Zenodo repository \citep{zenodo}. \STU\, can reproduce the data as the residuals do not feature any significant deviation from random Poisson noise. However, we tested more complex hotspot geometries to check whether it would bring any improvements.

\begin{table}[!ht]
    \centering
    \begin{tabular}{c|cc}
        \hline
        \hline
         Model & $\ln \mathcal{Z}$ & $\ln p( \text{d} |\bm{\theta}_{\text ML})$ \\
         \hline
         \hline
         \STU & $-34597.536$ & $-34562.457$ \\ 
         \STPDT & $-34584.879$ & $-34545.016$ \\ 
         \PDTU & $-34582.226$ & $-34535.223$ \\
         \hline
         \STPDT\, - $\sin(i) \sim \mathcal{U}(0,1)$ & $-34587.397$ & $-34545.048$ \\
         \STPDT\, - H$_{\text{p}}$ & $-34584.443$ & $-34544.061$ \\
         \hline
         \STPDT\, - $\Delta\phi_0=0.01$ & $-34688.143$ & $-34647.830$ \\
         \STPDT\, - $\Delta\phi_0=0.02$ & $-34627.362$  & $-34587.275$ \\
         \hline
         \STPDT\, - Restricted prior & $-34585.381$ & $-34545.088$ \\
         \STPDT\, - \textbf{Headline} & $-34585.589$ & $-34544.717$ \\
         \hline
         \hline
    \end{tabular}
    \caption{Log-evidence and maximum log-likelihood values for the different investigated models. Besides the  \STPDT\, Headline and restricted prior, they all share the same X-PSI and \textsc{MultiNest} settings. Models including a phase shift $\Delta \phi$ of the data are presented in Section~\ref{subsec:noise}.}
    \label{tab:model_metrics}
\end{table}

\subsection{Exploratory ST+PDT} \label{subsec:ST+PDT}

\textbf{Model definition} -- We now explore the \STPDT\, (\textit{Single Temperature - Protruding Double Temperature}) model. Compared to \STU, this model keeps one circular single temperature region (the ST region), while the second hotspot is replaced by two overlapping circular spots, each with its own non-null temperature (the PDT region). The spot being overlapped is referred to as ceding and the overlapping one as superseding. A PDT spot has four more parameters than a ST spot. They parametrize the ceding spot size, relative position to the superseding spot and temperature while the superseding spot is parametrized like an ST hot region. 

\textbf{Geometry and modes} -- The inference resulted in four modes, all with a geometry similar to that of \STU: one hot region at the equator and another at the pole, with a North/South degeneracy. Although the respective colatitude of the ST and PDT regions is not constrained a priori, the new PDT region ends up being always located on either pole, while the ST region is always at the equator.

The ST region parameters have similar posterior distributions as the equatorial ST region in the \STU\, model, with a well constrained phase ($\phi_p = 0.410\pm0.006$), size ($\zeta_p = 6.8\pm1.4\,^\circ$), and temperature ($\log_{10}(T_p \, \rm{[K]}) = 6.01\pm0.02$), but a poor colatitude localization ($\theta_p = 89\pm16\,^\circ$).

The PDT region has two possible configurations that explain the data equally well, each with a North/South degeneracy. While both PDT configurations are made of a large and rather cold region (with angular radius  $\zeta \sim 25^\circ$ and temperature $T\sim 10^{5.7}\,$K) along with another small and hot region ($\zeta \sim 4^\circ$ and $T\sim 10^{6.2}\,$K), these regions can be either ceding or superseding components. In both configurations, the small and hot regions arrange themselves on the surface to approximately reproduce a similar surface and position of the hotspot as the polar hot region from the \STU\, model (see~Section~\ref{subsec:Headline} where a visual representation of similar configurations is provided).

Compared to the posterior predictive \STU\, bolometric pulse, the sharpness of the main peak is slightly increased to better reproduce the data. However, the new PDT is not mainly responsible for this pulse and has settled into a configuration where it partly contributes to the main peak while being fully responsible for the weak peak. 

\textbf{NICER Background} -- The inferred background is similar to that of \STU\, as it reaches the registered NICER signal at $2\,$keV and still features the peaks at $\sim 400$\,eV and $\sim577\,$eV to account for the phase-independent peaks. 

\textbf{Inferred parameters} -- The inferred radius of $R_{\rm eq}^\STPDT = 10.2^{+1.0}_{-0.8}\,$km is higher than for \STU\, by $\sim$1.2\,km, but is still compatible at $1\sigma$ (Figure~\ref{fig:RM_differentModels}). However, the radius values also depend on the mode, where the hot superseding configuration corresponds to a radius lower than the hot ceding configuration by $\sim$0.5\,km. Comparing mass-radius posteriors, it seems that the \STPDT\, distribution corresponds to the high radius tail of \STU. However, this tail was not classified as a distinct \STU\, mode by \textsc{MultiNest}. The full \STPDT\, posterior is available in the Zenodo repository \citep{zenodo}.

\textbf{Model performance metrics} -- The \STPDT\, residuals do not feature any deviation from random Poisson noise and is able to explain well the data. The log-evidence of this model is decisively better than the value of the \STU\, model using the Kass and Raftery criterion \cite[][here $\Delta \ln \mathcal{Z} \simeq 15.31$]{kass_bayes_1995}. Moreover, the max-likelihood also shows strong improvement. All values are reported in Table~\ref{tab:model_metrics}. This makes \STPDT\, a more suitable model than \STU\, to explain the data. 

\textbf{Supplementary runs} -- We relaxed the orbital angular momentum and spin alignment hypothesis (see Section~\ref{subsec:priors}) for one \STPDT\, run by leaving the inclination angle unrestricted. It resulted in three modes, each replicating distinctive geometric spot configuration among the four modes that were seen in the exploratory \STPDT, but this time with slightly lower inclinations. All modes have inclination and radius constraints consistent with the reference exploratory \STPDT\, model, with respective inferred $\sin{i}$ values of $0.978^{+0.015}_{-0.027}$, $0.892^{+0.063}_{-0.130}$ and $0.910^{+0.059}_{-0.113}$. This indicates that the assumption of alignment of the spin axis with the orbital momentum is not problematic in this case. This run was more computationally expensive while resulting in similar, although not as tight, radius and geometry constraints. Hence, this configuration was only tested for this model.

To assess the effects of a different atmosphere model, an alternative ST+PDT exploratory run was carried out using the partially ionized \texttt{NSX} hydrogen model (see Section~\ref{subsec:priors} for the choice of this atmospheric model). We retrieve the same four geometric modes as for the fully ionized atmosphere case with slightly higher, but compatible, equatorial radius constraints (with difference in the median values $\Delta R_{\rm eq} \simeq 0.38\,$km). The log-evidence with the partially ionized atmosphere model is better, but not significantly \cite[$\Delta \ln Z \simeq 0.44$][]{kass_bayes_1995}. Hence this atmosphere model gives results consistent with those of the fully ionized model, which we use for the rest of the inferences.

\subsection{Exploratory PDT-U} \label{subsec:PDT-U}

\textbf{Model definition} -- A \PDTU\, model (\textit{Protruding Double Temperature - Unshared}) is composed of two PDT hotspots. Although the \STPDT\ results were convincing, it was tested to investigate the possible metric gains that a second PDT hotspot would bring. Note that this model is a much more computationally intensive one than \STU\, and \STPDT.

\textbf{Geometry and modes} -- The inference results in two configurations, each with two North/South degenerate modes, still with one region at the pole and the other at the equator. As for \STPDT, the \PDTU\, model has a configuration where the ceding region is hotter and smaller than the associated superseding one. However, this region is now located at the equator while it was the polar region for \STPDT. Moreover, the geometric parameters posterior distributions are still similar to those of both \STU\, and \STPDT. The polar hot region, in particular, is fully consistent with the \STPDT\, posterior in the hot superseding configuration (see Section~\ref{subsec:ST+PDT} for more details on this configuration).

\textbf{NICER Background} -- The inferred background of this model is, again, similar to that of \STU. It still shows an excess around the peaks in the spectrum and reaches the registered NICER signal at $2.0$\,keV.

\begin{figure}[!t]
    \centering
    \includegraphics[width=\linewidth]{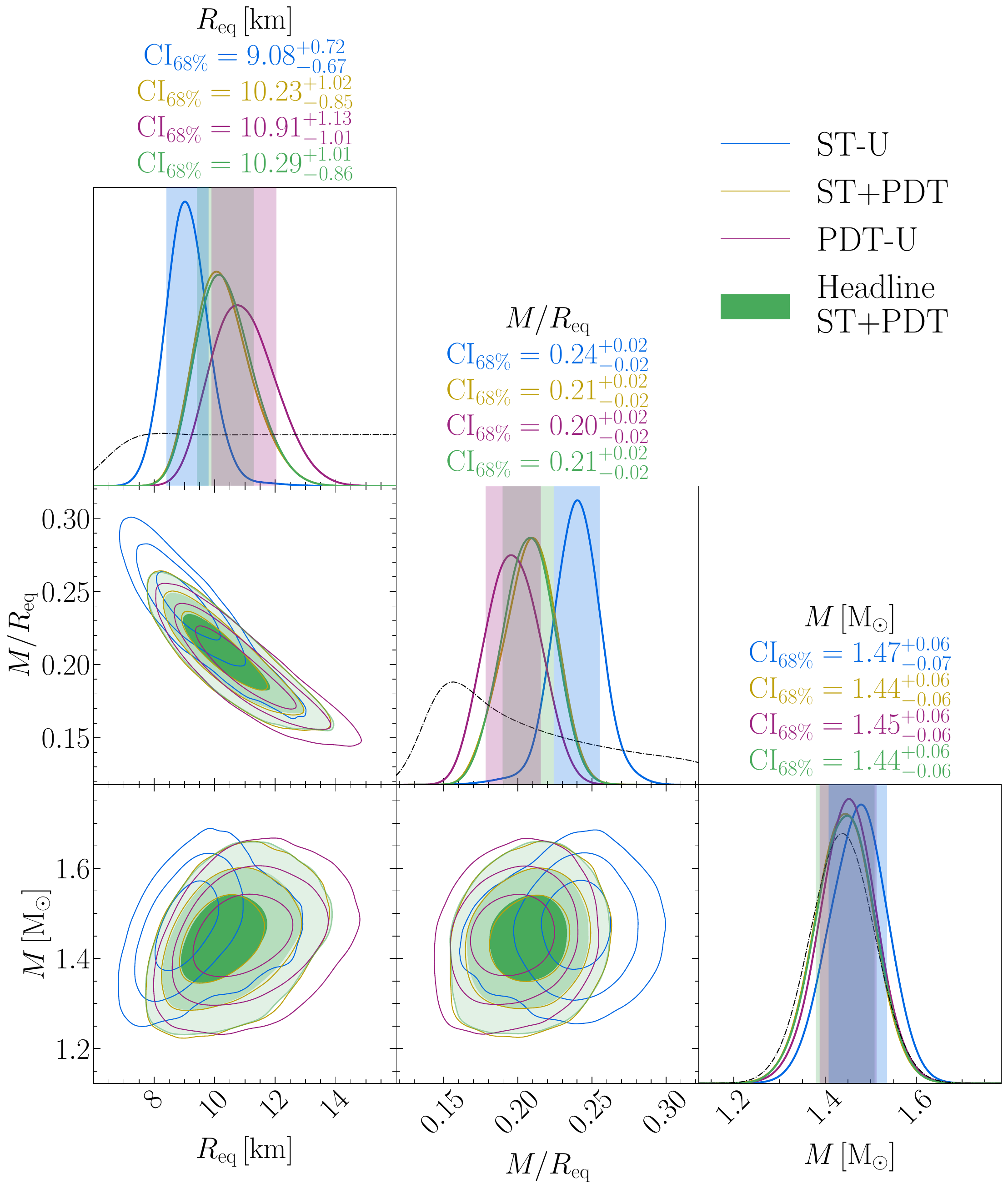}
    \caption{Posterior distributions of the mass, radius, and compactness for the runs presented in Section~\ref{sec:inference}. Dash-dotted black lines on the diagonal plots show the marginalized prior distributions, shared across all runs. The shaded vertical bands show the $68.3\%$ credible intervals. The contours in the 2D posteriors show the $68.3\%$, $95.4\%$, and $99.7\%$ credible regions, which are filled for the Headline result.}
    \label{fig:RM_differentModels}
\end{figure}

\textbf{Inferred parameters} -- The inferred \PDTU\, radius $R_{\rm eq}^\PDTU=10.9^{+1.1}_{-1.0}\,$km is compatible with the \STPDT\, one, with a difference in median values of $\Delta R_{\rm eq} \simeq 0.7\,\text{km} \,$ ($< 1\sigma$, Figure~\ref{fig:RM_differentModels}), which makes them consistent with one another.

Interestingly, for \PDTU, all individual modes yield a higher $R_{\rm eq}$ and $N_{\rm H}$ than the combined distribution. This was not the case for \STU\, or \STPDT. This happens when the  distributions of the different modes intersect only for lower values of $R_{\rm eq}$ and $N_{\rm H}$ in some parts of the parameter space. Such a phenomenon was already observed when combining inference results from different bursts in \cite{kini_constraining_2024}.

The posterior distribution of the angular radii of both superseding hotspots spans rather low values around $\zeta \sim 3\pm1\degree$, with the best-fit parameters reaching even lower values with sizes of $\zeta \simeq 1 \degree - 1.7 \degree$ across both superseding hotspots in both modes. The full \PDTU\, posterior is available in the Zenodo repository \citep{zenodo}.

\textbf{Model performance metrics} -- Although \PDTU\, has a better log-evidence and maximum likelihood than \STPDT\, ($\Delta \ln \mathcal{Z} = 2.65$, Table~\ref{tab:model_metrics}), indicative that \PDTU\, performs better, such a small difference is not decisive according to the Kass and Raftery criteria \citep{kass_bayes_1995}. \cite{vinciguerra_x-psi_2023} previously found that a log-evidence difference of a few units could not tell models apart: in that work, the log-evidence of a more complex model was $\sim 1.5$ units greater than for a simpler model, although the data were generated by the simpler model. It is a possibility that the evidence of \PDTU\, increases, although not significantly, by accumulating evidence in the region of the parameter space where it reproduces \STPDT.

Finally, \PDTU\, is much more computationally expensive to run, while resulting in similar geometric and radius constraints, which are our main parameters of interest. Therefore, for our high resolution run we opt to continue with the simpler \STPDT\, model, which already reproduces well the data. Further explanation on this choice is given in Section~\ref{sec:Validation}.

\subsection{Headline ST+PDT} \label{subsec:Headline}

\textbf{Model definition} -- The Headline run was carried out with the \STPDT\, model, with improved X-PSI resolution settings and sampling hyper-parameters (see values in Appendix~\ref{app:compare_resolution} and \ref{app:headline}) and still with the multi-modal variant of \textsc{MultiNest}. Because the omnipresence of the North/South degeneracy across previous runs, which gave similar radius constraints, slowed down the computation times and made thorough sampling harder, geometric priors were constrained to remove this degeneracy. This was effectively done by forcing the polar PDT region to lie on the North hemisphere of the NS. A test run with the exploratory \STPDT\, showed that this procedure speeds up computation by a factor of $\sim$2 while resulting in similar geometric and radius constraints, with a difference of radius of $\sim30$\,m, the same order of the $\sim15$\,m Monte Carlo standard error, an uncertainty inherent to sampling procedures. Moreover, restricting the prior volume allows for more thorough sampling, making it harder to miss high likelihood regions.

The choice of the North hemisphere, instead of the South hemisphere, for the location of the PDT hot region allows for more conservative uncertainties on the radius. This is because the two North PDT configurations of the exploratory \STPDT\, have individual median radii that are, respectively, the lower and higher values among all the modes found in this exploratory run, with uncertainties greater or equal to that of their southern counterpart.

\begin{figure*}[!ht]
    \centering
    \includegraphics[width=0.49\linewidth]{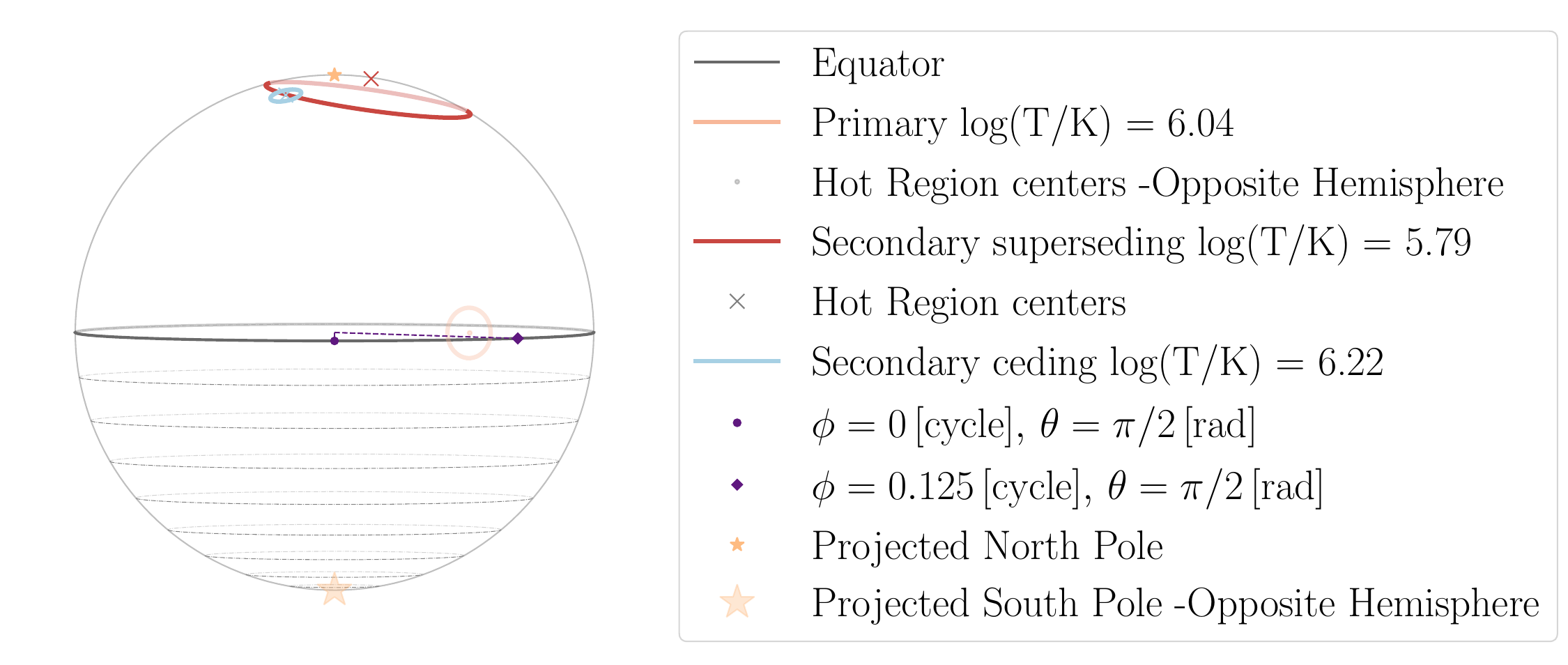}
    \includegraphics[width=0.49\linewidth]{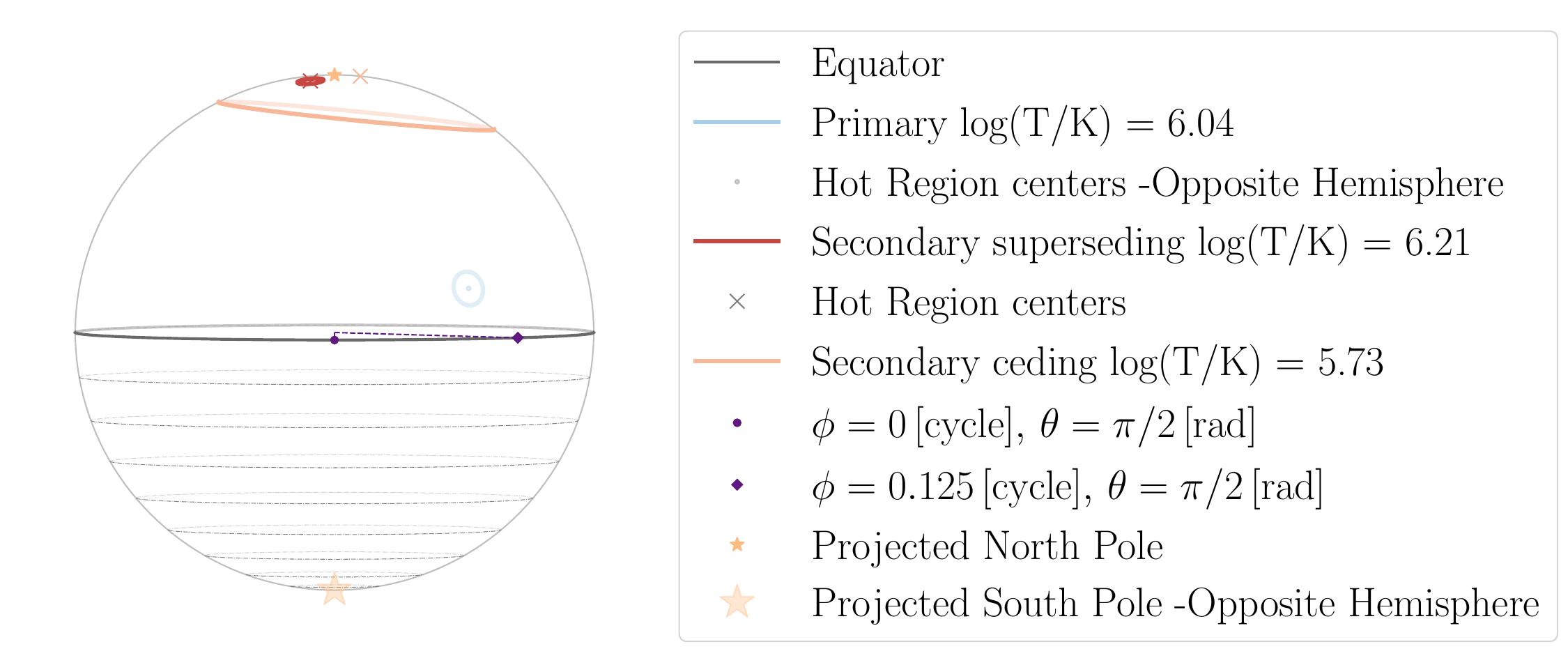}
    \caption{Representation of the best fit geometries, as seen from Earth, for the two main modes uncovered by the Headline \STPDT\, run. The pulsar is represented without gravitational or relativistic effects at phase zero of the data. The hotspots located behind the star are plotted with dimmed colors. The hotspot color changes from blue to red, from the hottest to the coldest temperatures. The equatorial ST hotspot can be seen at phase $\phi \sim 0.4$ (behind the star) for both modes. The polar PDT hotspot has two possible configurations, where the smaller and hotter components can be either ceding or superseding components (see Section~\ref{subsec:ST+PDT}). The right geometry is the overall best fit geometry. These representations do not capture the full posterior distributions of the geometric parameters and should be considered with care.}
    \label{fig:BestFitGeometries}
\end{figure*}

\begin{figure*}[!ht]
    \centering
    \includegraphics[width=\linewidth]{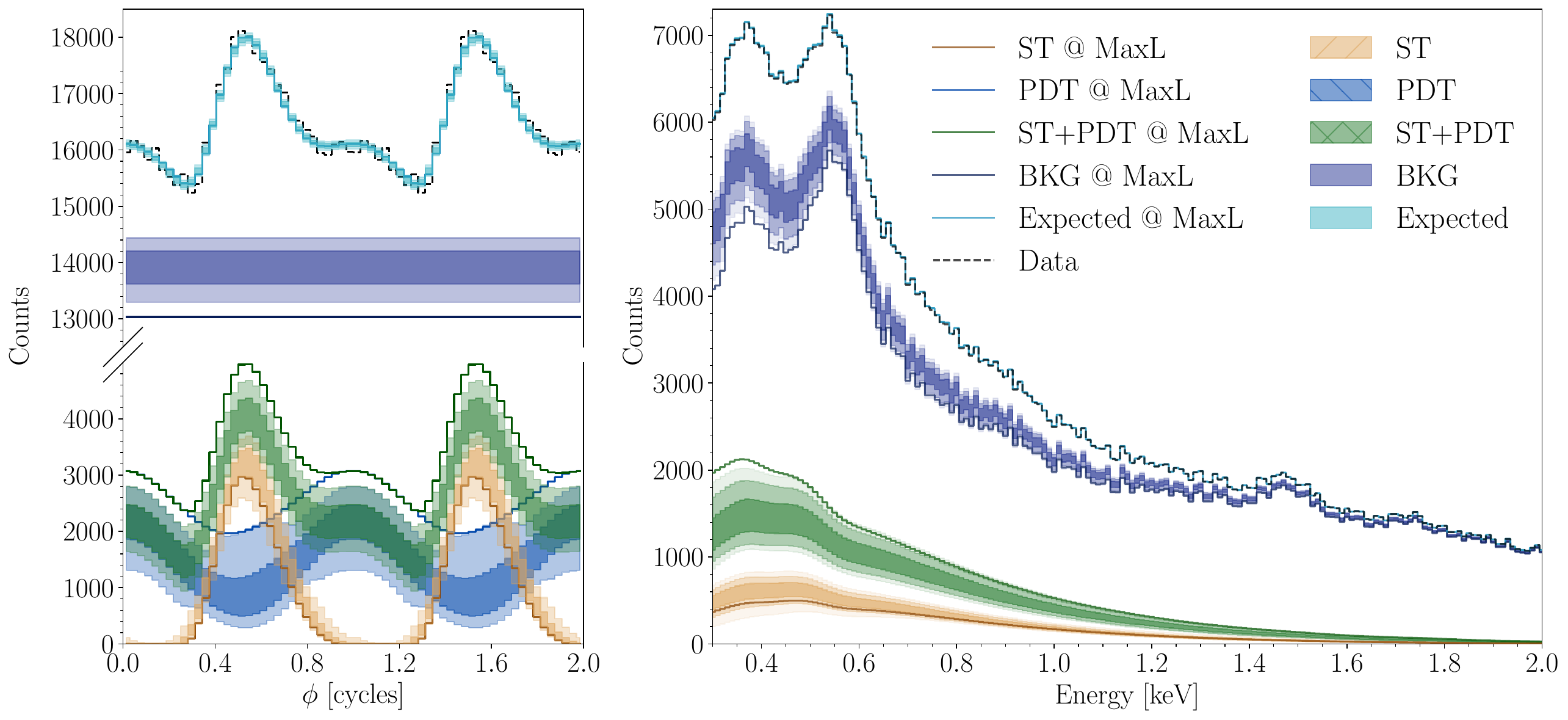}
    \caption{Posterior predictive distribution and best fit model of the bolometric pulse on the left panel and of the phase-averaged spectrum on the right panel. The posterior predictive distribution represents the posterior probability distribution of the observables. The contribution of the individual model components are shown in orange for ST,  blue for PDT (left panel only) and purple for the background (BKG). The combined model components are in green for ST+PDT, and in light blue for the total (Expected, i.e., ST+PDT+BKG). The shaded regions, from darker to lighter, represent, respectively, the $68\%$, $95\%$ (and $99\%$ for the right panel only) credible interval computed over $5000$ samples. The solid lines represent the maximum likelihood model and the black dashed line the data. The left panel has a discontinuous y-axis and the horizontal line at $\sim13,000$ counts represent the phase-independent maximum likelihood background. On the right panel, the expected distribution, maximum likelihood and data all overlap. Note that the sampling procedure does not fit individually the ST, PDT, or BKG components but their sum (see Section~\ref{subsec:Headline} for details).}
    \label{fig:Headline_posterior predictive}
\end{figure*}

\textbf{Geometry and modes} -- This analysis uncovered four modes, two of which were already previously found in the exploratory run. The two new modes both having significantly lower log-evidence ($ \Delta \ln \mathcal{Z} \le 5.7 $) are hence disfavored and have only a barely visible impact on the total posterior distribution. These modes also appeared in the run testing the setup to break the degeneracy and are not discussed here, but their presence itself reveals that \textsc{MultiNest} finds even low evidence modes. The two modes existing in the headline run have nearly the same distributions as previously found in the exploratory \STPDT\, run. The best fit geometry of each mode is shown on Figure~\ref{fig:BestFitGeometries}. Analogous to the exploratory \STPDT, the PDT spot is solely responsible for the weak peak and partially contributes to the main peak (Figure~\ref{fig:Headline_posterior predictive}). 

\textbf{NICER Background} -- The background is similar to the one of exploratory runs. However, the deviation of the maximum likelihood background from the inferred distribution median is $\sim$3$\sigma$ at energies below $\sim 0.45$\,keV, and is decreasing with increasing  energy (right panel of Figure~\ref{fig:Headline_posterior predictive}). However, note that we do not fit the individual ST, PDT, or background components individually but their sum and the best fit geometry are consistent with the total spectrum expected distribution, which also has smaller uncertainties than individual ST, PDT or background contributions (Figure~\ref{fig:Headline_posterior predictive}). 

The uncertainties in the contributions of the different components arise from uncertainties in the NICER source spectrum, which is mostly dependent on the XMM spectrum and cross-calibration factors (see Section~\ref{subsec:Background}). As the latter are poorly constrained and the count statistics of the XMM observation is low, the accuracy of the background constraint is limited. 

In practice, the best fit parameter vector of this headline \STPDT\, run has ARF scaling factors (see Section~\ref{subsec:response}) lying on the edge of the prior distribution. This is compensated by the background marginalization, resulting in a background spectrum on the lower edge of the distribution along with a high likelihood value.

Hence, although it is consistent with the posterior geometric distribution, the best fit geometry should be considered with care. Whenever possible, the full posterior distributions of the geometric parameters should be used instead. The maximum a-posteriori parameter vector of the individual modes could also be considered. These can be retrieved in Appendix~\ref{app:headline} (for the geometric parameters) or from the Zenodo repository \cite[for the full parameter vector,][]{zenodo}.

\textbf{Inferred parameters} -- For this headline run, we infer the radius of PSR~J0614$-$3329 to be $R_{\rm eq}=10.29^{+1.01}_{-0.86}$\,km for a mass of $M=1.44 ^{+0.06}_{-0.07} \, M_{\odot}$. This is consistent with the exploratory \STPDT\, run and both exploratory \STU\, and \PDTU\, runs.

The inferred distribution of $N_{\rm H}=0.36^{+0.43}_{-0.25} \times10^{20}\,\text{cm}^{-2}$ reaches the lower bound of the prior (see Appendix \ref{app:headline} for the posterior distribution of a subset of parameters, the full Headline \STPDT\, posterior is available in the Zenodo repository from \citealt{zenodo}). This value is compatible with the \cite{hi4pi_collaboration_hi4pi_2016} value, the latter being also compatible with $0$.

A diagnostic to test the convergence of \textsc{MultiNest} is to see whether increasing sampling accuracy affects the posterior distribution and evidence \citep{higson2019diagnostic}. Here, the posterior is similar to that of the LR run with the same restricted priors, although uncertainties are slightly larger. This indicates that sampling has likely converged.

\textbf{Model performance metrics} -- As evidence is proportional to the prior, which was restricted for this Headline run in order to remove the North/South degeneracy, the new log-evidence can not be directly compared to the exploratory run value. Instead, it should be compared to the exploratory \STPDT\, run with restricted priors value (see Table~\ref{tab:model_metrics}), whose log-evidence is consistent within $3\sigma$ with the Headline value.

The posterior predictive bolometric pulse (see left panel of Figure~\ref{fig:Headline_posterior predictive}) fits well the data. The PDT spot has higher flux and more uncertainty associated to its inferred spectrum than the ST spot (see right panel of Figure~\ref{fig:Headline_posterior predictive}). This is not due to the multi-modality and happens for both configurations, indicating that the PDT parameters are more difficult to constrain than the ST ones in our case.

\pagebreak

\section{Validation} \label{sec:Validation}
In this section, we describe the different steps taken to assess the validity of the results from our exploratory and Headline runs.
 
\subsection{X-PSI resolution}\label{subsec:XPSIresolution}

As we used LR settings for the exploratory runs and mixed resolution (XR) settings for the Headline run (see Appendix~\ref{app:compare_resolution} for more information), we investigate the effects of the settings configuration on sampling.

\begin{figure}[!t]
    \centering
    \includegraphics[width=\linewidth]{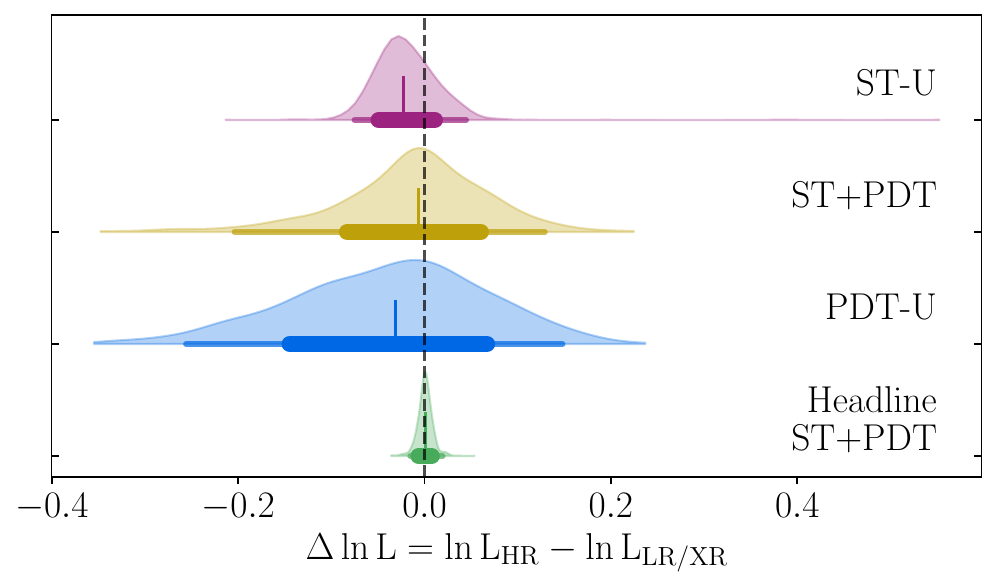}
    \caption{Histogram of $\Delta \ln\text{L}$ for the three exploratory \STU, \STPDT, and \PDTU\, runs and for the headline \STPDT\, run. The short colored vertical lines represent the median of each distribution. The thick (thin) horizontal lines at the base of the distributions represent the $68$\% ($95\%$) equal-tailed credible intervals. For visualization purpose, a dashed vertical line centered on $\Delta \ln\text{L} = 0$ is plotted.}
    \label{fig:DeltaLogL}
\end{figure}

The effects of the LR settings on the likelihood were investigated by drawing $1000$ samples from the posterior of the models presented in Section~\ref{sec:inference} and computing a likelihood with high resolution (HR) settings for each sample, which was then compared against the likelihood with LR (exploratory) or XR (headline) settings used during sampling. The $\Delta \ln\text{L} = \ln\text{L}_{\rm HR} - \ln\text{L}_{\rm LR/XR}$ distributions are plotted in Figure~\ref{fig:DeltaLogL}. The distributions are all compatible with $0.0$ at the $1\sigma$ level. The spread, which measures the accuracy of the likelihood, increases with model complexity, reaching $\sigma \sim 0.1$ for \PDTU. This is because, under the same X-PSI resolution settings, we observe a higher likelihood accuracy for ST regions compared to PDT regions, as the former exhibit a simpler geometry. While the accuracy (i.e., a maximum value of $|\Delta \ln\text{L}|$) needed for precise likelihood computation is difficult to determine, we deem these values low compared to the range of values taken by the likelihood during sampling, $\sim \mathcal{O}(10^4)$. Moreover, a log-evidence difference of $\gtrsim 5$ would be needed to claim a significant superiority of a model over another, which is $50$ times more than our reported accuracy. Hence we could not find any systematic bias between LR and HR likelihoods for the posterior geometries, although we can conclude that \PDTU\, likelihoods were less accurate than those of the other models.

Even if their likelihoods under HR settings are higher, configurations with more complex, thin, elongated structures with biased LR settings likelihoods could be missed while sampling. However, these $\Delta \ln\text{L}$ distributions of LR runs support LR being accurate in the posterior space of geometries for all \STU, \STPDT\, and \PDTU, and their posteriors are consistent with a local maximum of the HR model. Nevertheless, we did not demonstrate that LR settings yield accurate likelihoods over the whole prior space of geometries and it is hence not possible from such considerations to conclude that a complex structure was not missed. It is possible that, somewhere in the prior space, there is a better fitting geometric configuration for which LR likelihoods are not accurate and biased toward lower values. Such a region could be unknowingly missed during sampling and would appear when using high X-PSI resolution. The Headline \STPDT\, model was run with a higher X-PSI resolution and sampling accuracy to account for this possibility and converged toward the same solution as the exploratory run (see Section~\ref{subsec:Headline}). This indicates that it could also be the case for \PDTU, which can be confirmed with a higher resolution \PDTU\, run.

For the Headline \STPDT\, model, we find a Gaussian-like distribution of $\Delta \ln\text{L}$ with $\mu_{\Delta \ln\text{L}} \simeq 8 \times 10^{-4} $ and $\sigma_{\Delta \ln\text{L}} \sim 8 \times 10^{-3} $ (Figure~\ref{fig:DeltaLogL}). As these values are one order of magnitude lower than for LR runs, we deem this resolution increase sufficient. Moreover, it would be too computationally expensive, given current computational resources, to further increase the resolution (see Appendix ~\ref{app:compare_resolution}).

\subsection{Noise realization}\label{subsec:noise}

The Poisson noise in the NICER data can have an effect on the inferred parameters \citep{vinciguerra_x-psi_2023}. This phenomenon has also been investigated here by testing different binning of the phases for the PSR~J0614$-$3329 data. Even though the phase measurement of all collected photons is very precise ($\sim 2\,\mu$s), the data used for likelihood computation (Figure~\ref{fig:J0614_NICER}) is binned in $1/32=0.03125$ rotation cycles (about $100\, \mu$s wide) bins to save on computational time. By uniformly shifting the phase of individual NICER photons by a small amount ($< 1\,$ bin) before binning, most photons will remain in the same bins, and some will end up in the neighboring bin, therefore resulting in slightly different bolometric pulse profiles. This way, we mimic different noise realizations of the binned pulse profile, which, however, all share the same spectrum and XMM data. Using shifts of $0.01$ and $0.02$ cycles, that are smaller than the bin width, we ran LR runs with the preferred \STPDT\, model.

\begin{figure}[!t]
    \centering
    \includegraphics[width=\linewidth]{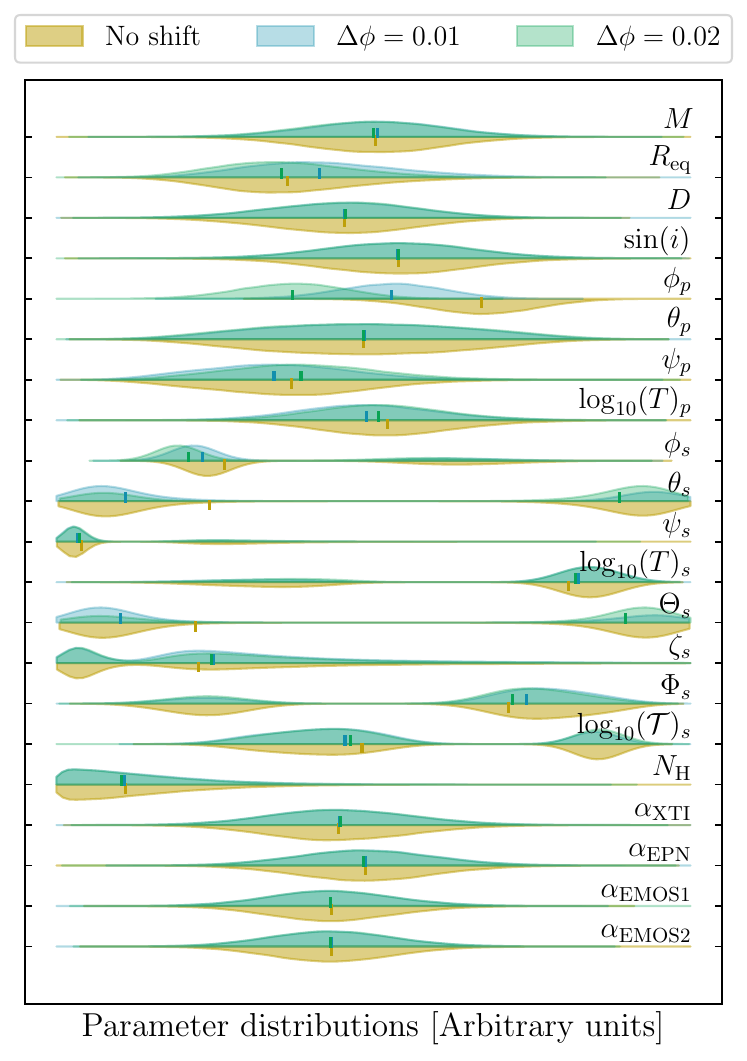}
    \caption{One dimensional marginal distributions of all the parameters for the \STPDT\, exploratory runs with different phase shifts applied to the input data. The top of the "violin" distributions (green and blue) show the results from the inference on shifted data and the bottom (yellow) is for the reference \STPDT\, run with no shift. Beside the spot phases $\phi_p$ and $\phi_s$, all runs have highly similar posteriors. For visualization purposes, the units along the x-axis are arbitrary. The parameter nomenclature and descriptions can be found in Appendix \ref{app:headline}.}
    \label{fig:compare_shifts}
\end{figure}

All runs resulted in similar posterior distributions for all the parameters, except for the phase of the hot regions (Figure~\ref{fig:compare_shifts}). As expected, the observed shift of the inferred hot region phases $\phi_p$ and $\phi_s$ matches exactly the shift applied to the data (although this is more difficult to visualize for the secondary hot region, as $\phi_s$ has two modes with large phase separation). However, one mode is missing for the $0.02$ cycles shift, which indicates that the parameter space is not fully sampled. On the other hand, no other mode was found during these analyses, which indicates that sampling likely did not miss any significant supplementary mode because of noise realization.

The log-evidences and maximum likelihoods have strong dependence on the binning of the data. This dependence (given in Table~\ref{tab:model_metrics}) is greater than the difference in metrics between all the \STU, \STPDT, and \PDTU\, models.

This check allows to ensure that the solutions inferred by the previous exploratory runs are not artifacts from noise realization being over-fitted, but rather are consistent across different noise realizations.

\subsection{Model comparison}\label{subsec:model_comparison}

As it was not trivial to differentiate \STPDT\, and \PDTU\, from log-evidence values alone (see Section~\ref{subsec:PDT-U}), we turned to the \texttt{arviz} package \citep{vehtari_practical_2017}, which provides model comparison techniques taking into account the complexity of the models and comparing their predictive power. This requires a point-wise likelihood, i.e., a likelihood value for each individual bin. This product is, however, not available in our case, since our inferences use channel-wise background marginalized likelihoods, which is not applicable bin by bin (see Section~\ref{subsec:Background}). Instead, we use a simple Poisson likelihood and specify as supplementary input parameters the background parameter values.

\begin{figure}[!t]
    \centering
    \includegraphics[width=\linewidth]{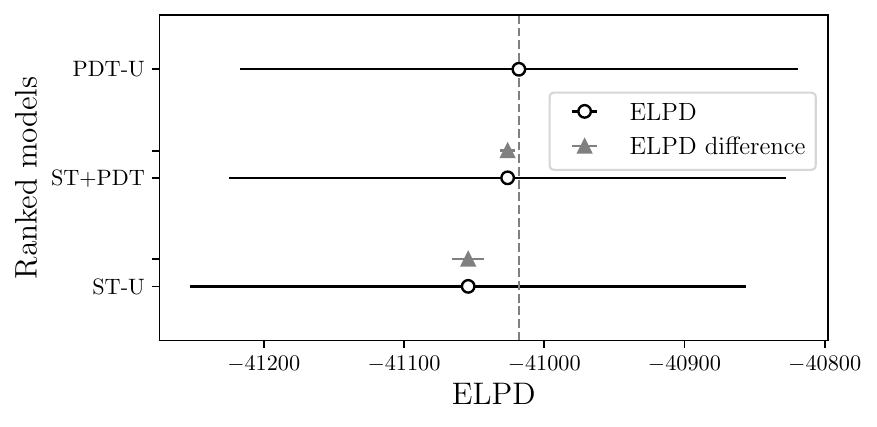}
    \caption{Graphic model comparison between \STU, \STPDT\, and \PDTU\, for $10^4$ posterior samples. The black circles and bars represent ELPD and its uncertainty. The gray triangles for \STU\, and \STPDT\, represent the same ELPD values, but their error bars display the uncertainty on the ELPD differences with respect to the \PDTU\, one.}
    \label{fig:model_comparison}
\end{figure}

Using $10^4$ samples, the variance of the expected log point-wise predictive density (ELPD) is large for all models, making them all compatible, but the variance of difference of ELPD is not as large (Figure~\ref{fig:model_comparison}). The significance of these deviations can be measured with two metrics (Leave-One-Out and Weighted Akaike Information Criteria), both of which favor \PDTU\, over \STPDT, but not significantly, while \STU\, is totally disfavored. This is in line with our previous conclusions using evidence. However, using the ELPD metric likely introduces some bias in our case, as the Poisson likelihood is different from the background marginalized likelihood used during sampling. As this model comparison could not clearly disentangle \PDTU\, and the simpler \STPDT\, model, which both result in similar radius constraints, and given the lower computational time of \STPDT, we chose to use the latter for the headline model.

\section{Discussion} \label{sec:discussion}

\subsection{Comparison to other MSPs with PPM constraints}

The inferred radius of PSR~J0614$-$3329 is constrained at the $\sim \pm 9\%$ level ($68\%$ credible interval)\footnote{These numbers do not include systematic uncertainties, which are difficult to quantify. A rough estimate can be obtained by looking at the dispersion of radius values for different models. For instance, it is of the order of $\sim5\%$ for PSR~J0614$-$3329 and $\sim20\%$ for PSR~J0030$+$0451.}, which is not as tight as for PSR~J0437$-$4715 (about $\pm 7\%$, \citealt{choudhury_nicer_2024}), but is of the order of PSR~J0740$+$6620 (about $\pm 9\%$, \citealt{salmi_radius_2024}). The radius constraint of PSR~J1231$-$1411 \citep{salmi_J1231_2024} is harder to compare, since that study used different radius priors, some of which were EOS informed, causing the tighter constraint. Some individual models for PSR~J0030$+$0451 achieved lower $R_{\rm eq}$ uncertainty thanks to higher photon statistics, resulting from a deeper exposure and a brighter source  \citep{riley_nicer_2019,vinciguerra_updated_2024}. 

While \cite{vinciguerra_updated_2024} incorporated background constraints in the PSR~J0030$+$0451 analysis, \cite{riley_nicer_2019} did not. The former resulted in different models fitting the data equally well, partially because of the lack of a mass prior. Our result is consistent with the \STPDT\, solution for PSR~J0030$+$0451 from \cite{vinciguerra_updated_2024}. However,  our inferred radius is not directly comparable to that of the \PDTU\, solution from PSR~J0030$+$0451, as this solution points toward a mass of $\sim$1.7\,\Msun\ instead of $\sim$1.4\,\Msun. 

Interestingly, the measured radius is compatible at the $1\sigma$ level with the inferred radius of PSR~J0437$-$4715, a MSP of a similar mass \citep{choudhury_nicer_2024} and  with the indirectly inferred\footnote{GW events carry information on tidal deformability. Extracting radius information from this information is an EOS-dependent process.} mass-radius contours for the NSs whose merger produced the GW170817 signal \citep{abbott_gw170817_2018}. 

The PSR~J0614$-$3329 constraints presented in this work are deemed highly reliable as the inferred parameters are consistent across all different geometric models, for different X-PSI resolutions and different prior assumptions tested, such as the spin-orbital momentum alignment and atmospheric ionization state. These results all have a median radius in the $9-11$\,km range. They all share a similar non-antipodal geometry, with one hot region on the pole, which is always partially visible, and a hot region near the equator, going in and out of view and causing the main peak in the profile.

This consistency across all runs in our analysis could be explained by the tight priors on mass and inclination. PSR~J0030$+$0451 and PSR~J1231$-$1411 are missing such constraints, making the inferred radius and mass ranges different from model to model. Although PSR~J0740$+$6620 has such constraints and $2.5$ times more exposure than PSR~J0614$-$3329, it is also twice as far and $\sim7$ times fainter.

PSR~J0437$-$4715 is very bright with a tight prior on its mass, inclination and distance \citep{reardon_neutron_2024} and also has little change in measured radius across models \citep{choudhury_nicer_2024}, but an AGN is in the field of view and a 3C50-informed background was used, both introducing possible bias. Moreover, PSR~J0437$-$4715 is seen at a $137.5^\circ$ inclination, which may allow for a completely different hotspot configuration when changing models. This does not seem to be the case for PSR~J0614$-$3329, which is seen edge-on. Overall, PSR~J0614$-$3329 likely owes its radius measurement stability to its tight prior from radio timing, along with the fact that it is seen edge-on.

\subsection{Implications of PSR~J0614$-$3329 on the dense matter EOS}\label{subsec:Implications}

With the new PPM-derived mass-radius posteriors of PSR~J0614$-$3329 in hand, a subsequent EOS inference is necessary to understand the impacts of this new measurement on the inferred dense matter EOS and NS properties. To perform an EOS inference, we build on the work of \cite{rutherford_constraining_2024}. Starting from the Baym-Pethick-Sutherland crust EOS \citep{Baym71} for densities below $\approx0.5 n_0$, where $n_0=0.16\, \mathrm{fm}^{-3}$ is the nuclear saturation density, we use the EOS band from next-to-next-to-next-to-leading order (N$^3$LO) chiral effective field theory ($\chi$EFT) calculations up to $1.5 n_0$ \citep{Keller2023}. Beyond this density, we use two different high-density EOS parameterizations: one based on a piecewise-polytropic model (PP, \citealt{Hebeler2013}) and one based on the speed of sound in the NS (CS, \citealt{Greif19}). This results in a very large prior range for the EOS. For details on how the dense matter EOS models are constructed and their prior assumptions, we refer to \citet[][and references therein]{rutherford_constraining_2024}.

We obtain posterior EOS distributions through sampling with \textsc{MultiNest}. This entails first evaluating the priors based on allowed EOSs in the $\chi$EFT range and for a broad range of parameters that span the PP or CS extensions. The likelihood incorporates the mass-radius constraints of MSPs obtained with PPM and mass-tidal deformability from GW events. This framework is implemented using the open-source dense matter EOS inference code NEOST\footnote{\url{https://github.com/xpsi-group/neost}} (\texttt{v2.1.0}, \citealt{Raaijmakers24}). NEOST has been used for NICER EOS constraints and other dense matter inferences with prerelease versions having been employed in \cite{Greif19, Raaijmakers19,Raaijmakers20,Raaijmakers21, Rutherford:2022xeb}.

We compare our new EOS inferences to the ``New'' scenario from \cite{rutherford_constraining_2024}. This scenario is defined by the combination of the \textsc{ST-U} solution of PSR~J0740$+$6620 from \cite{salmi_radius_2024}, the \textsc{ST+PDT} solution of PSR~J0030$+$0451 from \cite{vinciguerra_updated_2024}, the \textsc{CST+PDT} solution of PSR~J0437$-$4715 from \cite{choudhury_nicer_2024}, and the mass-tidal deformability posteriors of GW170817 and GW190425 from \cite{PhysRevX.9.011001} and \cite{abbott_gw190425_2020}, respectively. We investigate the effect of adding the new mass-radius result of PSR~J0614$-$3329 to this data set. Interestingly, with the new result from PSR~J0614$-$3329 included, the EOS inference results would be very similar if only MSPs with mass and inclination priors were accounted for and PSR~J0030$+$0451 (\textsc{ST+PDT}) left out (see the Zenodo repository of \cite{zenodo} for figures and samples of the inferred EOS in that case). 

\begin{figure*}[!ht]
    \centering
    \includegraphics[width=\linewidth]{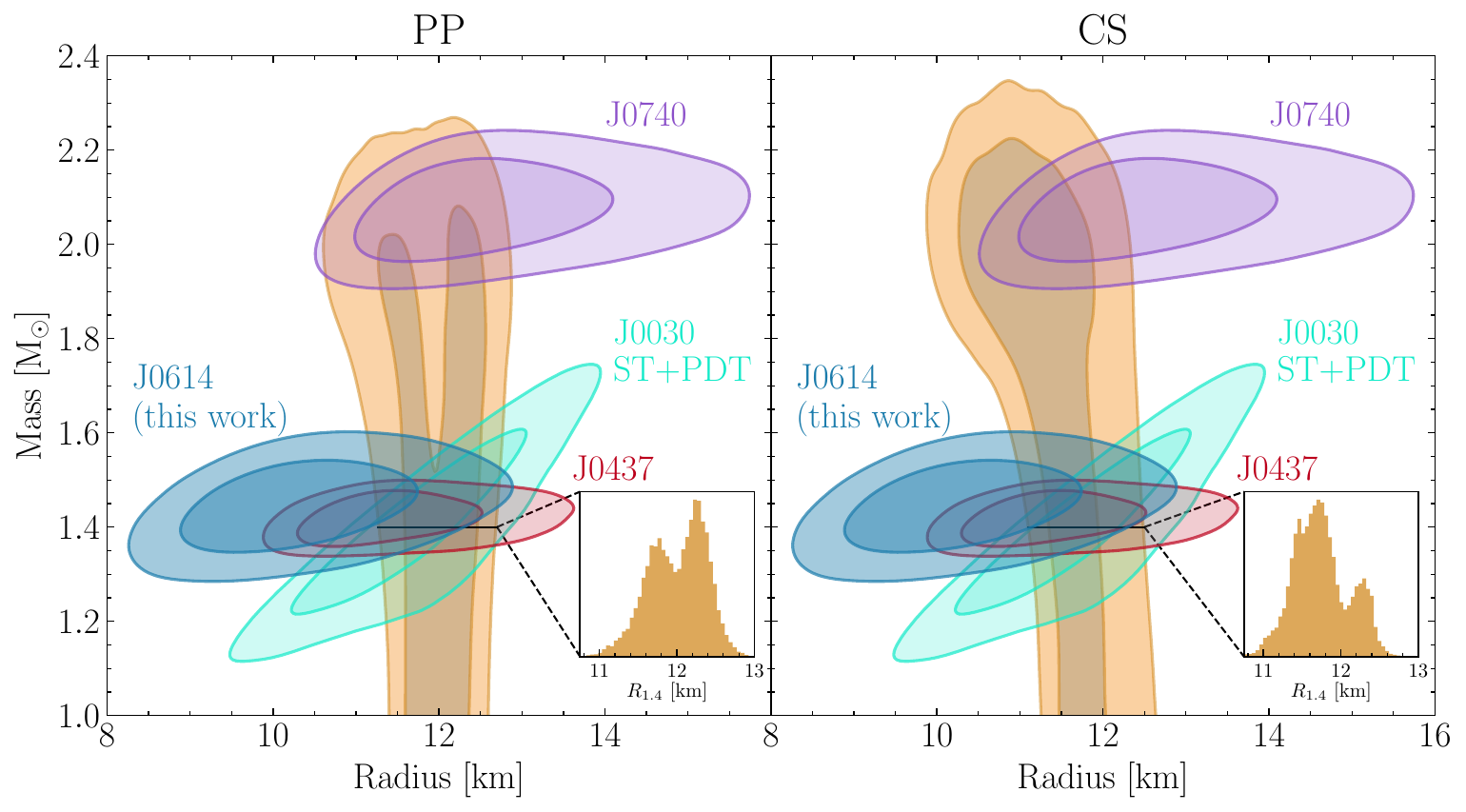}
    \caption{Mass–radius posterior distributions in golden obtained from the EOS inference using the N$^3$LO $\chi$EFT band up to $1.5 n_0$ and the PP (left panel) or CS extensions (right panel). The EOS inference includes the mass-radius likelihood from PSR~J0614$-$3329 obtained in this work in blue, PSR~J0740$+$6620 in purple, PSR~J0437$-$4715 in red, the \STPDT\, model of PSR~J0030$+$0451 in cyan, and the GW170817 constraints from the tidal deformability (for details without PSR~J0614$-$3329 see the ``New'' scenario in \citealt{rutherford_constraining_2024}). The dark (light) regions give the 68\% (95\%) credible regions. The insets show the radius distribution for a $1.4 \, M_\odot$ star, indicated by the black horizontal lines.}
    \label{fig:MR_plane}
\end{figure*}

The EOS inference yields mass-radius posterior distributions (Figure~\ref{fig:MR_plane}) centered around 12\,km with maximum masses $\lesssim 2.4 \, M_\odot$ for both the PP and CS extensions. Our results for the radii of 1.4 and $2.0\,M_\odot$ NSs, as well as the maximum mass configurations are also reported in Table~\ref{tab:pressure_CI}. These inferred mass-radius posteriors are less than one standard deviation away from \cite{rutherford_constraining_2024} ``New'' data scenario, although they shift toward lower radii by $\sim 300\,$m, with a minimal impact on the maximum mass for both PP and CS models. This slight shift is not only due to the addition of PSR~J0614$-$3329, but also to the fact that the inferred radius of PSR~J0614$-$3329 is consistent with those of PSR~J0030$+$0451 (\textsc{ST+PDT}) and PSR~J0437$-$4715.

\begin{figure}[!ht]
    \centering
    \includegraphics[width=\linewidth]{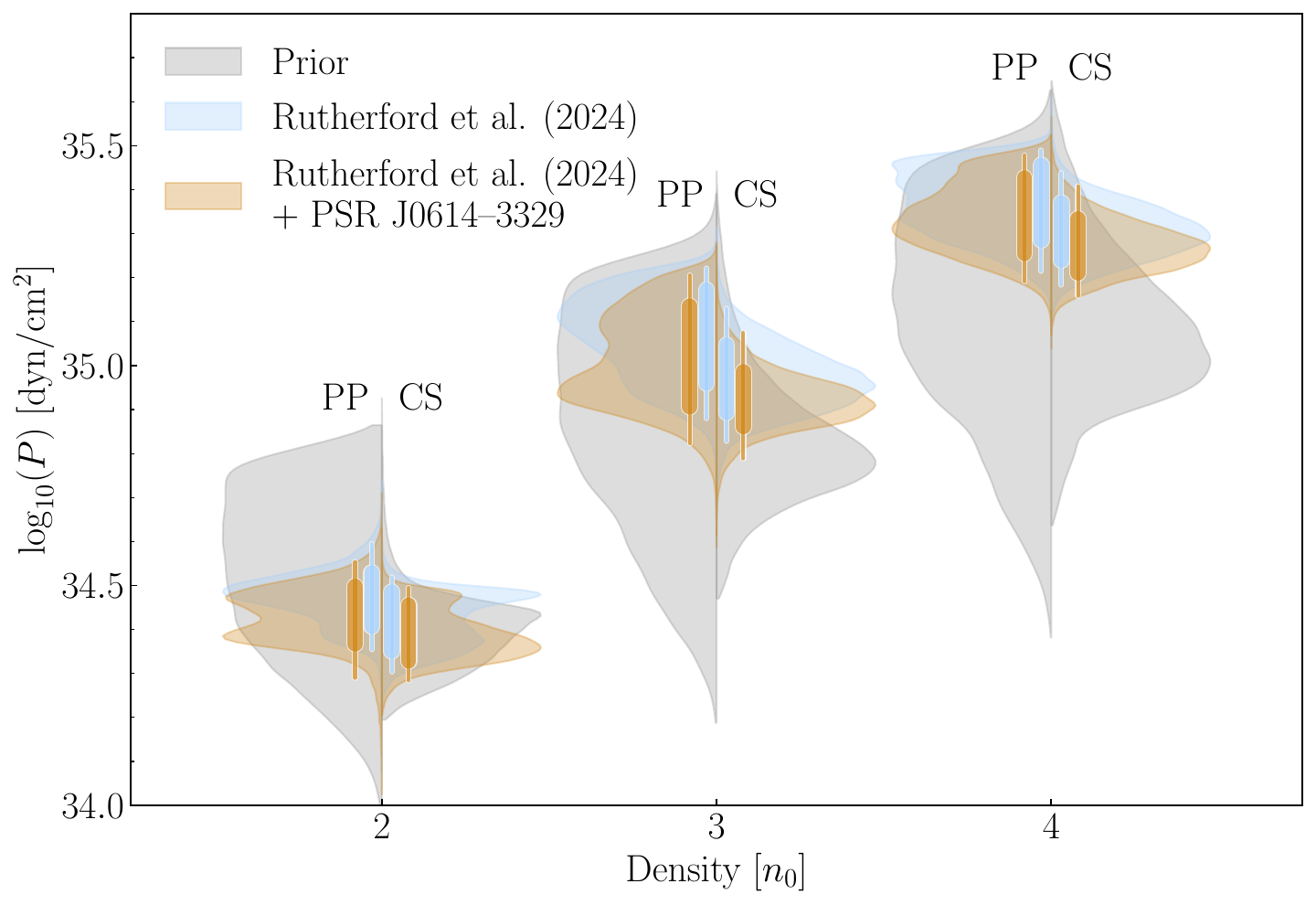}
    \caption{Prior (gray) and posterior distributions of the pressure at $2, 3$, and $4 n_0$ using the EOS inference based on the N$^3$LO $\chi$EFT band up to $1.5 n_0$ and the PP (left violins) and CS extensions (right violins). We compare the posterior distributions without the new NICER results given by \citet{rutherford_constraining_2024} (blue) and including the mass-radius constraint from PSR~J0614$-$3329 obtained in this work (golden). The vertical thick (thin) lines mark the equal-tailed 68\% (95\%) credible intervals (see Table~\ref{tab:pressure_CI} for associated values).}
    \label{fig:EOS_posterior}
\end{figure}

At the 68\% credible interval, the PP model posteriors show a bimodal-like structure. This is also seen in both the PP and CS models for the distribution of the radius $R_{1.4}$ of a $1.4 \, M_\odot$ NS in the insets of Figure~\ref{fig:MR_plane}. This bimodal-like structure is already present in the posterior mass-radius distributions without the inclusion of PSR~J0614$-$3329, but is less pronounced for the PP model and evenly peaked below and above 12\,km for the CS model \citep[see Figure 7 of][]{rutherford_constraining_2024}. These bimodal-like distributions are the result of larger radii (stiffer) EOS being favored by both the prior and the large measured mass of PSR~J0740$+$6620 plus smaller radii (softer) EOS being favored by both PSR~J0437$-$4715, PSR~J0614$-$3329, GW170817 data, and, marginally, the ST+PDT contour of PSR~J0030$+$0451, which overlaps with the mass-radius contours of the latter sources. Including PSR~J0614$-$3329, which offers more posterior support for mass-radius samples below 12\,km, make this bimodal-like structure slightly less pronounced for the CS model and more pronounced for the PP model. This suggests that the inclusion of PSR~J0614$-$3329 slightly softens the EOS. This trend is confirmed in the pressure posterior distributions at $2, 3$ and $4 n_0$ shown in Figure~\ref{fig:EOS_posterior} (see also the values in Table~\ref{tab:pressure_CI}), where the pressure is pushed to slightly lower values, while still being on the stiffer end of the priors, confirming a pattern seen in \cite{Raaijmakers21} and \cite{rutherford_constraining_2024}. As expected, the bimodal-like feature is also present in the pressure posterior distributions, particularly at $2 n_0$.

Adding PSR~J0614$-$3329 allows for slightly more negative difference of the radii of 2.0 and $1.4\,M_\odot$ NSs,  $\Delta R = R_{2.0} - R_{1.4}$. This can be explained by a softening of the EOS, as it would reduce more prominently the posterior radius around 2.0 \Msun~than at 1.4 \Msun. This effect can be seen on the lower radius branch of the bimodal-like structure for PP. Hence, the EOS slightly softening with the inclusion of PSR~J0614$-$3329 is further confirmed. These values, along with other key quantities, are provided on Table~\ref{tab:pressure_CI}. 

\begin{table}[!ht]
\caption{Results from the posterior EOS distributions using the N$^3$LO $\chi$EFT band up to $1.5 n_0$ and the PP (middle column) or CS extensions (right column) including the new mass-radius constraints from PSR~J0614$-$3329 (corresponding to the gold regions in Figures \ref{fig:MR_plane} and \ref{fig:EOS_posterior}). }
\begin{ruledtabular}
\begin{tabular}{ l c c }
& PP & CS \\
\hline
$R_{1.4}$\, [km] & $12.05^{+0.56}_{-0.79}$ & $11.71^{+0.71}_{-0.63}$ \\
$R_{2.0}$\, [km] & $11.99^{+0.85}_{-1.25}$ & $11.20^{+1.05}_{-0.94}$ \\
$\Delta R$\, [km] & $-0.10^{+0.41}_{-0.75}$ & $-0.56^{+0.60}_{-0.73}$ \\
\hline
$M_{\rm TOV}$\, [\Msun] & $2.13^{+0.13}_{-0.18}$ & $2.05^{+0.24}_{-0.16}$ \\
$R_{\rm TOV}$\, [km] & $11.72^{+1.13}_{-1.33}$ & $10.65^{+1.30}_{-0.87}$ \\
\hline
$\log_{10} \varepsilon_{c, \rm{TOV}}$\, [g/cm$^3$] & $15.11^{+0.27}_{-0.20}$ & $15.43^{+0.05}_{-0.14}$ \\
$n_{c,\rm{TOV}}/n_0$\, & $4.11^{+2.53}_{-1.31}$ & $7.17^{+1.24}_{-1.42}$ \\
$\log_{10} P_{c, \rm{TOV}}$\, [dyn/cm$^2$] & $35.61^{+0.35}_{-0.32}$ & $35.94^{+0.23}_{-0.39}$ \\
\hline
$\log_{10} P (2n_0)$\, [dyn/cm$^2$] & $34.43^{+0.12}_{-0.14}$ & $34.38^{+0.12}_{-0.09}$ \\
$\log_{10} P (3n_0)$\, [dyn/cm$^2$] & $35.01^{+0.20}_{-0.18}$ & $34.92^{+0.16}_{-0.13}$ \\
$\log_{10} P (4n_0)$\, [dyn/cm$^2$] & $35.33^{+0.14}_{-0.14}$ & $35.27^{+0.14}_{-0.11}$ \\
\end{tabular}
\end{ruledtabular}
    \begin{tablenotes}
      \item \textbf{Notes} - $R_{1.4}$ and $R_{2.0}$ are the radii of a $1.4$\,\Msun \, and $2.0$\,\Msun \,NS, with $\Delta R = R_{2.0} - R_{1.4}$. $M_{\rm TOV}$ and $R_{\rm TOV}$ are the mass and radius of a non-rotating, maximum mass NS, with $\varepsilon_{c, \rm{TOV}}$, $n_{c,\rm{TOV}}$ and $P_{c, \rm{TOV}}$ being its central energy density, density, and pressure, respectively. We also give the pressure $P$ at 2, 3, and $4 n_0$ (see Figure~\ref{fig:EOS_posterior} for associated distributions). For this table only, to compare with \cite{rutherford_constraining_2024}, median values with equal-tailed $95\%$ credible intervals are given.
    \end{tablenotes}
\label{tab:pressure_CI}
\end{table}

In addition, if instead of the Headline \STPDT, the \PDTU\, geometry had been used for PSR~J0614$-$3329, we expect that the new EOS inferences would still have stronger support for softer EOS than the ``New'' scenario in \cite{rutherford_constraining_2024}, while the reported mass-radius contours at $68\%$ and $95\%$ confidence intervals would change notably less from the ``New'' scenario than in the \STPDT\, case, as mass-radius posteriors from the exploratory \PDTU\, run mostly overlap the PSR~J0437$-$4715 ones (Figure~\ref{fig:RM_differentModels}).

\subsection{Caveats}

During our analyses, we did not try the \textit{Concentric Single Temperature} (CST) or \textit{Concentric Double Temperature} (CDT) hot region models (see \citealt{riley_nicer_2019}). These are made of two concentric overlapping circular spots of different temperatures. The configuration of such models can be reproduced by the PDT model, where both spot centers align, which would result in a $\zeta$ parameter (angular distance between ceding and superseding components centers) to be close to $0$. For the \PDTU\, model, we found $\zeta \sim 29\pm14 ^\circ$ for both hot regions, while for the Headline \STPDT, the mode closest to such a configuration had $\zeta \sim 4.5^{+3.0}_{-1.5} \,^\circ$. In the latter case, although $\zeta$ is small it is only compatible with 0 at $\sim 3 \sigma$.

We also did not test any \textit{Protruding Single Temperature} (PST) hot region (a PDT region whose superseding region does not emit thermal radiation). The PST configuration can be reproduced by the PDT model, which  would appear in the posterior as an unconstrained distribution of the superseding component temperature, ranging from the lowest prior value up to $\sim 10^{5.4}\,\rm{K}$, where the superseding region starts to emit significant amount of radiation. No such structure was found in the posteriors of the \STPDT\, and \PDTU\, models.

Although the PST, CST and CDT geometric models did not appear as specific solutions of a PDT region, it would, however, be interesting to test them in further analyses of this source.

This issue of model choice is central in PPM, as it affects inferred radii constraints (see for instance \citealt{vinciguerra_updated_2024}). Choosing an overly complex model can result in biased radius measurements with overestimated evidence, as it might be the case with \PDTU. Here, we chose for our headline result to use the \STPDT\, model over the \PDTU\, one, which has better evidence, although not significantly. Thorough model comparison was undertaken, but we could not definitively tell whether the \PDTU\, model was preferred or not (see Section~\ref{subsec:model_comparison}). The similarity and consistency of radii and geometry constraints from \PDTU\, and \STPDT, as well as \STPDT\, being the simplest model fitting the data well, motivated the choice of the latter for our headline run. Our limited computation resources also weighed in on this choice. Further X-PSI analyses of this source could be carried out with a high resolution \PDTU\, model.

Although many tests have been done in Section~\ref{sec:Validation} to assess the efficiency of the sampling and the accuracy of the likelihood, it is impossible to assert the completeness of the sampling. There is, however, no strong evidence supporting the incomplete sampling case and the procedure seems to have effectively converged as the headline \STPDT\, run yielded similar results as the exploratory run. In retrospect, this supports our choice of LR for exploratory runs (Section \ref{subsec:xpsi}), which proved to be a sufficiently high resolution as we found no discrepancies between the exploratory and Headline \STPDT\, runs. Using higher resolution X-PSI settings for the Headline \STPDT\, was, however, required to look for complex geometries that could have been missed during sampling. These higher resolution X-PSI settings were picked judiciously so that computation time was not too heavily affected while improving accuracy by one order of magnitude (Appendix \ref{app:compare_resolution}). Moreover, we also showed that the LR exploratory run posterior samples have likelihoods consistent with that of a high resolution likelihood (Section~\ref{subsec:XPSIresolution}) and that our inference is resilient to noise (Section~\ref{subsec:noise}). 

\subsection{Different modes and surface geometries}

As the inferred distributions appear to have different modes whose distributions do not overlap at $> 5\sigma $ in some region of the parameter space, using the multi-modal version of \textsc{MultiNest} increases the acceptance rate of samples, which can help to reduce computation time while retaining sampling accuracy \citep{shaw_efficient_2007,MultiNest_2009}. The inferred radii differ for the two possible headline geometries (Figure~\ref{fig:BestFitGeometries}), which individually have different radius values and lower uncertainties (see posterior distributions in Appendix~\ref{app:headline}). Hence, breaking this multi-modality and converging on one single mode would reduce the total uncertainty in the radius measurement by a few hundred meters.

Physically motivated hot region priors can help favoring one geometry over the other by ruling out geometric configurations \citep{Kalapotharakos_geometry_2009, Das_PPM_GRMHD_2025}. Such priors can also help differentiate between \STPDT\, and \PDTU\, models.

Future improvements in the available data with more NICER exposure, more accurate priors on distance, inclination and mass from radio pulsar timing \citep[e.g., with the Square Kilometer Array,][]{ska_2025_densematter} can also help break this multi-modality. As the XMM exposure is only $\sim14\,$ks long\footnote{To put this in context with previous analyses, the MOS1/2 exposures were $\sim 18\,$ks for PSR~J0740$+$6620 \citep{salmi_radius_2022}, $\sim 120\,$ks long for PSR~J0030$+$0451 \citep{vinciguerra_updated_2024}, and $\sim 180\,$ks long for PSR~J0437$-$4715 \citep{choudhury_nicer_2024}.}, more XMM data would be valuable to further constrain the background, one of the main limitations of our study. This will in turn help to tighten the uncertainties on the inferred radius. Higher quality data obtained with next-generation X-ray observatories such as eXTP \citep[$\sim 2030$,][]{Li_eXTP_2025} and NewAthena \citep[$\sim 2037$,][]{cruise_newathena_2024} will also be of great help for this endeavor. These observatories will also decrease uncertainties of individual modes thanks to their large effective area, time resolution and constrained background.

\section{Conclusion} \label{sec:conclusion}

In this work, we have presented the radius measurement of PSR~J0614$-$3329 with PPM using Bayesian inference on joint NICER and XMM data. Our analyses benefited from tight priors on both mass and inclination. The results were consistent across tested models, all revealing a similar geometry composed of polar and equatorial hot regions. The inferred radii for different models were all consistent at the 68\% confidence level and the final measured radius is $R_{\rm eq}=10.29^{+1.01}_{-0.86}\,$km for a mass of $M=1.44^{+0.06}_{-0.07}\,M_\odot$ using an \STPDT\, model. This radius range is consistent with GW observation of GW170817 as well as with PPM measurements of PSR~J0437$-$4715 and the \STPDT\, model of PSR~J0030$+$0451, all of which are in the same mass range.

Various tests were carried out to assess the validity of different inferences. It was found that (1) the effect of LR likelihood on sampling for the exploratory runs was limited, (2) the inferred parameter posterior distributions were not affected by noise realization, and (3) model comparison disfavored \STU\, and did not significantly favor \PDTU\, over \STPDT. As inferred radii and geometries were consistent, we chose the simpler \STPDT\, model, which explains well the data, to run the Headline result.  

The effect of the tight inclination prior and atmosphere ionization state on the inferred radius and geometry for the \STPDT\, model was also tested. These show little dependency of inferred parameters on these assumptions, although the inclination prior helped to better constrain the geometry and the radius. Along with the consistent results across all different geometric models, these make the constraints obtained in this study resilient to various hypotheses and systematics.

Further EOS analyses of the impact of this radius measurement result in a slight softening of the EOS compared to our previous work \citep{rutherford_constraining_2024}, as well as shifting the allowed mass-radius
region toward lower radii by $\sim 300$\,m for both high-density EOS extensions, making them compatible with previous analyses within less than one standard deviation. For the maximum mass, this new radius constraint has little impact.

This radius measurement could be narrowed down in the future with better geometry priors to break degeneracies, with improved X-ray data from XMM, NewAthena, or eXTP to better constrain the background or with a tighter prior on the mass from radio timing.

\vspace{1cm}
\textit{\large Acknowledgments}: 
The authors acknowledge fruitful discussion with Vivek~Krishnan and Michael~Kramer about the pulsar mass measurements from radio timing with MeerKAT. The authors thank the referee for their suggestions, which contributed to improve this article. This study has been partially supported through the grant EUR TESS N°ANR-18-EURE-0018 in the framework of the Programme des Investissements d'Avenir. This work was performed using HPC resources from CALMIP (Grant 2016-19056). Part of this work has been supported by the CEFIPRA grant IFC/F5904-B/2018 and ANR-20-CE31-0010 (MORPHER). All researchers from IRAP acknowledge the support of the CNES. Portions of this work performed at NRL were supported by NASA. D.C., B.D. and A.L.W. acknowledge support from ERC Consolidator grant No.~865768 AEONS (PI: Watts). M.H., Y.K. and A.L.W. acknowledge support from the NWO grant ENW-XL OCENW.XL21.XL21.038 \textit{Probing the phase diagram of Quantum Chromodynamics} (PI: Watts). The work of M.M., A.S., and I.S. was supported by the European Research Council (ERC) under the European Union’s Horizon 2020 research and innovation program (Grant Agreement No. 101020842). N.R.~acknowledges support from NASA grant No.80NSSC22K0092. S.B.~acknowledges funding from NASA grants 80NSSC20K0275 and 80NSSC22K0728. The work presented is based in part on observations with XMM-Newton, an ESA Science Mission with instruments and contributions directly funded by ESA Member states and NASA. This research has made use of data products and software provided by the High Energy Astrophysics Science Archive Research Center (HEASARC), which is a service of the Astrophysics Science Division at NASA/GSFC and the High Energy Astrophysics Division of the Smithsonian Astrophysical Observatory.  This work has relied on NASA's Astrophysics Data System (ADS) bibliographic services and the ArXiv. 

\vspace{5mm}
\textit{\large Facilities}: NICER, XMM-Newton

\vspace{5mm}
\textit{\large Softwares}: ArviZ~\citep{kumar_arviz_2019}, 
         Cython~\citep{cython2011},
         fgivenx~\citep{fgivenx},
         GetDist~\citep[][\url{https://github.com/cmbant/getdist}]{Lewis19},
         GNU~Scientific~Library~\citep[GSL;][]{Gough:2009},
         HEASoft \citep{2014ascl.soft08004N},         IPython~\citep{IPython2007},
         Jupyter~\citep{Kluyver:2016aa},
         Matplotlib~\citep{Hunter:2007,matplotlibv2},
         MPI~\citep{MPI},
         MPI for Python~\citep{mpi4py},
         \textsc{MultiNest}~\citep{MultiNest_2009},
         NEOST~\texttt{v2.1.0} (\url{https://github.com/xpsi-group/neost}; \citealt{Raaijmakers24}),
         nestcheck~\citep{higson2018nestcheck,higson2018sampling, higson2019diagnostic},
         NumPy~\citep{Numpy2011},
         OpenMP~\citep{openmp},
         \textsc{PyMultiNest}~\citep{PyMultiNest},
         Python/C~language~\citep{python2007},
         SAS \citep{2004ASPC..314..759G}
         SciPy~\citep{Scipy},
         X-PSI~\texttt{v2.2.7} (\url{https://github.com/xpsi-group/xpsi}; \citealt{xpsi})

\appendix

\section{Test of X-PSI resolution}\label{app:compare_resolution}

To limit the computation time needed for our Headline result, the effects of the different X-PSI settings on likelihood computation time and accuracy were tested on a set of $1000$ samples from the posterior of the exploratory \STPDT\, model (see Section~\ref{subsec:ST+PDT}). Our goal is to attain higher accuracy while limiting computation time.

The different settings whose effects on likelihood accuracy were studied are given in Table~\ref{tab:XPSI_settings}. \texttt{sqrt\_num\_cells} is the square root of the approximate number of cells used to represent a spot and \texttt{num\_rays} is the number of emission angle grid points for which light bending angles are precomputed. When the hot region has two temperature components (ceding and superseding), \texttt{min\_sqrt\_num\_cells} and \texttt{max\_sqrt\_num\_cells} are, respectively, the minimum and maximum values of \texttt{sqrt\_num\_cells} for the individual components. \texttt{num\_leaves} and \texttt{num\_energies} are, respectively, the sizes of the regularly spaced arrays of phase and energy over which the model is computed. Folding with the instrument response and integrating over the data phase bins, we get a modeled observation for a set of parameters, which is compared to the data to compute the likelihood. Increasing any of these values improves the accuracy of the likelihood computation.

\begin{figure}[!t]
    \centering
    \includegraphics[width=\linewidth]{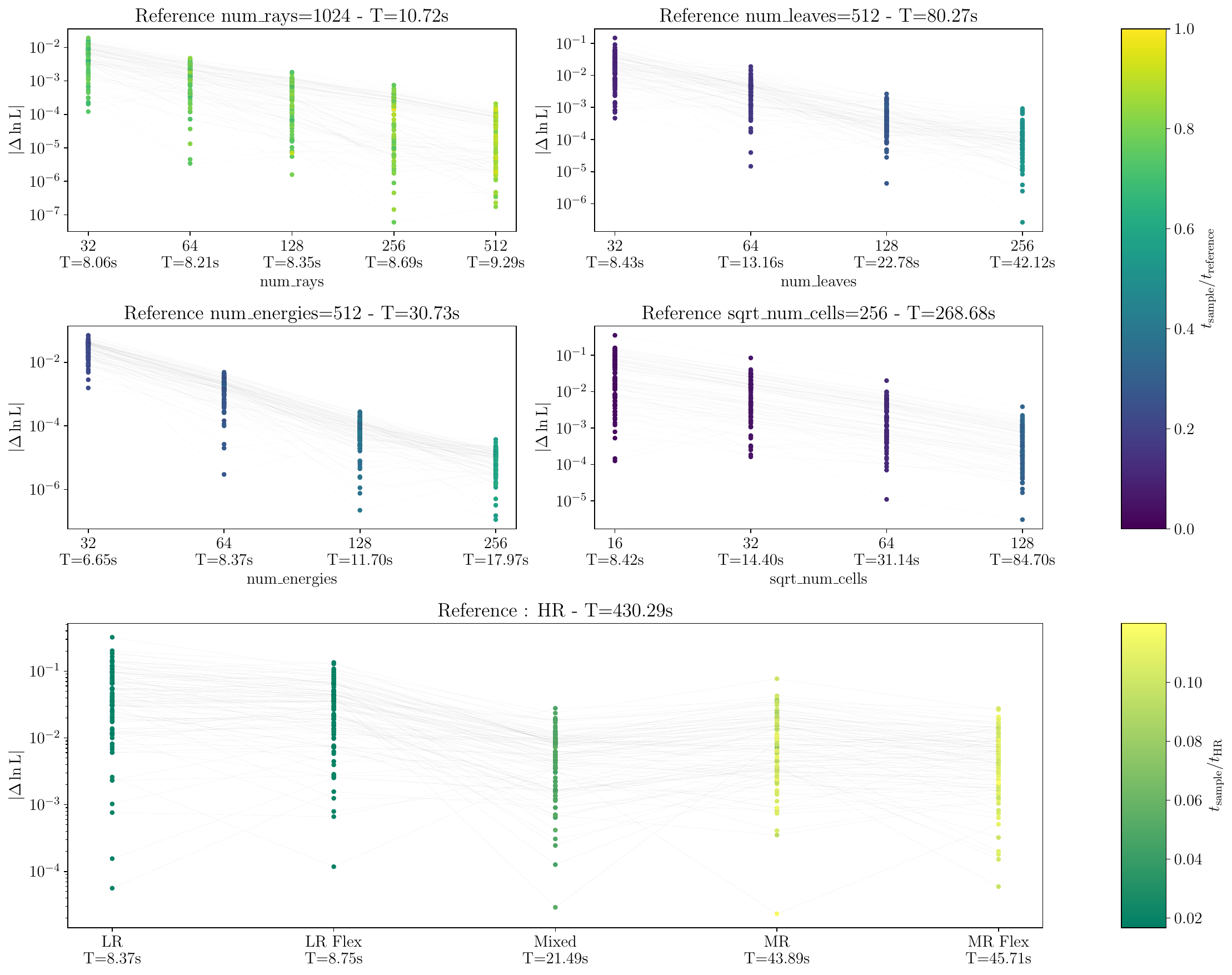}
    \caption{Comparison of likelihood accuracies and computation times for different settings configurations. The first two rows show variation of individual settings, all other settings being fixed to their LR values, while the last row shows the variation of different settings relative to a HR configuration. Along with the computation time totaled across all samples, the abscissae axes represent the used X-PSI setting value for the first two rows and is arbitrary for the final one. The ordinate axes give the likelihood accuracy $|\Delta \ln\text{L}|$ with respect to a reference with the highest setting value tested, whose configuration is in the title of each panel. The colorbars represent the ratio of computation time of each sample to the reference setting, with a single colorbar shared for the first two rows and a different one for the last row. Gray lines link together points obtained with the same posterior sample. }
    \label{fig:compareXPSIparams}
\end{figure}

We first investigated the effect of individual X-PSI settings, all other settings being fixed to their LR value. For varying setting values, we measured and registered the computation time along with the likelihood value for each sample in our set. These are then compared to the highest resolution setting tested. Results are plotted on the first two rows of Figure~\ref{fig:compareXPSIparams}.

From these investigations, we can see that the computation time has a linear dependence on \texttt{num\_leaves} and a superlinear dependence \texttt{sqrt\_num\_cells}, while it is sublinear for the other two settings. Moreover, increasing \texttt{sqrt\_num\_cells} does not improved the likelihood accuracy as much as it does for the other settings. It is hence important to keep this setting at values tailored to the problem in order to reduce computation time. For instance, ST regions need less \texttt{sqrt\_num\_cells} than PDT ones for an accurate likelihood computation, as the latter is more geometrically complex, especially in the case of thin and elongated structures. Having \texttt{min\_sqrt\_num\_cells} and \texttt{max\_sqrt\_num\_cells} different than \texttt{sqrt\_num\_cells} also allows for more accurate model computation with PDT geometries by using an adaptive mesh to represent the hot regions.

Different X-PSI resolutions, with different settings given in Table~\ref{tab:XPSI_settings}, were then investigated in a similar manner. Note that the LR settings were used for the exploratory runs and that the XR settings are different for ST and PDT. Moreover, the additional Flex settings configurations allow \texttt{min\_sqrt\_num\_cells} and \texttt{max\_sqrt\_num\_cells} to be different to \texttt{sqrt\_num\_cells}. Results for the different configurations can be found on the last row of Figure~\ref{fig:compareXPSIparams}. 

\begin{table*}[!ht]
    \centering
    \begin{tabular}{l|ccc|cc|cc}
         \hline
         \hline
         Settings & LR & MR & HR & LR~Flex & MR~Flex & XR - ST & XR - PDT \\
         \hline
         \texttt{num\_rays} & 100 & 512 & 640 & 100 & 512 & 512 & 512 \\
         \texttt{num\_leaves} & 32 & 64 & 128 & 32 & 64 & 64 & 64 \\
         \texttt{num\_energies} & 64 & 128 & 256 & 64 & 128 & 64 & 64 \\
         \texttt{sqrt\_num\_cells} & 16 & 32 & 64 & 16 & 32 & 16 & 32 \\
         \texttt{min\_sqrt\_num\_cells} & 16 & 32 & 64 & 10 & 16 & 16 & 16 \\
         \texttt{max\_sqrt\_num\_cells} & 16 & 32 & 64 & 32 & 80 & 16 & 80 \\
         \hline 
         \hline
    \end{tabular}
    \caption{Values of the different X-PSI settings, each column displays a different likelihood resolution. Low resolution (LR), medium resolution (MR), and high resolution (HR) values are consistent with resolutions from \cite{choudhury_exploring_2024}, besides for \texttt{min\_sqrt\_num\_cells} and \texttt{max\_sqrt\_num\_cells}. Flex configuration allow for values of \texttt{min\_sqrt\_num\_cells} and \texttt{max\_sqrt\_num\_cells} to be different to that of \texttt{sqrt\_num\_cells}. There is a single mixed resolution (XR) configuration that has different settings for ST and PDT regions.}
    \label{tab:XPSI_settings}
\end{table*}

The LR~Flex configuration does not give strong accuracy improvements and is slower than the previously used LR. All Mixed, MR, and MR~Flex give comparable likelihood accuracies. Among these, the Mixed configuration gives computation time $\sim 2$ times faster than the others. Hence, this was used for the Headline result.  

\pagebreak

\section{Headline result}\label{app:headline}

\begin{figure}
    \centering
    \includegraphics[width=\linewidth]{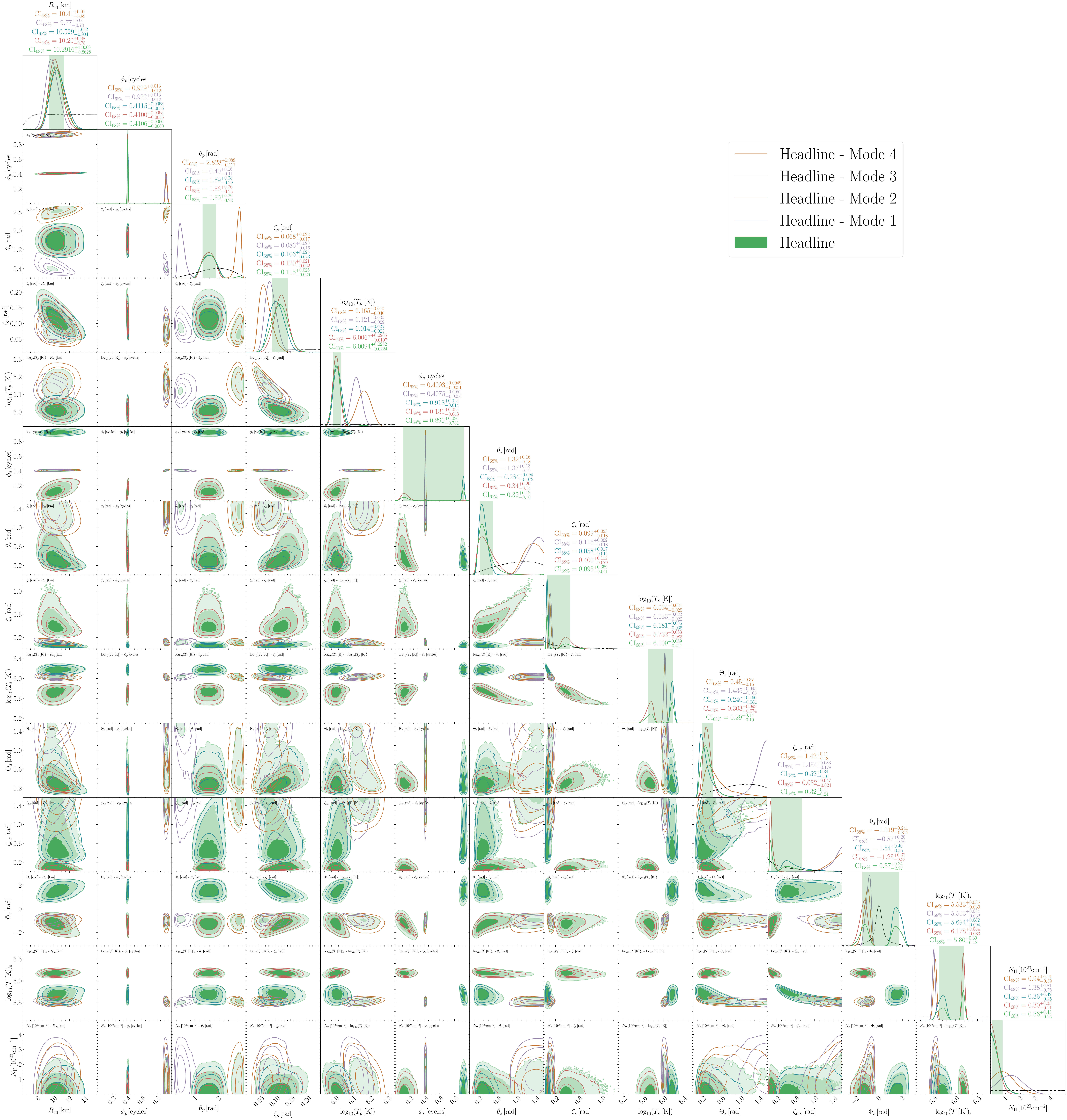}
    \caption{Posterior distribution of the radius, hotspot geometric parameters and hydrogen column density for the Headline \STPDT\, run. Individual mode distributions (red, blue, purple, orange) are plotted along with the full distribution (green). The left plot from Figure~\ref{fig:BestFitGeometries} corresponds to Mode 1 and the right plot to Mode 2. The dashed-dotted black lines on the diagonal plots show the marginalized prior distributions, shared across all runs. The shaded vertical bands show the $68.3\%$ credible intervals. The contours in the 2D posteriors show the $68.3\%$, $95.4\%$, and $99.7\%$ credible regions, which are filled for the mode-averaged distribution only. The full posterior distribution, including all $21$ parameters, can be found in the Zenodo repository \citep{zenodo}.}
    \label{fig:Headline_posterior}
\end{figure}

The Headline \STPDT\, run summary is given in Table~\ref{tab:Headline_result}. The posterior distribution of a subset of parameters from the headline \STPDT\,run, with individual modes detailed, is given in Figure~\ref{fig:Headline_posterior}. The figure of the full posterior distributions, including all $21$ parameters, can be found in the Zenodo repository \citep{zenodo}. 

\begin{table*}[!ht]
    \centering
    \caption{Summary table for the Headline run \STPDT.}
    \begin{tabular}{llllll}
    \hline \hline 
        Parameter & Description & Prior PDF & $\widehat{\rm CI}_{68\%}$ & $\widehat{\rm ML}$ & $\widehat{D}_{\rm KL}$ \\
        \hline
        $P$ [ms] & Coordinate spin period& fixed &  & \\
        \hline
        $M$ [$M_\odot$] & Gravitational mass & $\mathcal{N}(1.44,0.07)$ & $1.444^{+0.063}_{-0.065}$ & 1.548 & 0.016 \\
        $R_{\rm eq}$ [km] & Coordinate equatorial radius& $\mathcal{U}(6, 16)$ & $10.29^{+1.01}_{-0.86}$ & 12.61 & 1.330 \\
        $D$ [kpc] & Earth distance& $\mathcal{N}(0.595,0.045)$ & $0.607^{+0.039}_{-0.042}$ & 0.6685 & 0.064 \\
        $\sin ( i )$& Sine of angle from line of sight to spin axis & $\mathcal{N}(0.999559,0.000117)$ & $0.9996\pm0.0001$ & 0.999575 & 0.004 \\
        \hline
        & Compactness condition & $R_{\rm polar}/r_g(M) > 3$ & & &\\
        & Surface gravity condition & $ 13.7 \le \log_{10} g_{\rm eff} \le 15.0$ & & &\\
        \hline
        $\phi_p$ [cycles]  & p region - initial phase & $\mathcal{U}(0,1)$ & $0.4106\pm0.0060$ & 0.412 & 5.213 \\
        $\theta_p$ [rad]  & p region - center colatitude & $\cos(\theta_p) \sim \mathcal{U}(-1,1)$ & $1.59^{+0.29}_{-0.28}$ & 1.424 & 0.887 \\
         $\zeta_p$ [rad]  & p region - angular radius \tnote{1} & $\mathcal{U}(0,\pi / 2 )$ & $0.115^{+0.025}_{-0.026}$ & 0.067 & 3.271 \\
        $\log_{10}(T_p \text{[K]})$ & p region - effective temperature& $\mathcal{U}(5.1,6.8)$ & $6.009^{+0.025}_{-0.022}$ & 6.044 & 3.972 \\
        $\phi_s$ [cycles]  & s region - initial phase  $^{\rm a}$ & $\mathcal{U}(0,1)$ & $0.890^{+0.036}_{-0.781}$ & 0.919 & 2.119 \\
        $\theta_s$ [rad]  & s region, superseding - center colatitude& $\cos(\theta_s) \sim \mathcal{U}(0,1)$ & $0.32^{+0.18}_{-0.10}$ & 0.191 & 2.254 \\
        $\zeta_s$ [rad]  & s region, superseding - angular radius& $\mathcal{U}(0,\pi / 2 )$ & $0.093^{+0.359}_{-0.041}$ & 0.051 & 1.115 \\
        $\log_{10}(T_s \text{[K]})$ & s region, superseding - effective temperature& $\mathcal{U}(5.1,6.8)$ & $6.109^{+0.089}_{-0.417}$ & 6.209 & 1.911 \\
        $\Theta_s$ [rad]  & s region, ceding - center colatitude& $\cos(\Theta_s) \sim \mathcal{U}(0,1)$ & $0.29^{+0.14}_{-0.10}$ & 0.10 & 2.665 \\
        $\zeta_{c,s}$ [rad]  & s region, ceding - angular radius& $\mathcal{U}(0,\pi/2)$ & $0.32^{+0.41}_{-0.24}$ & 0.56 & 0.385 \\
        $\Phi_s$ [rad]  & Azimuthal offset between the s components& $\mathcal{U}(-\pi,\pi)$& $0.87^{+0.84}_{-2.27}$ & 2.05 & 2.519 \\
        $\log_{10}(T_{s,c} \text{[K]})$ & s region, ceding - effective temperature& $\mathcal{U}(5.1,6.8)$ & $5.80^{+0.39}_{-0.18}$ & 5.73 & 1.846 \\
        \hline
        & Non overlapping hot regions condition & & & &\\
        \hline
        $N_{\rm H} [10^{20} \text{cm}^{-2}]$ & Interstellar neutral Hydrogen column density & $\mathcal{U}(0.001,10)$ & $0.36^{+0.43}_{-0.25}$ & 0.124 & 3.064 \\
        $\alpha_{\text{NICER}} $ & NICER effective area scaling factor & $\mathcal{N}(1.0,0.104)$ & $0.968^{+0.098}_{-0.091}$ & 1.17 & 0.071 \\
        $\alpha_{\text{EPN}} $ & PN effective area scaling factor & $\mathcal{N}(1.0,0.104)$ & $0.955^{+0.069}_{-0.067}$ & 0.867 & 0.3230 \\
        $\alpha_{\text{EMOS1}} $ & MOS1 effective area scaling factor & $\mathcal{N}(1.0,0.104)$ & $1.055^{+0.076}_{-0.075}$ & 0.982 & 0.292 \\
        $\alpha_{\text{EMOS2}} $ & MOS2 effective area scaling factor & $\mathcal{N}(1.0,0.104)$ & $1.019^{+0.076}_{-0.074}$ & 0.911 & 0.142 \\
        \hline \hline
        & Sampling information & & & &\\
        \hline
        & Number of parameters: $21$ & & & &\\
        & Number of live points: $2 \times10^4$ & & & &\\
        & Sampling efficiency: $0.05$ & & & &\\
        & Evidence tolerance: $0.1$ & & & &\\
        & Likelihood evaluations: $248\,129\,437$ & & & &\\
        & Estimated evidence: $\ln \mathcal{Z} = 34585.589 \pm 0.041$ & & & &\\
        & Computation time: $70\,207$ CPU hours & & & &\\
        & on Intel Skylake 6140 at 2.3 Ghz with 36 cores & & & &\\
        \hline \hline
    \end{tabular}
    \label{tab:Headline_result}
        \begin{tablenotes}
        \item \textsc{Notes} - We report the prior PDFs, the median with the surrounding $68.3\%$ credible intervals $\widehat{\rm CI}_{68\%}$, the maximum likelihood \\ parameter vector $\widehat{\rm ML}$, and the Kullback–Leibler divergence $\widehat{D}_{\rm KL}$ in bits representing prior-to-posterior information gain.
        \item $^{\rm a}$ Unlike most previous analyses, the secondary spot phase is not shifted by $0.5$ cycles.
        \end{tablenotes}
\end{table*}

\pagebreak

\bibliography{sample631}{}

@dataset{zenodo,
  author       = {Mauviard, Lucien and
                  Guillot, Sebastien and
                  Salmi, Tuomo and
                  Choudhury, Devarshi and
                  Dorsman, Bas and
                  González-Caniulef, Denis and
                  Hoogkamer, Mariska and
                  Huppenkothen, Daniela and
                  Kazantsev, Christine and
                  Kini, Yves and
                  Olive, Jean-François and
                  Stammler, Pierre and
                  Watts, Anna and
                  Mendes, Melissa and
                  Rutherford, Nathan and
                  Schwenk, Achim and
                  Svensson, Isak and
                  Bogdanov, Slavko and
                  Kerr, Matthew and
                  Ray, Paul and
                  Guillemot, Lucas and
                  Cognard, Ismaël and
                  Theureau, Gilles},
  title        = {Reproduction Package for: "A NICER view of the 1.4 solar masses edge-on pulsar PSR J0614-3329"},
  month        = jun,
  year         = 2025,
  publisher    = {Zenodo},
  version      = {2.0.0},
  doi          = {10.5281/zenodo.15603405},
  url          = {https://doi.org/10.5281/zenodo.15603405},
}

@article{Chatziioannou:2024tjq,
       author = {{Chatziioannou}, Katerina and {Cromartie}, H. Thankful and {Gandolfi}, Stefano and {Tews}, Ingo and {Radice}, David and {Steiner}, Andrew W. and {Watts}, Anna L.},
        title = "{Neutron stars and the dense matter equation of state: from microscopic theory to macroscopic observations}",
      journal = {arXiv e-prints},
         year = 2024,
        month = jul,
          eid = {arXiv:2407.11153},
        pages = {arXiv:2407.11153},
          doi = {10.48550/arXiv.2407.11153},
archivePrefix = {arXiv},
       eprint = {2407.11153},
 primaryClass = {nucl-th},
       adsurl = {https://ui.adsabs.harvard.edu/abs/2024arXiv240711153C}
}

@article{Watts:2016uzu,
       author = {{Watts}, Anna L. and {Andersson}, Nils and {Chakrabarty}, Deepto and {Feroci}, Marco and {Hebeler}, Kai and {Israel}, Gianluca and {Lamb}, Frederick K. and {Miller}, M. Coleman and {Morsink}, Sharon and {{\"O}zel}, Feryal and {Patruno}, Alessandro and {Poutanen}, Juri and {Psaltis}, Dimitrios and {Schwenk}, Achim and {Steiner}, Andrew W. and {Stella}, Luigi and {Tolos}, Laura and {van der Klis}, Michiel},
        title = "{Colloquium: Measuring the neutron star equation of state using x-ray timing}",
      journal = {Reviews of Modern Physics},
         year = 2016,
        month = apr,
       volume = {88},
       number = {2},
          eid = {021001},
        pages = {021001},
          doi = {10.1103/RevModPhys.88.021001},
archivePrefix = {arXiv},
       eprint = {1602.01081},
 primaryClass = {astro-ph.HE},
       adsurl = {https://ui.adsabs.harvard.edu/abs/2016RvMP...88b1001W}
}

@article{Tews:2012fj,
       author = {{Tews}, I. and {Kr{\"u}ger}, T. and {Hebeler}, K. and {Schwenk}, A.},
        title = "{Neutron Matter at Next-to-Next-to-Next-to-Leading Order in Chiral Effective Field Theory}",
      journal = {\prl},
         year = 2013,
        month = jan,
       volume = {110},
       number = {3},
          eid = {032504},
        pages = {032504},
          doi = {10.1103/PhysRevLett.110.032504},
archivePrefix = {arXiv},
       eprint = {1206.0025},
 primaryClass = {nucl-th},
       adsurl = {https://ui.adsabs.harvard.edu/abs/2013PhRvL.110c2504T}
}

@article{Tews:2024owl,
       author = {{Tews}, I. and {Somasundaram}, R. and {Lonardoni}, D. and {G{\"o}ttling}, H. and {Seutin}, R. and {Carlson}, J. and {Gandolfi}, S. and {Hebeler}, K. and {Schwenk}, A.},
        title = "{Neutron matter from local chiral EFT interactions at large cutoffs}",
      journal = {arXiv e-prints},
         year = 2024,
        month = jul,
          eid = {arXiv:2407.08979},
        pages = {arXiv:2407.08979},
          doi = {10.48550/arXiv.2407.08979},
archivePrefix = {arXiv},
       eprint = {2407.08979},
 primaryClass = {nucl-th},
       adsurl = {https://ui.adsabs.harvard.edu/abs/2024arXiv240708979T}
}

@article{Koehn:2024set,
       author = {{Koehn}, Hauke and {Rose}, Henrik and {Pang}, Peter T.~H. and {Somasundaram}, Rahul and {Reed}, Brendan T. and {Tews}, Ingo and {Abac}, Adrian and {Komoltsev}, Oleg and {Kunert}, Nina and {Kurkela}, Aleksi and {Coughlin}, Michael W. and {Healy}, Brian F. and {Dietrich}, Tim},
        title = "{From Existing and New Nuclear and Astrophysical Constraints to Stringent Limits on the Equation of State of Neutron-Rich Dense Matter}",
      journal = {Physical Review X},
         year = 2025,
        month = apr,
       volume = {15},
       number = {2},
          eid = {021014},
        pages = {021014},
          doi = {10.1103/PhysRevX.15.021014},
archivePrefix = {arXiv},
       eprint = {2402.04172},
 primaryClass = {astro-ph.HE},
       adsurl = {https://ui.adsabs.harvard.edu/abs/2025PhRvX..15b1014K}
}

@ARTICLE{Greif19,
   author = {{Greif}, S.~K. and {Raaijmakers}, G. and {Hebeler}, K. and {Schwenk}, A. and 
	{Watts}, A.~L.},
    title = "{Equation of state sensitivities when inferring neutron star and dense matter properties}",
  journal = {MNRAS},
archivePrefix = "arXiv",
   eprint = {1812.08188},
 primaryClass = "astro-ph.HE",
     year = 2019,
   volume = 485,
    pages = {5363-5376},
      doi = {10.1093/mnras/stz654},
   adsurl = {http://adsabs.harvard.edu/abs/2019MNRAS.485.5363G}
}

@ARTICLE{Raaijmakers19,
       author = {{Raaijmakers}, G. and {Riley}, T.~E. and {Watts}, A.~L. and {Greif}, S.~K. and {Morsink}, S.~M. and {Hebeler}, K. and {Schwenk}, A. and {Hinderer}, T. and {Nissanke}, S. and {Guillot}, S. and {Arzoumanian}, Z. and {Bogdanov}, S. and {Chakrabarty}, D. and {Gendreau}, K.~C. and {Ho}, W.~C.~G. and {Lattimer}, J.~M. and {Ludlam}, R.~M. and {Wolff}, M.~T.},
        title = "{A Nicer View of PSR J0030+0451: Implications for the Dense Matter Equation of State}",
      journal = {\apjl},
         year = 2019,
        month = dec,
       volume = {887},
       number = {1},
          eid = {L22},
        pages = {L22},
          doi = {10.3847/2041-8213/ab451a},
archivePrefix = {arXiv},
       eprint = {1912.05703},
 primaryClass = {astro-ph.HE},
       adsurl = {https://ui.adsabs.harvard.edu/abs/2019ApJ...887L..22R}
}

@ARTICLE{Raaijmakers20,
       author = {{Raaijmakers}, G. and {Greif}, S.~K. and {Riley}, T.~E. and {Hinderer}, T. and {Hebeler}, K. and {Schwenk}, A. and {Watts}, A.~L. and {Nissanke}, S. and {Guillot}, S. and {Lattimer}, J.~M. and {Ludlam}, R.~M.},
        title = "{Constraining the Dense Matter Equation of State with Joint Analysis of NICER and LIGO/Virgo Measurements}",
      journal = {\apjl},
         year = 2020,
        month = apr,
       volume = {893},
       number = {1},
          eid = {L21},
        pages = {L21},
          doi = {10.3847/2041-8213/ab822f},
archivePrefix = {arXiv},
       eprint = {1912.11031},
 primaryClass = {astro-ph.HE},
       adsurl = {https://ui.adsabs.harvard.edu/abs/2020ApJ...893L..21R}
}

@ARTICLE{Raaijmakers21,
       author = {{Raaijmakers}, G. and {Greif}, S.~K. and {Hebeler}, K. and {Hinderer}, T. and {Nissanke}, S. and {Schwenk}, A. and {Riley}, T.~E. and {Watts}, A.~L. and {Lattimer}, J.~M. and {Ho}, W.~C.~G.},
        title = "{Constraints on the Dense Matter Equation of State and Neutron Star Properties from NICER's Mass-Radius Estimate of PSR J0740+6620 and Multimessenger Observations}",
      journal = {\apjl},
         year = 2021,
        month = sep,
       volume = {918},
       number = {2},
          eid = {L29},
        pages = {L29},
          doi = {10.3847/2041-8213/ac089a},
archivePrefix = {arXiv},
       eprint = {2105.06981},
 primaryClass = {astro-ph.HE},
       adsurl = {https://ui.adsabs.harvard.edu/abs/2021ApJ...918L..29R}
}

@ARTICLE{Raaijmakers24,
       author = {{Raaijmakers}, Geert and {Rutherford}, Nathan and {Timmerman}, Patrick and {Salmi}, Tuomo and {Watts}, Anna L. and {Prescod-Weinstein}, Chanda and {Svensson}, Isak and {Mendes}, Melissa},
        title = "{NEoST: A Python package for nested sampling of the neutron star equation of state}",
      journal = {The Journal of Open Source Software},
         year = 2025,
        month = jan,
       volume = {10},
       number = {105},
          eid = {6003},
        pages = {6003},
          doi = {10.21105/joss.06003},
       adsurl = {https://ui.adsabs.harvard.edu/abs/2025JOSS...10.6003R}
}

@article{Rutherford:2022xeb,
       author = {{Rutherford}, Nathan and {Raaijmakers}, Geert and {Prescod-Weinstein}, Chanda and {Watts}, Anna},
        title = "{Constraining bosonic asymmetric dark matter with neutron star mass-radius measurements}",
      journal = {\prd},
         year = 2023,
        month = may,
       volume = {107},
       number = {10},
          eid = {103051},
        pages = {103051},
          doi = {10.1103/PhysRevD.107.103051},
archivePrefix = {arXiv},
       eprint = {2208.03282},
 primaryClass = {astro-ph.HE},
       adsurl = {https://ui.adsabs.harvard.edu/abs/2023PhRvD.107j3051R}
}

@article{Keller2023,
       author = {{Keller}, J. and {Hebeler}, K. and {Schwenk}, A.},
        title = "{Nuclear Equation of State for Arbitrary Proton Fraction and Temperature Based on Chiral Effective Field Theory and a Gaussian Process Emulator}",
      journal = {\prl},
         year = 2023,
        month = feb,
       volume = {130},
       number = {7},
          eid = {072701},
        pages = {072701},
          doi = {10.1103/PhysRevLett.130.072701},
archivePrefix = {arXiv},
       eprint = {2204.14016},
 primaryClass = {nucl-th},
       adsurl = {https://ui.adsabs.harvard.edu/abs/2023PhRvL.130g2701K}
}

@article{Hebeler2013,
       author = {{Hebeler}, K. and {Lattimer}, J.~M. and {Pethick}, C.~J. and {Schwenk}, A.},
        title = "{Equation of State and Neutron Star Properties Constrained by Nuclear Physics and Observation}",
      journal = {\apj},
         year = 2013,
        month = aug,
       volume = {773},
       number = {1},
          eid = {11},
        pages = {11},
          doi = {10.1088/0004-637X/773/1/11},
archivePrefix = {arXiv},
       eprint = {1303.4662},
 primaryClass = {astro-ph.SR},
       adsurl = {https://ui.adsabs.harvard.edu/abs/2013ApJ...773...11H}
}

@ARTICLE{Baym71,
       author = {{Baym}, Gordon and {Pethick}, Christopher and {Sutherland}, Peter},
        title = "{The Ground State of Matter at High Densities: Equation of State and Stellar Models}",
      journal = {\apj},
         year = 1971,
        month = dec,
       volume = {170},
        pages = {299},
          doi = {10.1086/151216},
       adsurl = {https://ui.adsabs.harvard.edu/abs/1971ApJ...170..299B}
}

@ARTICLE{xpsi,
       author = {{Riley}, Thomas E. and {Choudhury}, Devarshi and {Salmi}, Tuomo and {Vinciguerra}, Serena and {Kini}, Yves and {Dorsman}, Bas and {Watts}, Anna L. and {Huppenkothen}, Daniela and {Guillot}, Sebastien},
        title = "{X-PSI: A Python package for neutron star X-ray pulse simulation and inference}",
      journal = {The Journal of Open Source Software},
         year = 2023,
        month = feb,
       volume = {8},
       number = {82},
          eid = {4977},
        pages = {4977},
          doi = {10.21105/joss.04977},
       adsurl = {https://ui.adsabs.harvard.edu/abs/2023JOSS....8.4977R}
}

@article{fgivenx,
       author = {{Handley}, Will},
        title = "{fgivenx: A Python package for functional posterior plotting}",
      journal = {The Journal of Open Source Software},
         year = 2018,
        month = aug,
       volume = {3},
       number = {28},
        pages = {849},
          doi = {10.21105/joss.00849},
archivePrefix = {arXiv},
       eprint = {1908.01711},
 primaryClass = {astro-ph.IM},
       adsurl = {https://ui.adsabs.harvard.edu/abs/2018JOSS....3..849H}
}

@article{higson2019diagnostic,
       author = {{Higson}, Edward and {Handley}, Will and {Hobson}, Michael and {Lasenby}, Anthony},
        title = "{NESTCHECK: diagnostic tests for nested sampling calculations}",
      journal = {\mnras},
         year = 2019,
        month = feb,
       volume = {483},
       number = {2},
        pages = {2044-2056},
          doi = {10.1093/mnras/sty3090},
archivePrefix = {arXiv},
       eprint = {1804.06406},
 primaryClass = {stat.CO},
       adsurl = {https://ui.adsabs.harvard.edu/abs/2019MNRAS.483.2044H}
}

@article{higson2018sampling,
       author = {{Higson}, Edward and {Handley}, Will and {Hobson}, Mike and {Lasenby}, Anthony},
        title = "{Sampling Errors in Nested Sampling Parameter Estimation}",
      journal = {Bayesian Analysis},
         year = 2018,
        month = mar,
       volume = {13},
       number = {3},
        pages = {873-896},
          doi = {10.1214/17-BA1075},
archivePrefix = {arXiv},
       eprint = {1703.09701},
 primaryClass = {stat.ME},
       adsurl = {https://ui.adsabs.harvard.edu/abs/2018BayAn..13..873H}
}

@article{higson2018nestcheck,
       author = {{Higson}, Edward},
        title = "{nestcheck: error analysis, diagnostic tests and plots for nested sampling calculations}",
      journal = {The Journal of Open Source Software},
         year = 2018,
        month = sep,
       volume = {3},
       number = {29},
        pages = {916},
          doi = {10.21105/joss.00916},
       adsurl = {https://ui.adsabs.harvard.edu/abs/2018JOSS....3..916H}
}

@ARTICLE{Lewis19,
    author = {{Lewis}, Antony},
    title = "{GetDist: a Python package for analysing Monte Carlo samples}",
    journal = {arXiv e-prints},
    year = 2019,
    month = oct,
    eid = {arXiv:1910.13970},
    pages = {arXiv:1910.13970},
    archivePrefix = {arXiv},
    eprint = {1910.13970},
    primaryClass = {astro-ph.IM},
    adsurl = {https://ui.adsabs.harvard.edu/abs/2019arXiv191013970L}
}

@ARTICLE{PyMultiNest,
    author = {{Buchner}, J. and {Georgakakis}, A. and {Nandra}, K. and {Hsu}, L. and {Rangel}, C. and {Brightman}, M. and {Merloni}, A. and {Salvato}, M. and {Donley}, J. and {Kocevski}, D.},
    title = "{X-ray spectral modelling of the AGN obscuring region in the CDFS: Bayesian model selection and catalogue}",
    journal = {\aap},
    archivePrefix = "arXiv",
    eprint = {1402.0004},
    primaryClass = "astro-ph.HE",
    year = 2014,
    month = apr,
    volume = 564,
    eid = {A125},
    pages = {A125},
    doi = {10.1051/0004-6361/201322971},
    adsurl = {http://adsabs.harvard.edu/abs/2014A\%26A...564A.125B}
}

@ARTICLE{MultiNest_2009,
    author = {{Feroz}, F. and {Hobson}, M.~P. and {Bridges}, M.},
    title = "{MULTINEST: an efficient and robust Bayesian inference tool for cosmology and particle physics}",
    journal = {\mnras},
    archivePrefix = "arXiv",
    eprint = {0809.3437},
    year = 2009,
    month = oct,
    volume = 398,
    pages = {1601-1614},
    doi = {10.1111/j.1365-2966.2009.14548.x},
    adsurl = {http://adsabs.harvard.edu/abs/2009MNRAS.398.1601F}
}

@article{buchner_ultranest_2021,
       author = {{Buchner}, Johannes},
        title = "{UltraNest - a robust, general purpose Bayesian inference engine}",
      journal = {The Journal of Open Source Software},
         year = 2021,
        month = apr,
       volume = {6},
       number = {60},
          eid = {3001},
        pages = {3001},
          doi = {10.21105/joss.03001},
archivePrefix = {arXiv},
       eprint = {2101.09604},
 primaryClass = {stat.CO},
       adsurl = {https://ui.adsabs.harvard.edu/abs/2021JOSS....6.3001B}
}

@INCOLLECTION{Kluyver:2016aa,
       author = {{Kluyver}, Thomas and {Ragan-Kelley}, Benjain and {P{\'e}rez}, Fernando and {Granger}, Brian and {Bussonnier}, Matthias and {Frederic}, Jonathan and {Kelley}, Kyle and {Hamrick}, Jessica and {Grout}, Jason and {Corlay}, Sylvain and {Ivanov}, Paul and {Avila}, Dami{\'a}n and {Abdalla}, Safia and {Willing}, Carol and {Jupyter Development Team}},
        title = "{Jupyter Notebooks{\textemdash}a publishing format for reproducible computational workflows}",
    booktitle = {IOS Press},
         year = 2016,
        pages = {87-90},
          doi = {10.3233/978-1-61499-649-1-87},
       adsurl = {https://ui.adsabs.harvard.edu/abs/2016ppap.book...87K}
}

@ARTICLE{IPython2007,
       author = {{Perez}, Fernando and {Granger}, Brian E.},
        title = "{IPython: A System for Interactive Scientific Computing}",
      journal = {Computing in Science and Engineering},
         year = 2007,
        month = jan,
       volume = {9},
       number = {3},
        pages = {21-29},
          doi = {10.1109/MCSE.2007.53},
       adsurl = {https://ui.adsabs.harvard.edu/abs/2007CSE.....9c..21P}
}

@misc{matplotlibv2,
       author = {{Droettboom}, Michael and {Caswell}, Thomas A and {Hunter}, John and {Firing}, Eric and {Hedegaard Nielsen}, Jens and {Lee}, Antony and {Sales De Andrade}, Elliott and {Varoquaux}, Nelle and {Stansby}, David and {Root}, Benjamin and {Elson}, Phil and {Dale}, Darren and {Lee}, Jae-Joon and {May}, Ryan and {Sepp{\"a}nen}, Jouni K. and {Klymak}, Jody and {McDougall}, Damon and {Straw}, Andrew and {Hobson}, Paul and {Cgohlke} and {Yu}, Tony S and {Ma}, Eric and {Vincent}, Adrien F. and {Silvester}, Steven and {Moad}, Charlie and {Katins}, Jan and {Kniazev}, Nikita and {Hoffmann}, Tim and {Ariza}, Federico and {W{\"u}rtz}, Peter},
        title = "{matplotlib/matplotlib v2.2.2}",
         year = 2018,
        month = mar,
          eid = {10.5281/zenodo.1202077},
          doi = {10.5281/zenodo.1202077},
      version = {v2.2.2},
    publisher = {Zenodo},
       adsurl = {https://ui.adsabs.harvard.edu/abs/2018zndo...1202077D}
}

@Article{Hunter:2007,
       author = {{Hunter}, John D.},
        title = "{Matplotlib: A 2D Graphics Environment}",
      journal = {Computing in Science and Engineering},
         year = 2007,
        month = may,
       volume = {9},
       number = {3},
        pages = {90-95},
          doi = {10.1109/MCSE.2007.55},
       adsurl = {https://ui.adsabs.harvard.edu/abs/2007CSE.....9...90H}
}

@article{mpi4py,
    title = {MPI for Python: Performance improvements and MPI-2 extensions},
    journal = {Journal of Parallel and Distributed Computing},
    volume = {68},
    number = {5},
    pages = {655-662},
    year = {2008},
    issn = {0743-7315},
    doi = {https://doi.org/10.1016/j.jpdc.2007.09.005},
    url = {https://www.sciencedirect.com/science/article/pii/S0743731507001712},
    author = {Lisandro Dalcín and Rodrigo Paz and Mario Storti and Jorge D’Elía}
}

@techreport{MPI,
    author = {Forum, Message P},
    title = {MPI: A Message-Passing Interface Standard},
    year = {1994},
    url = {https://www.mpi-forum.org/docs/mpi-1.0/mpi-10.ps},
    publisher = {University of Tennessee},
    address = {Knoxville, TN, USA}
}

@article{openmp,
    Author = {Dagum, Leonardo and Menon, Ramesh},
    Journal = {Computational Science \& Engineering, IEEE},
    Number = {1},
    Pages = {46--55},
    Publisher = {IEEE},
    Title = {OpenMP: an industry standard API for shared-memory programming},
    Volume = {5},
    Year = {1998}
}

@ARTICLE{Scipy,
  author  = {Virtanen, Pauli and Gommers, Ralf and Oliphant, Travis E. and
            Haberland, Matt and Reddy, Tyler and Cournapeau, David and
            Burovski, Evgeni and Peterson, Pearu and Weckesser, Warren and
            Bright, Jonathan and {van der Walt}, St{\'e}fan J. and
            Brett, Matthew and Wilson, Joshua and Millman, K. Jarrod and
            Mayorov, Nikolay and Nelson, Andrew R. J. and Jones, Eric and
            Kern, Robert and Larson, Eric and Carey, C J and
            Polat, {\.I}lhan and Feng, Yu and Moore, Eric W. and
            {VanderPlas}, Jake and Laxalde, Denis and Perktold, Josef and
            Cimrman, Robert and Henriksen, Ian and Quintero, E. A. and
            Harris, Charles R. and Archibald, Anne M. and
            Ribeiro, Ant{\^o}nio H. and Pedregosa, Fabian and
            {van Mulbregt}, Paul and {SciPy 1.0 Contributors}},
  title   = {{{SciPy} 1.0: Fundamental Algorithms for Scientific
            Computing in Python}},
  journal = {Nature Methods},
  year    = {2020},
  volume  = {17},
  pages   = {261--272},
  adsurl  = {https://rdcu.be/b08Wh},
  doi     = {10.1038/s41592-019-0686-2},
}

@ARTICLE{cython2011,
       author = {{Behnel}, Stefan and {Bradshaw}, Robert and {Citro}, Craig and {Dalcin}, Lisandro and {Seljebotn}, Dag Sverre and {Smith}, Kurt},
        title = "{Cython: The Best of Both Worlds}",
      journal = {Computing in Science and Engineering},
         year = 2011,
        month = mar,
       volume = {13},
       number = {2},
        pages = {31-39},
          doi = {10.1109/MCSE.2010.118},
       adsurl = {https://ui.adsabs.harvard.edu/abs/2011CSE....13b..31B}
}

@ARTICLE{Numpy2011,
       author = {{van der Walt}, St{\'e}fan and {Colbert}, S. Chris and {Varoquaux}, Ga{\"e}l},
        title = "{The NumPy Array: A Structure for Efficient Numerical Computation}",
      journal = {Computing in Science and Engineering},
         year = 2011,
        month = mar,
       volume = {13},
       number = {2},
        pages = {22-30},
          doi = {10.1109/MCSE.2011.37},
archivePrefix = {arXiv},
       eprint = {1102.1523},
 primaryClass = {cs.MS},
       adsurl = {https://ui.adsabs.harvard.edu/abs/2011CSE....13b..22V}
}

@book{Gough:2009,
    author = {Gough, Brian},
    title = {GNU Scientific Library Reference Manual - Third Edition},
    year = {2009},
    isbn = {0954612078},
    publisher = {Network Theory Ltd.},
    edition = {3rd}
}

@ARTICLE{Python2007,
       author = {{Oliphant}, Travis E.},
        title = "{Python for Scientific Computing}",
      journal = {Computing in Science and Engineering},
         year = 2007,
        month = jan,
       volume = {9},
       number = {3},
        pages = {10-20},
          doi = {10.1109/MCSE.2007.58},
       adsurl = {https://ui.adsabs.harvard.edu/abs/2007CSE.....9c..10O}
}

@article{luo_pint_2021,
       author = {{Luo}, Jing and {Ransom}, Scott and {Demorest}, Paul and {Ray}, Paul S. and {Archibald}, Anne and {Kerr}, Matthew and {Jennings}, Ross J. and {Bachetti}, Matteo and {van Haasteren}, Rutger and {Champagne}, Chloe A. and {Colen}, Jonathan and {Phillips}, Camryn and {Zimmerman}, Josef and {Stovall}, Kevin and {Lam}, Michael T. and {Jenet}, Fredrick A.},
        title = "{PINT: A Modern Software Package for Pulsar Timing}",
      journal = {\apj},
         year = 2021,
        month = apr,
       volume = {911},
       number = {1},
          eid = {45},
        pages = {45},
          doi = {10.3847/1538-4357/abe62f},
archivePrefix = {arXiv},
       eprint = {2012.00074},
 primaryClass = {astro-ph.IM},
       adsurl = {https://ui.adsabs.harvard.edu/abs/2021ApJ...911...45L}
}

@article{shaw_efficient_2007,
       author = {{Shaw}, J.~R. and {Bridges}, M. and {Hobson}, M.~P.},
        title = "{Efficient Bayesian inference for multimodal problems in cosmology}",
      journal = {\mnras},
         year = 2007,
        month = jul,
       volume = {378},
       number = {4},
        pages = {1365-1370},
          doi = {10.1111/j.1365-2966.2007.11871.x},
archivePrefix = {arXiv},
       eprint = {astro-ph/0701867},
 primaryClass = {astro-ph},
       adsurl = {https://ui.adsabs.harvard.edu/abs/2007MNRAS.378.1365S}
}

@article{kass_bayes_1995,
    author = {Robert E. Kass and Adrian E. Raftery},
    title = {Bayes Factors},
    journal = {Journal of the American Statistical Association},
    volume = {90},
    number = {430},
    pages = {773--795},
    year = {1995},
    publisher = {ASA Website},
    doi = {10.1080/01621459.1995.10476572},
    URL = {https://www.tandfonline.com/doi/abs/10.1080/01621459.1995.10476572},
    eprint = { https://www.tandfonline.com/doi/pdf/10.1080/01621459.1995.10476572}
}

@INPROCEEDINGS{skilling2004nested,
       author = {{Skilling}, John},
        title = "{Nested Sampling}",
    booktitle = {Bayesian Inference and Maximum Entropy Methods in Science and Engineering: 24th International Workshop on Bayesian Inference and Maximum Entropy Methods in Science and Engineering},
         year = 2004,
       editor = {{Fischer}, Rainer and {Preuss}, Roland and {Toussaint}, Udo Von},
       series = {American Institute of Physics Conference Series},
       volume = {735},
        month = nov,
    publisher = {AIP},
        pages = {395-405},
          doi = {10.1063/1.1835238},
       adsurl = {https://ui.adsabs.harvard.edu/abs/2004AIPC..735..395S}
}

@article{vehtari_practical_2017,
       author = {{Vehtari}, Aki and {Gelman}, Andrew and {Gabry}, Jonah},
        title = "{Practical Bayesian model evaluation using leave-one-out cross-validation and WAIC}",
      journal = {arXiv e-prints},
         year = 2015,
        month = jul,
          eid = {arXiv:1507.04544},
        pages = {arXiv:1507.04544},
          doi = {10.48550/arXiv.1507.04544},
archivePrefix = {arXiv},
       eprint = {1507.04544},
 primaryClass = {stat.CO},
       adsurl = {https://ui.adsabs.harvard.edu/abs/2015arXiv150704544V}
}

@ARTICLE{kumar_arviz_2019,
       author = {{Kumar}, Ravin and {Carroll}, Colin and {Hartikainen}, Ari and {Martin}, Osvaldo},
        title = "{ArviZ a unified library for exploratory analysis of Bayesian models in Python}",
      journal = {The Journal of Open Source Software},
         year = 2019,
        month = jan,
       volume = {4},
       number = {33},
          eid = {1143},
        pages = {1143},
          doi = {10.21105/joss.01143},
       adsurl = {https://ui.adsabs.harvard.edu/abs/2019JOSS....4.1143K}
}

@ARTICLE{Bogdanov2019a,
       author = {{Bogdanov}, Slavko and {Guillot}, Sebastien and {Ray}, Paul S. and {Wolff}, Michael T. and {Chakrabarty}, Deepto and {Ho}, Wynn C.~G. and {Kerr}, Matthew and {Lamb}, Frederick K. and {Lommen}, Andrea and {Ludlam}, Renee M. and {Milburn}, Reilly and {Montano}, Sergio and {Miller}, M. Coleman and {Baub{\"o}ck}, Michi and {{\"O}zel}, Feryal and {Psaltis}, Dimitrios and {Remillard}, Ronald A. and {Riley}, Thomas E. and {Steiner}, James F. and {Strohmayer}, Tod E. and {Watts}, Anna L. and {Wood}, Kent S. and {Zeldes}, Jesse and {Enoto}, Teruaki and {Okajima}, Takashi and {Kellogg}, James W. and {Baker}, Charles and {Markwardt}, Craig B. and {Arzoumanian}, Zaven and {Gendreau}, Keith C.},
        title = "{Constraining the Neutron Star Mass-Radius Relation and Dense Matter Equation of State with NICER. I. The Millisecond Pulsar X-Ray Data Set}",
      journal = {\apjl},
         year = 2019,
        month = dec,
       volume = {887},
       number = {1},
          eid = {L25},
        pages = {L25},
          doi = {10.3847/2041-8213/ab53eb},
archivePrefix = {arXiv},
       eprint = {1912.05706},
 primaryClass = {astro-ph.HE},
       adsurl = {https://ui.adsabs.harvard.edu/abs/2019ApJ...887L..25B}
}

@ARTICLE{bogdanov_constraining_2021,
       author = {{Bogdanov}, Slavko and {Dittmann}, Alexander J. and {Ho}, Wynn C.~G. and {Lamb}, Frederick K. and {Mahmoodifar}, Simin and {Miller}, M. Coleman and {Morsink}, Sharon M. and {Riley}, Thomas E. and {Strohmayer}, Tod E. and {Watts}, Anna L. and {Choudhury}, Devarshi and {Guillot}, Sebastien and {Harding}, Alice K. and {Ray}, Paul S. and {Wadiasingh}, Zorawar and {Wolff}, Michael T. and {Markwardt}, Craig B. and {Arzoumanian}, Zaven and {Gendreau}, Keith C.},
        title = "{Constraining the Neutron Star Mass-Radius Relation and Dense Matter Equation of State with NICER. III. Model Description and Verification of Parameter Estimation Codes}",
      journal = {\apjl},
         year = 2021,
        month = jun,
       volume = {914},
       number = {1},
          eid = {L15},
        pages = {L15},
          doi = {10.3847/2041-8213/abfb79},
archivePrefix = {arXiv},
       eprint = {2104.06928},
 primaryClass = {astro-ph.HE},
       adsurl = {https://ui.adsabs.harvard.edu/abs/2021ApJ...914L..15B}
}

@ARTICLE{guillot_nicer_2019,
       author = {{Guillot}, Sebastien and {Kerr}, Matthew and {Ray}, Paul S. and {Bogdanov}, Slavko and {Ransom}, Scott and {Deneva}, Julia S. and {Arzoumanian}, Zaven and {Bult}, Peter and {Chakrabarty}, Deepto and {Gendreau}, Keith C. and {Ho}, Wynn C.~G. and {Jaisawal}, Gaurava K. and {Malacaria}, Christian and {Miller}, M. Coleman and {Strohmayer}, Tod E. and {Wolff}, Michael T. and {Wood}, Kent S. and {Webb}, Natalie A. and {Guillemot}, Lucas and {Cognard}, Ismael and {Theureau}, Gilles},
        title = "{NICER X-Ray Observations of Seven Nearby Rotation-powered Millisecond Pulsars}",
      journal = {\apjl},
         year = 2019,
        month = dec,
       volume = {887},
       number = {1},
          eid = {L27},
        pages = {L27},
          doi = {10.3847/2041-8213/ab511b},
archivePrefix = {arXiv},
       eprint = {1912.05708},
 primaryClass = {astro-ph.HE},
       adsurl = {https://ui.adsabs.harvard.edu/abs/2019ApJ...887L..27G}
}

@ARTICLE{remillard_empirical_2022,
       author = {{Remillard}, Ronald A. and {Loewenstein}, Michael and {Steiner}, James F. and {Prigozhin}, Gregory Y. and {LaMarr}, Beverly and {Enoto}, Teruaki and {Gendreau}, Keith C. and {Arzoumanian}, Zaven and {Markwardt}, Craig and {Basak}, Arkadip and {Stevens}, Abigail L. and {Ray}, Paul S. and {Altamirano}, Diego and {Buisson}, Douglas J.~K.},
        title = "{An Empirical Background Model for the NICER X-Ray Timing Instrument}",
      journal = {\aj},
         year = 2022,
        month = mar,
       volume = {163},
       number = {3},
          eid = {130},
        pages = {130},
          doi = {10.3847/1538-3881/ac4ae6},
archivePrefix = {arXiv},
       eprint = {2105.09901},
 primaryClass = {astro-ph.IM},
       adsurl = {https://ui.adsabs.harvard.edu/abs/2022AJ....163..130R}
}

@INPROCEEDINGS{gendreau_neutron_2016,
       author = {{Gendreau}, Keith C. and {Arzoumanian}, Zaven and {Adkins}, Phillip W. and {Albert}, Cheryl L. and {Anders}, John F. and {Aylward}, Andrew T. and {Baker}, Charles L. and {Balsamo}, Erin R. and {Bamford}, William A. and {Benegalrao}, Suyog S. and {Berry}, Daniel L. and {Bhalwani}, Shiraz and {Black}, J. Kevin and {Blaurock}, Carl and {Bronke}, Ginger M. and {Brown}, Gary L. and {Budinoff}, Jason G. and {Cantwell}, Jeffrey D. and {Cazeau}, Thoniel and {Chen}, Philip T. and {Clement}, Thomas G. and {Colangelo}, Andrew T. and {Coleman}, Jerry S. and {Coopersmith}, Jonathan D. and {Dehaven}, William E. and {Doty}, John P. and {Egan}, Mark D. and {Enoto}, Teruaki and {Fan}, Terry W. and {Ferro}, Deneen M. and {Foster}, Richard and {Galassi}, Nicholas M. and {Gallo}, Luis D. and {Green}, Chris M. and {Grosh}, Dave and {Ha}, Kong Q. and {Hasouneh}, Monther A. and {Heefner}, Kristofer B. and {Hestnes}, Phyllis and {Hoge}, Lisa J. and {Jacobs}, Tawanda M. and {J{\o}rgensen}, John L. and {Kaiser}, Michael A. and {Kellogg}, James W. and {Kenyon}, Steven J. and {Koenecke}, Richard G. and {Kozon}, Robert P. and {LaMarr}, Beverly and {Lambertson}, Mike D. and {Larson}, Anne M. and {Lentine}, Steven and {Lewis}, Jesse H. and {Lilly}, Michael G. and {Liu}, Kuochia Alice and {Malonis}, Andrew and {Manthripragada}, Sridhar S. and {Markwardt}, Craig B. and {Matonak}, Bryan D. and {Mcginnis}, Isaac E. and {Miller}, Roger L. and {Mitchell}, Alissa L. and {Mitchell}, Jason W. and {Mohammed}, Jelila S. and {Monroe}, Charles A. and {Montt de Garcia}, Kristina M. and {Mul{\'e}}, Peter D. and {Nagao}, Louis T. and {Ngo}, Son N. and {Norris}, Eric D. and {Norwood}, Dwight A. and {Novotka}, Joseph and {Okajima}, Takashi and {Olsen}, Lawrence G. and {Onyeachu}, Chimaobi O. and {Orosco}, Henry Y. and {Peterson}, Jacqualine R. and {Pevear}, Kristina N. and {Pham}, Karen K. and {Pollard}, Sue E. and {Pope}, John S. and {Powers}, Daniel F. and {Powers}, Charles E. and {Price}, Samuel R. and {Prigozhin}, Gregory Y. and {Ramirez}, Julian B. and {Reid}, Winston J. and {Remillard}, Ronald A. and {Rogstad}, Eric M. and {Rosecrans}, Glenn P. and {Rowe}, John N. and {Sager}, Jennifer A. and {Sanders}, Claude A. and {Savadkin}, Bruce and {Saylor}, Maxine R. and {Schaeffer}, Alexander F. and {Schweiss}, Nancy S. and {Semper}, Sean R. and {Serlemitsos}, Peter J. and {Shackelford}, Larry V. and {Soong}, Yang and {Struebel}, Jonathan and {Vezie}, Michael L. and {Villasenor}, Joel S. and {Winternitz}, Luke B. and {Wofford}, George I. and {Wright}, Michael R. and {Yang}, Mike Y. and {Yu}, Wayne H.},
        title = "{The Neutron star Interior Composition Explorer (NICER): design and development}",
    booktitle = {Space Telescopes and Instrumentation 2016: Ultraviolet to Gamma Ray},
         year = 2016,
       editor = {{den Herder}, Jan-Willem A. and {Takahashi}, Tadayuki and {Bautz}, Marshall},
       series = {Society of Photo-Optical Instrumentation Engineers (SPIE) Conference Series},
       volume = {9905},
        month = jul,
          eid = {99051H},
        pages = {99051H},
          doi = {10.1117/12.2231304},
       adsurl = {https://ui.adsabs.harvard.edu/abs/2016SPIE.9905E..1HG}
}

@ARTICLE{ransom_three_2011,
       author = {{Ransom}, S.~M. and {Ray}, P.~S. and {Camilo}, F. and {Roberts}, M.~S.~E. and {{\c{C}}elik}, {\"O}. and {Wolff}, M.~T. and {Cheung}, C.~C. and {Kerr}, M. and {Pennucci}, T. and {DeCesar}, M.~E. and {Cognard}, I. and {Lyne}, A.~G. and {Stappers}, B.~W. and {Freire}, P.~C.~C. and {Grove}, J.~E. and {Abdo}, A.~A. and {Desvignes}, G. and {Donato}, D. and {Ferrara}, E.~C. and {Gehrels}, N. and {Guillemot}, L. and {Gwon}, C. and {Harding}, A.~K. and {Johnston}, S. and {Keith}, M. and {Kramer}, M. and {Michelson}, P.~F. and {Parent}, D. and {Saz Parkinson}, P.~M. and {Romani}, R.~W. and {Smith}, D.~A. and {Theureau}, G. and {Thompson}, D.~J. and {Weltevrede}, P. and {Wood}, K.~S. and {Ziegler}, M.},
        title = "{Three Millisecond Pulsars in Fermi LAT Unassociated Bright Sources}",
      journal = {\apjl},
         year = 2011,
        month = jan,
       volume = {727},
       number = {1},
          eid = {L16},
        pages = {L16},
          doi = {10.1088/2041-8205/727/1/L16},
archivePrefix = {arXiv},
       eprint = {1012.2862},
 primaryClass = {astro-ph.HE},
       adsurl = {https://ui.adsabs.harvard.edu/abs/2011ApJ...727L..16R}
}

@ARTICLE{reardon_gravitational-wave_2023,
       author = {{Reardon}, Daniel J. and {Zic}, Andrew and {Shannon}, Ryan M. and {Di Marco}, Valentina and {Hobbs}, George B. and {Kapur}, Agastya and {Lower}, Marcus E. and {Mandow}, Rami and {Middleton}, Hannah and {Miles}, Matthew T. and {Rogers}, Axl F. and {Askew}, Jacob and {Bailes}, Matthew and {Bhat}, N.~D. Ramesh and {Cameron}, Andrew and {Kerr}, Matthew and {Kulkarni}, Atharva and {Manchester}, Richard N. and {Nathan}, Rowina S. and {Russell}, Christopher J. and {Os{\l}owski}, Stefan and {Zhu}, Xing-Jiang},
        title = "{The Gravitational-wave Background Null Hypothesis: Characterizing Noise in Millisecond Pulsar Arrival Times with the Parkes Pulsar Timing Array}",
      journal = {\apjl},
         year = 2023,
        month = jul,
       volume = {951},
       number = {1},
          eid = {L7},
        pages = {L7},
          doi = {10.3847/2041-8213/acdd03},
archivePrefix = {arXiv},
       eprint = {2306.16229},
 primaryClass = {astro-ph.HE},
       adsurl = {https://ui.adsabs.harvard.edu/abs/2023ApJ...951L...7R}
}

@ARTICLE{bassa_cool_2016,
       author = {{Bassa}, C.~G. and {Antoniadis}, J. and {Camilo}, F. and {Cognard}, I. and {Koester}, D. and {Kramer}, M. and {Ransom}, S.~R. and {Stappers}, B.~W.},
        title = "{Cool white dwarf companions to four millisecond pulsars}",
      journal = {\mnras},
         year = 2016,
        month = feb,
       volume = {455},
       number = {4},
        pages = {3806-3813},
          doi = {10.1093/mnras/stv2607},
archivePrefix = {arXiv},
       eprint = {1511.01319},
 primaryClass = {astro-ph.HE},
       adsurl = {https://ui.adsabs.harvard.edu/abs/2016MNRAS.455.3806B}
}

@ARTICLE{shamohammadi_meerkat_2024,
       author = {{Shamohammadi}, M. and {Bailes}, M. and {Flynn}, C. and {Reardon}, D.~J. and {Shannon}, R.~M. and {Buchner}, S. and {Cameron}, A.~D. and {Camilo}, F. and {Corongiu}, A. and {Geyer}, M. and {Kramer}, M. and {Miles}, M. and {Spiewak}, R.},
        title = "{MeerKAT Pulsar Timing Array parallaxes and proper motions}",
      journal = {\mnras},
         year = 2024,
        month = may,
       volume = {530},
       number = {1},
        pages = {287-306},
          doi = {10.1093/mnras/stae016},
archivePrefix = {arXiv},
       eprint = {2401.06963},
 primaryClass = {astro-ph.HE},
       adsurl = {https://ui.adsabs.harvard.edu/abs/2024MNRAS.530..287S}
}

@ARTICLE{miles_meerkat_2025,
       author = {{Miles}, Matthew T. and {Shannon}, Ryan M. and {Reardon}, Daniel J. and {Bailes}, Matthew and {Champion}, David J. and {Geyer}, Marisa and {Gitika}, Pratyasha and {Grunthal}, Kathrin and {Keith}, Michael J. and {Kramer}, Michael and {Kulkarni}, Atharva D. and {Nathan}, Rowina S. and {Parthasarathy}, Aditya and {Porayko}, Nataliya K. and {Singha}, Jaikhomba and {Theureau}, Gilles and {Abbate}, Federico and {Buchner}, Sarah and {Cameron}, Andrew D. and {Camilo}, Fernando and {Moreschi}, Beatrice E. and {Shaifullah}, Golam and {Shamohammadi}, Mohsen and {Krishnan}, Vivek Venkatraman},
        title = "{The MeerKAT Pulsar Timing Array: the 4.5-yr data release and the noise and stochastic signals of the millisecond pulsar population}",
      journal = {\mnras},
         year = 2025,
        month = jan,
       volume = {536},
       number = {2},
        pages = {1467-1488},
          doi = {10.1093/mnras/stae2572},
archivePrefix = {arXiv},
       eprint = {2412.01148},
 primaryClass = {astro-ph.HE},
       adsurl = {https://ui.adsabs.harvard.edu/abs/2025MNRAS.536.1467M}
}

@ARTICLE{riley_nicer_2019,
       author = {{Riley}, T.~E. and {Watts}, A.~L. and {Bogdanov}, S. and {Ray}, P.~S. and {Ludlam}, R.~M. and {Guillot}, S. and {Arzoumanian}, Z. and {Baker}, C.~L. and {Bilous}, A.~V. and {Chakrabarty}, D. and {Gendreau}, K.~C. and {Harding}, A.~K. and {Ho}, W.~C.~G. and {Lattimer}, J.~M. and {Morsink}, S.~M. and {Strohmayer}, T.~E.},
        title = "{A NICER View of PSR J0030+0451: Millisecond Pulsar Parameter Estimation}",
      journal = {\apjl},
         year = 2019,
        month = dec,
       volume = {887},
       number = {1},
          eid = {L21},
        pages = {L21},
          doi = {10.3847/2041-8213/ab481c},
archivePrefix = {arXiv},
       eprint = {1912.05702},
 primaryClass = {astro-ph.HE},
       adsurl = {https://ui.adsabs.harvard.edu/abs/2019ApJ...887L..21R}
}

@ARTICLE{riley_nicer_2021,
       author = {{Riley}, Thomas E. and {Watts}, Anna L. and {Ray}, Paul S. and {Bogdanov}, Slavko and {Guillot}, Sebastien and {Morsink}, Sharon M. and {Bilous}, Anna V. and {Arzoumanian}, Zaven and {Choudhury}, Devarshi and {Deneva}, Julia S. and {Gendreau}, Keith C. and {Harding}, Alice K. and {Ho}, Wynn C.~G. and {Lattimer}, James M. and {Loewenstein}, Michael and {Ludlam}, Renee M. and {Markwardt}, Craig B. and {Okajima}, Takashi and {Prescod-Weinstein}, Chanda and {Remillard}, Ronald A. and {Wolff}, Michael T. and {Fonseca}, Emmanuel and {Cromartie}, H. Thankful and {Kerr}, Matthew and {Pennucci}, Timothy T. and {Parthasarathy}, Aditya and {Ransom}, Scott and {Stairs}, Ingrid and {Guillemot}, Lucas and {Cognard}, Ismael},
        title = "{A NICER View of the Massive Pulsar PSR J0740+6620 Informed by Radio Timing and XMM-Newton Spectroscopy}",
      journal = {\apjl},
         year = 2021,
        month = sep,
       volume = {918},
       number = {2},
          eid = {L27},
        pages = {L27},
          doi = {10.3847/2041-8213/ac0a81},
archivePrefix = {arXiv},
       eprint = {2105.06980},
 primaryClass = {astro-ph.HE},
       adsurl = {https://ui.adsabs.harvard.edu/abs/2021ApJ...918L..27R}
}

@ARTICLE{miller_psr_2019,
       author = {{Miller}, M.~C. and {Lamb}, F.~K. and {Dittmann}, A.~J. and {Bogdanov}, S. and {Arzoumanian}, Z. and {Gendreau}, K.~C. and {Guillot}, S. and {Harding}, A.~K. and {Ho}, W.~C.~G. and {Lattimer}, J.~M. and {Ludlam}, R.~M. and {Mahmoodifar}, S. and {Morsink}, S.~M. and {Ray}, P.~S. and {Strohmayer}, T.~E. and {Wood}, K.~S. and {Enoto}, T. and {Foster}, R. and {Okajima}, T. and {Prigozhin}, G. and {Soong}, Y.},
        title = "{PSR J0030+0451 Mass and Radius from NICER Data and Implications for the Properties of Neutron Star Matter}",
      journal = {\apjl},
         year = 2019,
        month = dec,
       volume = {887},
       number = {1},
          eid = {L24},
        pages = {L24},
          doi = {10.3847/2041-8213/ab50c5},
archivePrefix = {arXiv},
       eprint = {1912.05705},
 primaryClass = {astro-ph.HE},
       adsurl = {https://ui.adsabs.harvard.edu/abs/2019ApJ...887L..24M}
}

@ARTICLE{miller_radius_2021,
       author = {{Miller}, M.~C. and {Lamb}, F.~K. and {Dittmann}, A.~J. and {Bogdanov}, S. and {Arzoumanian}, Z. and {Gendreau}, K.~C. and {Guillot}, S. and {Ho}, W.~C.~G. and {Lattimer}, J.~M. and {Loewenstein}, M. and {Morsink}, S.~M. and {Ray}, P.~S. and {Wolff}, M.~T. and {Baker}, C.~L. and {Cazeau}, T. and {Manthripragada}, S. and {Markwardt}, C.~B. and {Okajima}, T. and {Pollard}, S. and {Cognard}, I. and {Cromartie}, H.~T. and {Fonseca}, E. and {Guillemot}, L. and {Kerr}, M. and {Parthasarathy}, A. and {Pennucci}, T.~T. and {Ransom}, S. and {Stairs}, I.},
        title = "{The Radius of PSR J0740+6620 from NICER and XMM-Newton Data}",
      journal = {\apjl},
         year = 2021,
        month = sep,
       volume = {918},
       number = {2},
          eid = {L28},
        pages = {L28},
          doi = {10.3847/2041-8213/ac089b},
archivePrefix = {arXiv},
       eprint = {2105.06979},
 primaryClass = {astro-ph.HE},
       adsurl = {https://ui.adsabs.harvard.edu/abs/2021ApJ...918L..28M}
}

@ARTICLE{salmi_radius_2022,
       author = {{Salmi}, Tuomo and {Vinciguerra}, Serena and {Choudhury}, Devarshi and {Riley}, Thomas E. and {Watts}, Anna L. and {Remillard}, Ronald A. and {Ray}, Paul S. and {Bogdanov}, Slavko and {Guillot}, Sebastien and {Arzoumanian}, Zaven and {Chirenti}, Cecilia and {Dittmann}, Alexander J. and {Gendreau}, Keith C. and {Ho}, Wynn C.~G. and {Miller}, M. Coleman and {Morsink}, Sharon M. and {Wadiasingh}, Zorawar and {Wolff}, Michael T.},
        title = "{The Radius of PSR J0740+6620 from NICER with NICER Background Estimates}",
      journal = {\apj},
         year = 2022,
        month = dec,
       volume = {941},
       number = {2},
          eid = {150},
        pages = {150},
          doi = {10.3847/1538-4357/ac983d},
archivePrefix = {arXiv},
       eprint = {2209.12840},
 primaryClass = {astro-ph.HE},
       adsurl = {https://ui.adsabs.harvard.edu/abs/2022ApJ...941..150S}
}

@ARTICLE{salmi_atmospheric_2023,
       author = {{Salmi}, Tuomo and {Vinciguerra}, Serena and {Choudhury}, Devarshi and {Watts}, Anna L. and {Ho}, Wynn C.~G. and {Guillot}, Sebastien and {Kini}, Yves and {Dorsman}, Bas and {Morsink}, Sharon M. and {Bogdanov}, Slavko},
        title = "{Atmospheric Effects on Neutron Star Parameter Constraints with NICER}",
      journal = {\apj},
         year = 2023,
        month = oct,
       volume = {956},
       number = {2},
          eid = {138},
        pages = {138},
          doi = {10.3847/1538-4357/acf49d},
archivePrefix = {arXiv},
       eprint = {2308.09319},
 primaryClass = {astro-ph.HE},
       adsurl = {https://ui.adsabs.harvard.edu/abs/2023ApJ...956..138S}
}

@ARTICLE{vinciguerra_x-psi_2023,
       author = {{Vinciguerra}, Serena and {Salmi}, Tuomo and {Watts}, Anna L. and {Choudhury}, Devarshi and {Kini}, Yves and {Riley}, Thomas E.},
        title = "{X-PSI Parameter Recovery for Temperature Map Configurations Inspired by PSR J0030+0451}",
      journal = {\apj},
         year = 2023,
        month = dec,
       volume = {959},
       number = {1},
          eid = {55},
        pages = {55},
          doi = {10.3847/1538-4357/acf9a0},
archivePrefix = {arXiv},
       eprint = {2308.08409},
 primaryClass = {astro-ph.HE},
       adsurl = {https://ui.adsabs.harvard.edu/abs/2023ApJ...959...55V}
}

@ARTICLE{vinciguerra_updated_2024,
       author = {{Vinciguerra}, Serena and {Salmi}, Tuomo and {Watts}, Anna L. and {Choudhury}, Devarshi and {Riley}, Thomas E. and {Ray}, Paul S. and {Bogdanov}, Slavko and {Kini}, Yves and {Guillot}, Sebastien and {Chakrabarty}, Deepto and {Ho}, Wynn C.~G. and {Huppenkothen}, Daniela and {Morsink}, Sharon M. and {Wadiasingh}, Zorawar and {Wolff}, Michael T.},
        title = "{An Updated Mass-Radius Analysis of the 2017-2018 NICER Data Set of PSR J0030+0451}",
      journal = {\apj},
         year = 2024,
        month = jan,
       volume = {961},
       number = {1},
          eid = {62},
        pages = {62},
          doi = {10.3847/1538-4357/acfb83},
archivePrefix = {arXiv},
       eprint = {2308.09469},
 primaryClass = {astro-ph.HE},
       adsurl = {https://ui.adsabs.harvard.edu/abs/2024ApJ...961...62V}
}

@ARTICLE{salmi_radius_2024,
       author = {{Salmi}, Tuomo and {Choudhury}, Devarshi and {Kini}, Yves and {Riley}, Thomas E. and {Vinciguerra}, Serena and {Watts}, Anna L. and {Wolff}, Michael T. and {Arzoumanian}, Zaven and {Bogdanov}, Slavko and {Chakrabarty}, Deepto and {Gendreau}, Keith and {Guillot}, Sebastien and {Ho}, Wynn C.~G. and {Huppenkothen}, Daniela and {Ludlam}, Renee M. and {Morsink}, Sharon M. and {Ray}, Paul S.},
        title = "{The Radius of the High-mass Pulsar PSR J0740+6620 with 3.6 yr of NICER Data}",
      journal = {\apj},
         year = 2024,
        month = oct,
       volume = {974},
       number = {2},
          eid = {294},
        pages = {294},
          doi = {10.3847/1538-4357/ad5f1f},
archivePrefix = {arXiv},
       eprint = {2406.14466},
 primaryClass = {astro-ph.HE},
       adsurl = {https://ui.adsabs.harvard.edu/abs/2024ApJ...974..294S}
}

@ARTICLE{kini_constraining_2024,
       author = {{Kini}, Yves and {Salmi}, Tuomo and {Vinciguerra}, Serena and {Watts}, Anna L. and {Bilous}, Anna and {Galloway}, Duncan K. and {van der Wateren}, Emma and {Khalsa}, Guru Partap and {Bogdanov}, Slavko and {Buchner}, Johannes and {Suleimanov}, Valery},
        title = "{Constraining the properties of the thermonuclear burst oscillation source XTE J1814-338 through pulse profile modelling}",
      journal = {\mnras},
         year = 2024,
        month = dec,
       volume = {535},
       number = {2},
        pages = {1507-1525},
          doi = {10.1093/mnras/stae2398},
archivePrefix = {arXiv},
       eprint = {2405.10717},
 primaryClass = {astro-ph.HE},
       adsurl = {https://ui.adsabs.harvard.edu/abs/2024MNRAS.535.1507K}
}

@ARTICLE{dittmann_more_2024,
       author = {{Dittmann}, Alexander J. and {Miller}, M. Coleman and {Lamb}, Frederick K. and {Holt}, Isiah M. and {Chirenti}, Cecilia and {Wolff}, Michael T. and {Bogdanov}, Slavko and {Guillot}, Sebastien and {Ho}, Wynn C.~G. and {Morsink}, Sharon M. and {Arzoumanian}, Zaven and {Gendreau}, Keith C.},
        title = "{A More Precise Measurement of the Radius of PSR J0740+6620 Using Updated NICER Data}",
      journal = {\apj},
         year = 2024,
        month = oct,
       volume = {974},
       number = {2},
          eid = {295},
        pages = {295},
          doi = {10.3847/1538-4357/ad5f1e},
archivePrefix = {arXiv},
       eprint = {2406.14467},
 primaryClass = {astro-ph.HE},
       adsurl = {https://ui.adsabs.harvard.edu/abs/2024ApJ...974..295D}
}

@article{choudhury_nicer_2024,
       author = {{Choudhury}, Devarshi and {Salmi}, Tuomo and {Vinciguerra}, Serena and {Riley}, Thomas E. and {Kini}, Yves and {Watts}, Anna L. and {Dorsman}, Bas and {Bogdanov}, Slavko and {Guillot}, Sebastien and {Ray}, Paul S. and {Reardon}, Daniel J. and {Remillard}, Ronald A. and {Bilous}, Anna V. and {Huppenkothen}, Daniela and {Lattimer}, James M. and {Rutherford}, Nathan and {Arzoumanian}, Zaven and {Gendreau}, Keith C. and {Morsink}, Sharon M. and {Ho}, Wynn C.~G.},
        title = "{A NICER View of the Nearest and Brightest Millisecond Pulsar: PSR J0437{\textendash}4715}",
      journal = {\apjl},
         year = 2024,
        month = aug,
       volume = {971},
       number = {1},
          eid = {L20},
        pages = {L20},
          doi = {10.3847/2041-8213/ad5a6f},
archivePrefix = {arXiv},
       eprint = {2407.06789},
 primaryClass = {astro-ph.HE},
       adsurl = {https://ui.adsabs.harvard.edu/abs/2024ApJ...971L..20C}
}

@ARTICLE{choudhury_exploring_2024,
       author = {{Choudhury}, Devarshi and {Watts}, Anna L. and {Dittmann}, Alexander J. and {Miller}, M. Coleman and {Morsink}, Sharon M. and {Salmi}, Tuomo and {Vinciguerra}, Serena and {Bogdanov}, Slavko and {Guillot}, Sebastien and {Wolff}, Michael T. and {Arzoumanian}, Zaven},
        title = "{Exploring Waveform Variations among Neutron Star Ray-tracing Codes for Complex Emission Geometries}",
      journal = {\apj},
         year = 2024,
        month = nov,
       volume = {975},
       number = {2},
          eid = {202},
        pages = {202},
          doi = {10.3847/1538-4357/ad7255},
archivePrefix = {arXiv},
       eprint = {2406.07285},
 primaryClass = {astro-ph.HE},
       adsurl = {https://ui.adsabs.harvard.edu/abs/2024ApJ...975..202C}
}

@article{salmi_J1231_2024,
       author = {{Salmi}, Tuomo and {Deneva}, Julia S. and {Ray}, Paul S. and {Watts}, Anna L. and {Choudhury}, Devarshi and {Kini}, Yves and {Vinciguerra}, Serena and {Cromartie}, H. Thankful and {Wolff}, Michael T. and {Arzoumanian}, Zaven and {Bogdanov}, Slavko and {Gendreau}, Keith and {Guillot}, Sebastien and {Ho}, Wynn C.~G. and {Morsink}, Sharon M. and {Cognard}, Isma{\"e}l and {Guillemot}, Lucas and {Theureau}, Gilles and {Kerr}, Matthew},
        title = "{A NICER View of PSR J1231‑1411: A Complex Case}",
      journal = {\apj},
         year = 2024,
        month = nov,
       volume = {976},
       number = {1},
          eid = {58},
        pages = {58},
          doi = {10.3847/1538-4357/ad81d2},
archivePrefix = {arXiv},
       eprint = {2409.14923},
 primaryClass = {astro-ph.HE},
       adsurl = {https://ui.adsabs.harvard.edu/abs/2024ApJ...976...58S}
}

@PHDTHESIS{riley_neutron_2019,
       author = {{Riley}, Thomas Edward},
        title = "{Neutron star parameter estimation from a NICER perspective}",
       school = {University of Amsterdam, Netherlands},
         year = 2019,
        month = jan,
       adsurl = {https://ui.adsabs.harvard.edu/abs/2019PhDT........97R}
}

@ARTICLE{hoogkamer_cross-comparison_2025,
       author = {{Hoogkamer}, Mariska and {Kini}, Yves and {Salmi}, Tuomo and {Watts}, Anna L. and {Buchner}, Johannes},
        title = "{Cross-Comparison of Sampling Algorithms for Pulse Profile Modeling of PSR J0740+6620}",
      journal = {arXiv e-prints},
         year = 2025,
        month = feb,
          eid = {arXiv:2502.13682},
        pages = {arXiv:2502.13682},
          doi = {10.48550/arXiv.2502.13682},
archivePrefix = {arXiv},
       eprint = {2502.13682},
 primaryClass = {astro-ph.HE},
       adsurl = {https://ui.adsabs.harvard.edu/abs/2025arXiv250213682H}
}

@ARTICLE{Das_PPM_GRMHD_2025,
       author = {{Das}, Pushpita and {Salmi}, Tuomo and {Davelaar}, Jordy and {Porth}, Oliver and {Watts}, Anna},
        title = "{Pulse Profiles of Accreting Neutron Stars from GRMHD Simulations}",
      journal = {arXiv e-prints},
         year = 2024,
        month = nov,
          eid = {arXiv:2411.16528},
        pages = {arXiv:2411.16528},
          doi = {10.48550/arXiv.2411.16528},
archivePrefix = {arXiv},
       eprint = {2411.16528},
 primaryClass = {astro-ph.HE},
       adsurl = {https://ui.adsabs.harvard.edu/abs/2024arXiv241116528D}
}

@ARTICLE{chang_diffusive_2003,
       author = {{Chang}, Philip and {Bildsten}, Lars},
        title = "{Diffusive Nuclear Burning in Neutron Star Envelopes}",
      journal = {\apj},
         year = 2003,
        month = mar,
       volume = {585},
       number = {1},
        pages = {464-474},
          doi = {10.1086/345551},
archivePrefix = {arXiv},
       eprint = {astro-ph/0210218},
 primaryClass = {astro-ph},
       adsurl = {https://ui.adsabs.harvard.edu/abs/2003ApJ...585..464C}
}

@ARTICLE{chang_evolution_2004,
       author = {{Chang}, Philip and {Bildsten}, Lars},
        title = "{Evolution of Young Neutron Star Envelopes}",
      journal = {\apj},
         year = 2004,
        month = apr,
       volume = {605},
       number = {2},
        pages = {830-839},
          doi = {10.1086/382271},
archivePrefix = {arXiv},
       eprint = {astro-ph/0312589},
 primaryClass = {astro-ph},
       adsurl = {https://ui.adsabs.harvard.edu/abs/2004ApJ...605..830C}
}

@ARTICLE{reardon_neutron_2024,
       author = {{Reardon}, Daniel J. and {Bailes}, Matthew and {Shannon}, Ryan M. and {Flynn}, Chris and {Askew}, Jacob and {Bhat}, N.~D. Ramesh and {Chen}, Zu-Cheng and {Cury{\l}o}, Ma{\l}gorzata and {Feng}, Yi and {Hobbs}, George B. and {Kapur}, Agastya and {Kerr}, Matthew and {Liu}, Xiaojin and {Manchester}, Richard N. and {Mandow}, Rami and {Mishra}, Saurav and {Russell}, Christopher J. and {Shamohammadi}, Mohsen and {Zhang}, Lei and {Zic}, Andrew},
        title = "{The Neutron Star Mass, Distance, and Inclination from Precision Timing of the Brilliant Millisecond Pulsar J0437-4715}",
      journal = {\apjl},
         year = 2024,
        month = aug,
       volume = {971},
       number = {1},
          eid = {L18},
        pages = {L18},
          doi = {10.3847/2041-8213/ad614a},
archivePrefix = {arXiv},
       eprint = {2407.07132},
 primaryClass = {astro-ph.HE},
       adsurl = {https://ui.adsabs.harvard.edu/abs/2024ApJ...971L..18R}
}

@ARTICLE{alpar_new_1982,
       author = {{Alpar}, M.~A. and {Cheng}, A.~F. and {Ruderman}, M.~A. and {Shaham}, J.},
        title = "{A new class of radio pulsars}",
      journal = {\nat},
         year = 1982,
        month = dec,
       volume = {300},
       number = {5894},
        pages = {728-730},
          doi = {10.1038/300728a0},
       adsurl = {https://ui.adsabs.harvard.edu/abs/1982Natur.300..728A}
}

@ARTICLE{Kalapotharakos_geometry_2009,
       author = {{Kalapotharakos}, Constantinos and {Wadiasingh}, Zorawar and {Harding}, Alice K. and {Kazanas}, Demosthenes},
        title = "{The Multipolar Magnetic Field of the Millisecond Pulsar PSR J0030+0451}",
      journal = {\apj},
         year = 2021,
        month = feb,
       volume = {907},
       number = {2},
          eid = {63},
        pages = {63},
          doi = {10.3847/1538-4357/abcec0},
archivePrefix = {arXiv},
       eprint = {2009.08567},
 primaryClass = {astro-ph.HE},
       adsurl = {https://ui.adsabs.harvard.edu/abs/2021ApJ...907...63K}
}

@INPROCEEDINGS{watts_constraining_2019,
       author = {{Watts}, Anna L.},
        title = "{Constraining the neutron star equation of state using pulse profile modeling}",
    booktitle = {Xiamen-CUSTIPEN Workshop on the Equation of State of Dense Neutron-Rich Matter in the Era of Gravitational Wave Astronomy},
         year = 2019,
       series = {American Institute of Physics Conference Series},
       volume = {2127},
        month = jul,
    publisher = {AIP},
          eid = {020008},
        pages = {020008},
          doi = {10.1063/1.5117798},
archivePrefix = {arXiv},
       eprint = {1904.07012},
 primaryClass = {astro-ph.HE},
       adsurl = {https://ui.adsabs.harvard.edu/abs/2019AIPC.2127b0008W}
}

@ARTICLE{harding_pulsar_2001,
       author = {{Harding}, Alice K. and {Muslimov}, Alexander G.},
        title = "{Pulsar Polar Cap Heating and Surface Thermal X-Ray Emission. I. Curvature Radiation Pair Fronts}",
      journal = {\apj},
         year = 2001,
        month = aug,
       volume = {556},
       number = {2},
        pages = {987-1001},
          doi = {10.1086/321589},
archivePrefix = {arXiv},
       eprint = {astro-ph/0104146},
 primaryClass = {astro-ph},
       adsurl = {https://ui.adsabs.harvard.edu/abs/2001ApJ...556..987H}
}

@ARTICLE{ruderman_theory_1975,
       author = {{Ruderman}, M.~A. and {Sutherland}, P.~G.},
        title = "{Theory of pulsars: polar gaps, sparks, and coherent microwave radiation.}",
      journal = {\apj},
         year = 1975,
        month = feb,
       volume = {196},
        pages = {51-72},
          doi = {10.1086/153393},
       adsurl = {https://ui.adsabs.harvard.edu/abs/1975ApJ...196...51R}
}

@ARTICLE{arons_pair_1981,
       author = {{Arons}, J.},
        title = "{Pair creation above pulsar polar caps - Steady flow in the surface acceleration zone and polar CAP X-ray emission}",
      journal = {\apj},
         year = 1981,
        month = sep,
       volume = {248},
        pages = {1099-1116},
          doi = {10.1086/159239},
       adsurl = {https://ui.adsabs.harvard.edu/abs/1981ApJ...248.1099A}
}

@ARTICLE{harding_pulsar_2002,
       author = {{Harding}, Alice K. and {Muslimov}, Alexander G.},
        title = "{Pulsar Polar Cap Heating and Surface Thermal X-Ray Emission. II. Inverse Compton Radiation Pair Fronts}",
      journal = {\apj},
         year = 2002,
        month = apr,
       volume = {568},
       number = {2},
        pages = {862-877},
          doi = {10.1086/338985},
archivePrefix = {arXiv},
       eprint = {astro-ph/0112392},
 primaryClass = {astro-ph},
       adsurl = {https://ui.adsabs.harvard.edu/abs/2002ApJ...568..862H}
}

@ARTICLE{PhysRevX.9.011001,
       author = {{Abbott}, B.~P. and {Abbott}, R. and {Abbott}, T.~D. and {Acernese}, F. and {Ackley}, K. and {Adams}, C. and {Adams}, T. and {Addesso}, P. and {Adhikari}, R.~X. and {Adya}, V.~B. and {Affeldt}, C. and {Agarwal}, B. and {Agathos}, M. and {Agatsuma}, K. and {Aggarwal}, N. and {Aguiar}, O.~D. and {Aiello}, L. and {Ain}, A. and {Ajith}, P. and {Allen}, B. and {Allen}, G. and {Allocca}, A. and {Aloy}, M.~A. and {Altin}, P.~A. and {Amato}, A. and {Ananyeva}, A. and {Anderson}, S.~B. and {Anderson}, W.~G. and {Angelova}, S.~V. and {Antier}, S. and {Appert}, S. and {Arai}, K. and {Araya}, M.~C. and {Areeda}, J.~S. and {Ar{\`e}ne}, M. and {Arnaud}, N. and {Arun}, K.~G. and {Ascenzi}, S. and {Ashton}, G. and {Ast}, M. and {Aston}, S.~M. and {Astone}, P. and {Atallah}, D.~V. and {Aubin}, F. and {Aufmuth}, P. and {Aulbert}, C. and {AultONeal}, K. and {Austin}, C. and {Avila-Alvarez}, A. and {Babak}, S. and {Bacon}, P. and {Badaracco}, F. and {Bader}, M.~K.~M. and {Bae}, S. and {Baker}, P.~T. and {Baldaccini}, F. and {Ballardin}, G. and {Ballmer}, S.~W. and {Banagiri}, S. and {Barayoga}, J.~C. and {Barclay}, S.~E. and {Barish}, B.~C. and {Barker}, D. and {Barkett}, K. and {Barnum}, S. and {Barone}, F. and {Barr}, B. and {Barsotti}, L. and {Barsuglia}, M. and {Barta}, D. and {Bartlett}, J. and {Bartos}, I. and {Bassiri}, R. and {Basti}, A. and {Batch}, J.~C. and {Bawaj}, M. and {Bayley}, J.~C. and {Bazzan}, M. and {B{\'e}csy}, B. and {Beer}, C. and {Bejger}, M. and {Belahcene}, I. and {Bell}, A.~S. and {Beniwal}, D. and {Bensch}, M. and {Berger}, B.~K. and {Bergmann}, G. and {Bernuzzi}, S. and {Bero}, J.~J. and {Berry}, C.~P.~L. and {Bersanetti}, D. and {Bertolini}, A. and {Betzwieser}, J. and {Bhandare}, R. and {Bilenko}, I.~A. and {Bilgili}, S.~A. and {Billingsley}, G. and {Billman}, C.~R. and {Birch}, J. and {Birney}, R. and {Birnholtz}, O. and {Biscans}, S. and {Biscoveanu}, S. and {Bisht}, A. and {Bitossi}, M. and {Bizouard}, M.~A. and {Blackburn}, J.~K. and {Blackman}, J. and {Blair}, C.~D. and {Blair}, D.~G. and {Blair}, R.~M. and {Bloemen}, S. and {Bock}, O. and {Bode}, N. and {Boer}, M. and {Boetzel}, Y. and {Bogaert}, G. and {Bohe}, A. and {Bondu}, F. and {Bonilla}, E. and {Bonnand}, R. and {Booker}, P. and {Boom}, B.~A. and {Booth}, C.~D. and {Bork}, R. and {Boschi}, V. and {Bose}, S. and {Bossie}, K. and {Bossilkov}, V. and {Bosveld}, J. and {Bouffanais}, Y. and {Bozzi}, A. and {Bradaschia}, C. and {Brady}, P.~R. and {Bramley}, A. and {Branchesi}, M. and {Brau}, J.~E. and {Briant}, T. and {Brighenti}, F. and {Brillet}, A. and {Brinkmann}, M. and {Brisson}, V. and {Brockill}, P. and {Brooks}, A.~F. and {Brown}, D.~D. and {Brunett}, S. and {Buchanan}, C.~C. and {Buikema}, A. and {Bulik}, T. and {Bulten}, H.~J. and {Buonanno}, A. and {Buskulic}, D. and {Buy}, C. and {Byer}, R.~L. and {Cabero}, M. and {Cadonati}, L. and {Cagnoli}, G. and {Cahillane}, C. and {Bustillo}, J. Calder{\'o}n and {Callister}, T.~A. and {Calloni}, E. and {Camp}, J.~B. and {Canepa}, M. and {Canizares}, P. and {Cannon}, K.~C. and {Cao}, H. and {Cao}, J. and {Capano}, C.~D. and {Capocasa}, E. and {Carbognani}, F. and {Caride}, S. and {Carney}, M.~F. and {Carullo}, G. and {Diaz}, J. Casanueva and {Casentini}, C. and {Caudill}, S. and {Cavagli{\`a}}, M. and {Cavalier}, F. and {Cavalieri}, R. and {Cella}, G. and {Cepeda}, C.~B. and {Cerd{\'a}-Dur{\'a}n}, P. and {Cerretani}, G. and {Cesarini}, E. and {Chaibi}, O. and {Chamberlin}, S.~J. and {Chan}, M. and {Chao}, S. and {Charlton}, P. and {Chase}, E. and {Chassande-Mottin}, E. and {Chatterjee}, D. and {Chatziioannou}, K. and {Cheeseboro}, B.~D. and {Chen}, H.~Y. and {Chen}, X. and {Chen}, Y. and {Cheng}, H. -P. and {Chia}, H.~Y. and {Chincarini}, A.},
        title = "{Properties of the Binary Neutron Star Merger GW170817}",
      journal = {Physical Review X},
         year = 2019,
        month = jan,
       volume = {9},
       number = {1},
          eid = {011001},
        pages = {011001},
          doi = {10.1103/PhysRevX.9.011001},
archivePrefix = {arXiv},
       eprint = {1805.11579},
 primaryClass = {gr-qc},
       adsurl = {https://ui.adsabs.harvard.edu/abs/2019PhRvX...9a1001A}
}

@ARTICLE{abbott_gw170817_2018,
       author = {{Abbott}, B.~P. and {Abbott}, R. and {Abbott}, T.~D. and {Acernese}, F. and {Ackley}, K. and {Adams}, C. and {Adams}, T. and {Addesso}, P. and {Adhikari}, R.~X. and {Adya}, V.~B. and {Affeldt}, C. and {Agarwal}, B. and {Agathos}, M. and {Agatsuma}, K. and {Aggarwal}, N. and {Aguiar}, O.~D. and {Aiello}, L. and {Ain}, A. and {Ajith}, P. and {Allen}, B. and {Allen}, G. and {Allocca}, A. and {Aloy}, M.~A. and {Altin}, P.~A. and {Amato}, A. and {Ananyeva}, A. and {Anderson}, S.~B. and {Anderson}, W.~G. and {Angelova}, S.~V. and {Antier}, S. and {Appert}, S. and {Arai}, K. and {Araya}, M.~C. and {Areeda}, J.~S. and {Ar{\`e}ne}, M. and {Arnaud}, N. and {Arun}, K.~G. and {Ascenzi}, S. and {Ashton}, G. and {Ast}, M. and {Aston}, S.~M. and {Astone}, P. and {Atallah}, D.~V. and {Aubin}, F. and {Aufmuth}, P. and {Aulbert}, C. and {AultONeal}, K. and {Austin}, C. and {Avila-Alvarez}, A. and {Babak}, S. and {Bacon}, P. and {Badaracco}, F. and {Bader}, M.~K.~M. and {Bae}, S. and {Baker}, P.~T. and {Baldaccini}, F. and {Ballardin}, G. and {Ballmer}, S.~W. and {Banagiri}, S. and {Barayoga}, J.~C. and {Barclay}, S.~E. and {Barish}, B.~C. and {Barker}, D. and {Barkett}, K. and {Barnum}, S. and {Barone}, F. and {Barr}, B. and {Barsotti}, L. and {Barsuglia}, M. and {Barta}, D. and {Bartlett}, J. and {Bartos}, I. and {Bassiri}, R. and {Basti}, A. and {Batch}, J.~C. and {Bawaj}, M. and {Bayley}, J.~C. and {Bazzan}, M. and {B{\'e}csy}, B. and {Beer}, C. and {Bejger}, M. and {Belahcene}, I. and {Bell}, A.~S. and {Beniwal}, D. and {Bensch}, M. and {Berger}, B.~K. and {Bergmann}, G. and {Bernuzzi}, S. and {Bero}, J.~J. and {Berry}, C.~P.~L. and {Bersanetti}, D. and {Bertolini}, A. and {Betzwieser}, J. and {Bhandare}, R. and {Bilenko}, I.~A. and {Bilgili}, S.~A. and {Billingsley}, G. and {Billman}, C.~R. and {Birch}, J. and {Birney}, R. and {Birnholtz}, O. and {Biscans}, S. and {Biscoveanu}, S. and {Bisht}, A. and {Bitossi}, M. and {Bizouard}, M.~A. and {Blackburn}, J.~K. and {Blackman}, J. and {Blair}, C.~D. and {Blair}, D.~G. and {Blair}, R.~M. and {Bloemen}, S. and {Bock}, O. and {Bode}, N. and {Boer}, M. and {Boetzel}, Y. and {Bogaert}, G. and {Bohe}, A. and {Bondu}, F. and {Bonilla}, E. and {Bonnand}, R. and {Booker}, P. and {Boom}, B.~A. and {Booth}, C.~D. and {Bork}, R. and {Boschi}, V. and {Bose}, S. and {Bossie}, K. and {Bossilkov}, V. and {Bosveld}, J. and {Bouffanais}, Y. and {Bozzi}, A. and {Bradaschia}, C. and {Brady}, P.~R. and {Bramley}, A. and {Branchesi}, M. and {Brau}, J.~E. and {Briant}, T. and {Brighenti}, F. and {Brillet}, A. and {Brinkmann}, M. and {Brisson}, V. and {Brockill}, P. and {Brooks}, A.~F. and {Brown}, D.~D. and {Brunett}, S. and {Buchanan}, C.~C. and {Buikema}, A. and {Bulik}, T. and {Bulten}, H.~J. and {Buonanno}, A. and {Buskulic}, D. and {Buy}, C. and {Byer}, R.~L. and {Cabero}, M. and {Cadonati}, L. and {Cagnoli}, G. and {Cahillane}, C. and {Calder{\'o}n Bustillo}, J. and {Callister}, T.~A. and {Calloni}, E. and {Camp}, J.~B. and {Canepa}, M. and {Canizares}, P. and {Cannon}, K.~C. and {Cao}, H. and {Cao}, J. and {Capano}, C.~D. and {Capocasa}, E. and {Carbognani}, F. and {Caride}, S. and {Carney}, M.~F. and {Carullo}, G. and {Casanueva Diaz}, J. and {Casentini}, C. and {Caudill}, S. and {Cavagli{\`a}}, M. and {Cavalier}, F. and {Cavalieri}, R. and {Cella}, G. and {Cepeda}, C.~B. and {Cerd{\'a}-Dur{\'a}n}, P. and {Cerretani}, G. and {Cesarini}, E. and {Chaibi}, O. and {Chamberlin}, S.~J. and {Chan}, M. and {Chao}, S. and {Charlton}, P. and {Chase}, E. and {Chassande-Mottin}, E. and {Chatterjee}, D. and {Chatziioannou}, K. and {Cheeseboro}, B.~D. and {Chen}, H.~Y. and {Chen}, X. and {Chen}, Y. and {Cheng}, H. -P. and {Chia}, H.~Y. and {Chincarini}, A.},
        title = "{GW170817: Measurements of Neutron Star Radii and Equation of State}",
      journal = {\prl},
         year = 2018,
        month = oct,
       volume = {121},
       number = {16},
          eid = {161101},
        pages = {161101},
          doi = {10.1103/PhysRevLett.121.161101},
archivePrefix = {arXiv},
       eprint = {1805.11581},
 primaryClass = {gr-qc},
       adsurl = {https://ui.adsabs.harvard.edu/abs/2018PhRvL.121p1101A}
}

@ARTICLE{abbott_gw190425_2020,
       author = {{Abbott}, B.~P. and {Abbott}, R. and {Abbott}, T.~D. and {Abraham}, S. and {Acernese}, F. and {Ackley}, K. and {Adams}, C. and {Adhikari}, R.~X. and {Adya}, V.~B. and {Affeldt}, C. and {Agathos}, M. and {Agatsuma}, K. and {Aggarwal}, N. and {Aguiar}, O.~D. and {Aiello}, L. and {Ain}, A. and {Ajith}, P. and {Allen}, G. and {Allocca}, A. and {Aloy}, M.~A. and {Altin}, P.~A. and {Amato}, A. and {Anand}, S. and {Ananyeva}, A. and {Anderson}, S.~B. and {Anderson}, W.~G. and {Angelova}, S.~V. and {Antier}, S. and {Appert}, S. and {Arai}, K. and {Araya}, M.~C. and {Areeda}, J.~S. and {Ar{\`e}ne}, M. and {Arnaud}, N. and {Aronson}, S.~M. and {Arun}, K.~G. and {Ascenzi}, S. and {Ashton}, G. and {Aston}, S.~M. and {Astone}, P. and {Aubin}, F. and {Aufmuth}, P. and {AultONeal}, K. and {Austin}, C. and {Avendano}, V. and {Avila-Alvarez}, A. and {Babak}, S. and {Bacon}, P. and {Badaracco}, F. and {Bader}, M.~K.~M. and {Bae}, S. and {Baird}, J. and {Baker}, P.~T. and {Baldaccini}, F. and {Ballardin}, G. and {Ballmer}, S.~W. and {Bals}, A. and {Banagiri}, S. and {Barayoga}, J.~C. and {Barbieri}, C. and {Barclay}, S.~E. and {Barish}, B.~C. and {Barker}, D. and {Barkett}, K. and {Barnum}, S. and {Barone}, F. and {Barr}, B. and {Barsotti}, L. and {Barsuglia}, M. and {Barta}, D. and {Bartlett}, J. and {Bartos}, I. and {Bassiri}, R. and {Basti}, A. and {Bawaj}, M. and {Bayley}, J.~C. and {Baylor}, A.~C. and {Bazzan}, M. and {B{\'e}csy}, B. and {Bejger}, M. and {Belahcene}, I. and {Bell}, A.~S. and {Beniwal}, D. and {Benjamin}, M.~G. and {Berger}, B.~K. and {Bergmann}, G. and {Bernuzzi}, S. and {Berry}, C.~P.~L. and {Bersanetti}, D. and {Bertolini}, A. and {Betzwieser}, J. and {Bhandare}, R. and {Bidler}, J. and {Biggs}, E. and {Bilenko}, I.~A. and {Bilgili}, S.~A. and {Billingsley}, G. and {Birney}, R. and {Birnholtz}, O. and {Biscans}, S. and {Bischi}, M. and {Biscoveanu}, S. and {Bisht}, A. and {Bitossi}, M. and {Bizouard}, M.~A. and {Blackburn}, J.~K. and {Blackman}, J. and {Blair}, C.~D. and {Blair}, D.~G. and {Blair}, R.~M. and {Bloemen}, S. and {Bobba}, F. and {Bode}, N. and {Boer}, M. and {Boetzel}, Y. and {Bogaert}, G. and {Bondu}, F. and {Bonnand}, R. and {Booker}, P. and {Boom}, B.~A. and {Bork}, R. and {Boschi}, V. and {Bose}, S. and {Bossilkov}, V. and {Bosveld}, J. and {Bouffanais}, Y. and {Bozzi}, A. and {Bradaschia}, C. and {Brady}, P.~R. and {Bramley}, A. and {Branchesi}, M. and {Brau}, J.~E. and {Breschi}, M. and {Briant}, T. and {Briggs}, J.~H. and {Brighenti}, F. and {Brillet}, A. and {Brinkmann}, M. and {Brockill}, P. and {Brooks}, A.~F. and {Brooks}, J. and {Brown}, D.~D. and {Brunett}, S. and {Buikema}, A. and {Bulik}, T. and {Bulten}, H.~J. and {Buonanno}, A. and {Buskulic}, D. and {Buy}, C. and {Byer}, R.~L. and {Cabero}, M. and {Cadonati}, L. and {Cagnoli}, G. and {Cahillane}, C. and {Calder{\'o}n Bustillo}, J. and {Callister}, T.~A. and {Calloni}, E. and {Camp}, J.~B. and {Campbell}, W.~A. and {Canepa}, M. and {Cannon}, K.~C. and {Cao}, H. and {Cao}, J. and {Carapella}, G. and {Carbognani}, F. and {Caride}, S. and {Carney}, M.~F. and {Carullo}, G. and {Casanueva Diaz}, J. and {Casentini}, C. and {Caudill}, S. and {Cavagli{\`a}}, M. and {Cavalier}, F. and {Cavalieri}, R. and {Cella}, G. and {Cerd{\'a}-Dur{\'a}n}, P. and {Cesarini}, E. and {Chaibi}, O. and {Chakravarti}, K. and {Chamberlin}, S.~J. and {Chan}, M. and {Chao}, S. and {Charlton}, P. and {Chase}, E.~A. and {Chassande-Mottin}, E. and {Chatterjee}, D. and {Chaturvedi}, M. and {Chatziioannou}, K. and {Cheeseboro}, B.~D. and {Chen}, H.~Y. and {Chen}, X. and {Chen}, Y. and {Cheng}, H. -P. and {Cheong}, C.~K. and {Chia}, H.~Y. and {Chiadini}, F. and {Chincarini}, A. and {Chiummo}, A. and {Cho}, G. and {Cho}, H.~S.},
        title = "{GW190425: Observation of a Compact Binary Coalescence with Total Mass {\ensuremath{\sim}} 3.4 M$_{{\ensuremath{\odot}}}$}",
      journal = {\apjl},
         year = 2020,
        month = mar,
       volume = {892},
       number = {1},
          eid = {L3},
        pages = {L3},
          doi = {10.3847/2041-8213/ab75f5},
archivePrefix = {arXiv},
       eprint = {2001.01761},
 primaryClass = {astro-ph.HE},
       adsurl = {https://ui.adsabs.harvard.edu/abs/2020ApJ...892L...3A}
}

@ARTICLE{pavlov_mass--radius_1997,
       author = {{Pavlov}, G.~G. and {Zavlin}, V.~E.},
        title = "{Mass-to-Radius Ratio for the Millisecond Pulsar J0437-4715}",
      journal = {\apjl},
         year = 1997,
        month = nov,
       volume = {490},
       number = {1},
        pages = {L91-L94},
          doi = {10.1086/311007},
archivePrefix = {arXiv},
       eprint = {astro-ph/9709255},
 primaryClass = {astro-ph},
       adsurl = {https://ui.adsabs.harvard.edu/abs/1997ApJ...490L..91P}
}

@ARTICLE{zavlin_soft_1997,
       author = {{Zavlin}, V.~E. and {Pavlov}, G.~G.},
        title = "{Soft X-rays from polar caps of the millisecond pulsar J0437-4715}",
      journal = {\aap},
         year = 1998,
        month = jan,
       volume = {329},
        pages = {583-598},
          doi = {10.48550/arXiv.astro-ph/9708101},
archivePrefix = {arXiv},
       eprint = {astro-ph/9708101},
 primaryClass = {astro-ph},
       adsurl = {https://ui.adsabs.harvard.edu/abs/1998A&A...329..583Z}
}

@ARTICLE{Yang_inclination_2023,
       author = {{Yang}, Hao-ran and {Li}, Xiang-dong},
        title = "{Magnetic Inclination Evolution of Accreting Neutron Stars in Intermediate/Low-mass X-Ray Binaries}",
      journal = {\apj},
         year = 2023,
        month = mar,
       volume = {945},
       number = {1},
          eid = {2},
        pages = {2},
          doi = {10.3847/1538-4357/acba09},
archivePrefix = {arXiv},
       eprint = {2302.11243},
 primaryClass = {astro-ph.HE},
       adsurl = {https://ui.adsabs.harvard.edu/abs/2023ApJ...945....2Y}
}

@ARTICLE{bogdanov_nearest_2013,
       author = {{Bogdanov}, Slavko},
        title = "{The Nearest Millisecond Pulsar Revisited with XMM-Newton: Improved Mass-radius Constraints for PSR J0437-4715}",
      journal = {\apj},
         year = 2013,
        month = jan,
       volume = {762},
       number = {2},
          eid = {96},
        pages = {96},
          doi = {10.1088/0004-637X/762/2/96},
archivePrefix = {arXiv},
       eprint = {1211.6113},
 primaryClass = {astro-ph.HE},
       adsurl = {https://ui.adsabs.harvard.edu/abs/2013ApJ...762...96B}
}

@ARTICLE{ho_atmospheres_2001,
       author = {{Ho}, Wynn C.~G. and {Lai}, Dong},
        title = "{Atmospheres and spectra of strongly magnetized neutron stars}",
      journal = {\mnras},
         year = 2001,
        month = nov,
       volume = {327},
       number = {4},
        pages = {1081-1096},
          doi = {10.1046/j.1365-8711.2001.04801.x},
archivePrefix = {arXiv},
       eprint = {astro-ph/0104199},
 primaryClass = {astro-ph},
       adsurl = {https://ui.adsabs.harvard.edu/abs/2001MNRAS.327.1081H}
}

@ARTICLE{rutherford_constraining_2024,
       author = {{Rutherford}, Nathan and {Mendes}, Melissa and {Svensson}, Isak and {Schwenk}, Achim and {Watts}, Anna L. and {Hebeler}, Kai and {Keller}, Jonas and {Prescod-Weinstein}, Chanda and {Choudhury}, Devarshi and {Raaijmakers}, Geert and {Salmi}, Tuomo and {Timmerman}, Patrick and {Vinciguerra}, Serena and {Guillot}, Sebastien and {Lattimer}, James M.},
        title = "{Constraining the Dense Matter Equation of State with New NICER Mass{\textendash}Radius Measurements and New Chiral Effective Field Theory Inputs}",
      journal = {\apjl},
         year = 2024,
        month = aug,
       volume = {971},
       number = {1},
          eid = {L19},
        pages = {L19},
          doi = {10.3847/2041-8213/ad5f02},
archivePrefix = {arXiv},
       eprint = {2407.06790},
 primaryClass = {astro-ph.HE},
       adsurl = {https://ui.adsabs.harvard.edu/abs/2024ApJ...971L..19R}
}

@ARTICLE{drischler_chiral_2019,
       author = {{Drischler}, C. and {Hebeler}, K. and {Schwenk}, A.},
        title = "{Chiral Interactions up to Next-to-Next-to-Next-to-Leading Order and Nuclear Saturation}",
      journal = {\prl},
         year = 2019,
        month = feb,
       volume = {122},
       number = {4},
          eid = {042501},
        pages = {042501},
          doi = {10.1103/PhysRevLett.122.042501},
       adsurl = {https://ui.adsabs.harvard.edu/abs/2019PhRvL.122d2501D}
}

@ARTICLE{lattimer_equation_2016,
       author = {{Lattimer}, James M. and {Prakash}, Madappa},
        title = "{The equation of state of hot, dense matter and neutron stars}",
      journal = {\physrep},
         year = 2016,
        month = mar,
       volume = {621},
        pages = {127-164},
          doi = {10.1016/j.physrep.2015.12.005},
archivePrefix = {arXiv},
       eprint = {1512.07820},
 primaryClass = {astro-ph.SR},
       adsurl = {https://ui.adsabs.harvard.edu/abs/2016PhR...621..127L}
}

@ARTICLE{oertel_equations_2017,
       author = {{Oertel}, M. and {Hempel}, M. and {Kl{\"a}hn}, T. and {Typel}, S.},
        title = "{Equations of state for supernovae and compact stars}",
      journal = {Reviews of Modern Physics},
         year = 2017,
        month = jan,
       volume = {89},
       number = {1},
          eid = {015007},
        pages = {015007},
          doi = {10.1103/RevModPhys.89.015007},
archivePrefix = {arXiv},
       eprint = {1610.03361},
 primaryClass = {astro-ph.HE},
       adsurl = {https://ui.adsabs.harvard.edu/abs/2017RvMP...89a5007O}
}

@ARTICLE{burgio_neutron_2021,
       author = {{Burgio}, G.~F. and {Schulze}, H. -J. and {Vida{\~n}a}, I. and {Wei}, J. -B.},
        title = "{Neutron stars and the nuclear equation of state}",
      journal = {Progress in Particle and Nuclear Physics},
         year = 2021,
        month = sep,
       volume = {120},
          eid = {103879},
        pages = {103879},
          doi = {10.1016/j.ppnp.2021.103879},
archivePrefix = {arXiv},
       eprint = {2105.03747},
 primaryClass = {nucl-th},
       adsurl = {https://ui.adsabs.harvard.edu/abs/2021PrPNP.12003879B}
}

@ARTICLE{hi4pi_collaboration_hi4pi_2016,
       author = {{HI4PI Collaboration} and {Ben Bekhti}, N. and {Fl{\"o}er}, L. and {Keller}, R. and {Kerp}, J. and {Lenz}, D. and {Winkel}, B. and {Bailin}, J. and {Calabretta}, M.~R. and {Dedes}, L. and {Ford}, H.~A. and {Gibson}, B.~K. and {Haud}, U. and {Janowiecki}, S. and {Kalberla}, P.~M.~W. and {Lockman}, F.~J. and {McClure-Griffiths}, N.~M. and {Murphy}, T. and {Nakanishi}, H. and {Pisano}, D.~J. and {Staveley-Smith}, L.},
        title = "{HI4PI: A full-sky H I survey based on EBHIS and GASS}",
      journal = {\aap},
         year = 2016,
        month = oct,
       volume = {594},
          eid = {A116},
        pages = {A116},
          doi = {10.1051/0004-6361/201629178},
archivePrefix = {arXiv},
       eprint = {1610.06175},
 primaryClass = {astro-ph.GA},
       adsurl = {https://ui.adsabs.harvard.edu/abs/2016A&A...594A.116H}
}

@ARTICLE{he_correlation_2013,
       author = {{He}, C. and {Ng}, C. -Y. and {Kaspi}, V.~M.},
        title = "{The Correlation between Dispersion Measure and X-Ray Column Density from Radio Pulsars}",
      journal = {\apj},
         year = 2013,
        month = may,
       volume = {768},
       number = {1},
          eid = {64},
        pages = {64},
          doi = {10.1088/0004-637X/768/1/64},
archivePrefix = {arXiv},
       eprint = {1303.5170},
 primaryClass = {astro-ph.HE},
       adsurl = {https://ui.adsabs.harvard.edu/abs/2013ApJ...768...64H}
}

@ARTICLE{cruise_newathena_2024,
       author = {{Cruise}, Mike and {Guainazzi}, Matteo and {Aird}, James and {Carrera}, Francisco J. and {Costantini}, Elisa and {Corrales}, Lia and {Dauser}, Thomas and {Eckert}, Dominique and {Gastaldello}, Fabio and {Matsumoto}, Hironori and {Osten}, Rachel and {Petrucci}, Pierre-Olivier and {Porquet}, Delphine and {Pratt}, Gabriel W. and {Rea}, Nanda and {Reiprich}, Thomas H. and {Simionescu}, Aurora and {Spiga}, Daniele and {Troja}, Eleonora},
        title = "{The NewAthena mission concept in the context of the next decade of X-ray astronomy}",
      journal = {Nature Astronomy},
         year = 2025,
        month = jan,
       volume = {9},
        pages = {36-44},
          doi = {10.1038/s41550-024-02416-3},
archivePrefix = {arXiv},
       eprint = {2501.03100},
 primaryClass = {astro-ph.IM},
       adsurl = {https://ui.adsabs.harvard.edu/abs/2025NatAs...9...36C}
}

@ARTICLE{Li_eXTP_2025,
       author = {{Li}, Ang and {Watts}, Anna L. and {Zhang}, Guobao and {Guillot}, Sebastien and {Xu}, Yanjun and {Santangelo}, Andrea and {Zane}, Silvia and {Feng}, Hua and {Zhang}, Shuang-Nan and {Ge}, Mingyu and {Qi}, Liqiang and {Salmi}, Tuomo and {Dorsman}, Bas and {Miao}, Zhiqiang and {Tu}, Zhonghao and {Cavecchi}, Yuri and {Zhou}, Xia and {Zheng}, Xiaoping and {Wang}, Weihua and {Cheng}, Quan and {Liu}, Xuezhi and {Wei}, Yining and {Wang}, Wei and {Xu}, Yujing and {Weng}, Shanshan and {Zhu}, Weiwei and {Li}, Zhaosheng and {Shao}, Lijing and {Tuo}, Youli and {Dohi}, Akira and {Lyu}, Ming and {Liu}, Peng and {Yuan}, Jianping and {Wang}, Mingyang and {Zhang}, Wenda and {Li}, Zexi and {Tao}, Lian and {Zhang}, Liang and {Shen}, Hong and {Provid{\^e}ncia}, Constan{\c{c}}a and {Tolos}, Laura and {Patruno}, Alessandro and {Li}, Li and {Liu}, Guozhu and {Zhou}, Kai and {Chen}, Lie-Wen and {Fan}, Yizhong and {Kajino}, Toshitaka and {Lai}, Dong and {Li}, Xiangdong and {Meng}, Jie and {Tang}, Xiaodong and {Xiao}, Zhigang and {Xiong}, Shaolin and {Xu}, Renxin and {Zhou}, Shan-Gui and {Ballantyne}, David R. and {Fiorella Burgio}, G. and {Chenevez}, J{\'e}r{\^o}me and {Choudhury}, Devarshi and {Fantina}, Anthea F. and {Galloway}, Duncan K. and {Gulminelli}, Francesca and {Hebeler}, Kai and {Hoogkamer}, Mariska and {Kini}, Yves and {Kurkela}, Aleksi and {Linares}, Manuel and {Margueron}, J{\'e}r{\^o}me and {Mendes}, Melissa and {Oertel}, Micaela and {Papitto}, Alessandro and {Poutanen}, Juri and {Rea}, Nanda and {Schwenk}, Achim and {Svensson}, Isak and {Tsang}, David and {Vuorinen}, Aleksi and {Andersson}, Nils and {Miller}, M. Coleman and {Rezzolla}, Luciano and {Stone}, Jirina R. and {Thomas}, Anthony W.},
        title = "{Dense Matter in Neutron Stars with eXTP}",
      journal = {arXiv e-prints},
         year = 2025,
        month = jun,
          eid = {arXiv:2506.08104},
        pages = {arXiv:2506.08104},
archivePrefix = {arXiv},
       eprint = {2506.08104},
 primaryClass = {astro-ph.HE},
       adsurl = {https://ui.adsabs.harvard.edu/abs/2025arXiv250608104L}
}

@article{marshall_concordance_2021,
       author = {{Marshall}, Herman L. and {Chen}, Yang and {Drake}, Jeremy J. and {Guainazzi}, Matteo and {Kashyap}, Vinay L. and {Meng}, Xiao-Li and {Plucinsky}, Paul P. and {Ratzlaff}, Peter and {van Dyk}, David A. and {Wang}, Xufei},
        title = "{Concordance: In-flight Calibration of X-Ray Telescopes without Absolute References}",
      journal = {\aj},
         year = 2021,
        month = dec,
       volume = {162},
       number = {6},
          eid = {254},
        pages = {254},
          doi = {10.3847/1538-3881/ac230a},
archivePrefix = {arXiv},
       eprint = {2108.13476},
 primaryClass = {astro-ph.IM},
       adsurl = {https://ui.adsabs.harvard.edu/abs/2021AJ....162..254M}
}

@ARTICLE{ska_2025_densematter,
       author = {{Basu}, A. and {Graber}, V. and {Lower}, M.~E. and {Antonelli}, M. and {Antonopoulou}, D. and {Bagchi}, M. and {Char}, P. and {Freire}, P.~C.~C. and {Haskell}, B. and {Hu}, H. and {Jones}, D.~I. and {Mukhopadhyay}, B. and {Oertel}, M. and {Rea}, N. and {Sagun}, V. and {Shaw}, B. and {Singha}, J. and {Stappers}, B.~W. and {Thongmeearkom}, T. and {Watts}, A.~L. and {Weltevrede}, P. and {SKA Pulsar Science Working Group}},
        title = "{Probing neutron star interiors and the properties of cold ultra-dense matter with the SKA}",
      journal = {Accepted for publication in OJA},
         year = 2025,
          eid = {},
        pages = {},
}

@MISC{2014ascl.soft08004N,
       author = {{Nasa High Energy Astrophysics Science Archive Research Center (HEASARC)}},
        title = "{HEAsoft: Unified Release of FTOOLS and XANADU}",
 howpublished = {Astrophysics Source Code Library, record ascl:1408.004},
         year = 2014,
        month = aug,
          eid = {ascl:1408.004},
        pages = {ascl:1408.004},
archivePrefix = {ascl},
       eprint = {1408.004},
       adsurl = {https://ui.adsabs.harvard.edu/abs/2014ascl.soft08004N}
}

@INPROCEEDINGS{2004ASPC..314..759G,
       author = {{Gabriel}, C. and {Denby}, M. and {Fyfe}, D.~J. and {Hoar}, J. and {Ibarra}, A. and {Ojero}, E. and {Osborne}, J. and {Saxton}, R.~D. and {Lammers}, U. and {Vacanti}, G.},
        title = "{The XMM-Newton SAS - Distributed Development and Maintenance of a Large Science Analysis System: A Critical Analysis}",
    booktitle = {Astronomical Data Analysis Software and Systems (ADASS) XIII},
         year = 2004,
       editor = {{Ochsenbein}, Francois and {Allen}, Mark G. and {Egret}, Daniel},
       series = {Astronomical Society of the Pacific Conference Series},
       volume = {314},
        month = jul,
        pages = {759},
       adsurl = {https://ui.adsabs.harvard.edu/abs/2004ASPC..314..759G}
}
\bibliographystyle{aasjournal}

\end{document}